%% file: main.tex
\documentclass[12pt]{article}

\usepackage[letterpaper,margin=25mm]{geometry}
\usepackage[T1]{fontenc}
\usepackage{lmodern}
\usepackage{graphicx}
\usepackage{microtype}
\usepackage{amsmath,amssymb}
\usepackage[round]{natbib}
\usepackage{url}
\usepackage{xr-hyper}
\usepackage[colorlinks,bookmarksopen,bookmarksnumbered,citecolor=blue!60!black,urlcolor=blue!60!black,linkcolor=black]{hyperref}
\usepackage{lscape}
\usepackage[section]{placeins}
\usepackage{lineno}
\usepackage{threeparttable}
\usepackage{booktabs}
\usepackage{setspace}
\usepackage{tikz}
\usetikzlibrary{arrows.meta,shapes.geometric,calc}
\usepackage{xcolor}
\usepackage[disable]{todonotes}

\newif\ifdraftaids
\draftaidsfalse
\makeatletter
\newcommand{\tdmark}[1]{\ifdraftaids\textsuperscript{\textcolor{red}{\textbf{[#1]}}}%
	\@ifundefined{tdseen@#1}{\label{td:#1}\global\@namedef{tdseen@#1}{}}{}\fi}
\makeatother

\newcommand{\lnof}[1]{\pgfmathparse{2.302585*#1}\pgfmathprintnumber[fixed,precision=2]{\pgfmathresult}}
\newcommand{\pctof}[1]{\pgfmathparse{100*#1}\pgfmathprintnumber[fixed,precision=1]{\pgfmathresult}}

\input{analysis/out/numbers.tex}

\title{A kinematic rupture generator for multi-segment fault geometries}

\author{B. A. Bradley$^{1}$ and J. A. M. Faulkner$^{1}$\\[4pt]
	\normalsize $^{1}$Department of Civil and Environmental Engineering,\\
	\normalsize University of Canterbury, Christchurch, New Zealand\\
	\normalsize Corresponding author: brendon.bradley@canterbury.ac.nz}

\date{}

\begin{document}
\setlength{\emergencystretch}{3em}

\maketitle

\input{sections/00-abstract}

\onehalfspacing

\ifdraftaids
	\newpage
	\noindent\textbf{\textcolor{red}{Draft aids (removed in the submitted version):}} table of contents, and the size of each section in Earthquake Spectra publication units.
	\tableofcontents
	\input{draft/length_summary}
\fi

\newpage

\input{sections/01-introduction}

\input{sections/02-usecase}

\input{sections/03-framework}

\input{sections/04-srcmod-data}

\input{sections/05-moment-partitioning}

\input{sections/06-rupture-path}

\input{sections/07-jumping-and-delay}

\input{sections/08-procedure-and-uncertainty}

\input{sections/09-demonstration}

\input{sections/11-conclusions}

\input{sections/12-data-and-resources}

\input{sections/13-acknowledgements}

\bibliographystyle{abbrvnat}
\bibliography{paper}

\ifdraftaids
	\clearpage
	\input{sections/99-outstanding}
\fi

\end{document}

\maketitle

\noindent
This supplement provides supporting material for the manuscript
\emph{A kinematic rupture generator for multi-segment fault geometries}.
Section~\ref{sup:srcmod} documents the extraction of the calibration dataset from the SRCMOD database of finite-fault rupture models \citep{mai_srcmod_2014}.
Section~\ref{sup:theory} develops the scaling arguments that underlie the moment-partitioning model of manuscript Section~5, together with the calibration of the apportionment over the pooled SRCMOD population and the development of its stochastic perturbation.
Sections~\ref{sup:rupturePath} and~\ref{sup:jumping} give supporting detail for the rupture-path model of manuscript Section~6 and the jump location and time-delay model of manuscript Section~7, respectively.
Section~\ref{sup:stochastic} shows stochastic realizations of the rupture generated in manuscript Section~9.

\tableofcontents

\newpage

\section{Extraction of the calibration dataset from SRCMOD}
\label{sup:srcmod}

Table~1 of the manuscript summarizes the subsets of the SRCMOD database used to calibrate each aspect of the framework, and this section documents how they were obtained.
SRCMOD reports each plane of a source model as a ``segment''.
In the terminology of the manuscript (Section~3) each such plane is a \emph{component}, and a \emph{segment} is the physically distinct structure that one or more components represent; that terminology is used throughout this supplement.

\subsection{Source of the data and version used}

The calibration and evaluation presented in the manuscript draw upon the SRCMOD database of published finite-fault rupture models \citep{mai_srcmod_2014}.
All analysis was performed on the MATLAB-format models of the distribution dated 2~December~2024 \citep{mai_srcmod_2014}.
The retrieved models are redistributed, with attribution, in the archive described in the Data and Resources section of the manuscript, so that the analysis can be repeated on the identical input.
Of the \dataCatalogueModels{} event tags listed in the catalog accompanying that distribution, \dataTotalModels{} contain model data.

\subsection{Fields extracted from the source models}

Each model is read from the MATLAB distribution by the parsing code released with this paper.
Table~\ref{tab:sup-fields} lists every field, with its units, and distinguishes quantities read directly from those derived.
Where a field is reported under more than one name across contributed models, the alternatives are given together and the first present is used.

\begin{table}[htb!]
	\centering
	\caption{Fields read from each SRCMOD source model, and the quantities derived from them. Component-level fields carry a \texttt{segN} prefix in models that report more than one component and none otherwise.}
	\label{tab:sup-fields}
	\footnotesize
	\input{analysis/out/supplement/table_fields.tex}
\end{table}

\subsection{Discrimination of components from patches}
\label{sup:segvpatch}

A recurring ambiguity in the database is that some contributed models report each individual sub-fault patch as a separate component, so that a rupture on a single planar structure is represented as a large number of nominal components.
Such models carry no information about geometric segmentation and would, if retained, contaminate any statistic computed over ``multi-segment'' ruptures.
They are therefore excluded from every subset.

The exclusion is applied by a rule upon the reported component count: a model reporting more than \dataSegmentPatchThreshold{} components is taken to be reporting patches rather than components.
Of the \dataTotalModels{} models, \dataExcludedPatchModels{} exceed the threshold, and the smallest of them reports $\dataExcludedPatchMin{}$ components while the largest reports $\dataExcludedPatchMax{}$.
The excluded models are listed, with their reported component counts, in the released dataset.

\subsection{Curved segments reported as multiple components}
\label{sup:curved}

A second and less obvious ambiguity is that some inversions represent a single
\emph{curved} fault as a chain of planar facets, each reported as a separate
component.
Such a model is not multi-segment as it does not represent discontinuous physical structures.

Such models are separated by the segment rule of the manuscript itself, under which components
belong to the same segment when they lie within
\dataComponentSeparation{}\,km of one another and their dips agree to within
\dataComponentDipDelta{}$^\circ$.
The separation is the minimum distance between the portions of the plane surfaces of the two components that lie below a depth of 5~km, relaxed to just above the base of the shallower component where that is shallower than 5~km.

Of the \dataReportedMulti{} crustal models reporting more than one component,
\dataCurvedSingle{} form a single segment and \dataTrueMulti{} form two or
more.

\subsection{Trimming of rupture area}
\label{sup:trimming}

Finite-fault inversions are commonly parameterized over a fault plane larger than the region that actually slipped, so that the nominal rupture area over-estimates the area over which appreciable moment was released.
A trimming rule is therefore applied before rupture area is computed.

The rule adopted is that of \citet{somerville_characterizing_1999}, applied independently to each component.
Rows and columns are removed from the edges of the slip array one at a time, the edge removed at each step being that whose mean slip is smallest, and the process terminating when no edge has a mean slip below $30\%$ of the mean over the retained region.
The threshold is therefore relative to the surviving sub-array and is recomputed at each step, rather than being fixed with respect to the original array.

\subsection{Subsets used and event lists}
\label{sup:subsets}

The model subsets used in each calibration are summarized in Table~1 of the manuscript, and the jump-pair subset is enumerated in Table~\ref{tab:sup-events-jump-delay} below.
The corresponding lists for the magnitude--area and moment-ratio subsets are released with the distilled dataset rather than reproduced here due to their size (\dataMagnitudeAreaModels{} and \dataMultiSegment{} models respectively).

	{\footnotesize
		\input{analysis/out/supplement/table_events_jump_delay.tex}
	}

\subsection{Schema of the distilled dataset}

The distilled dataset released with this paper contains the quantities derived from the source models that are required to reproduce each figure, in a form that does not require the reader to re-run the parsing code or to obtain SRCMOD independently.

The dataset comprises three comma-separated files, one row per rupture, per component and per jump pair respectively, together with a schema document generated from the files themselves.
Their columns are given below.

\input{analysis/out/supplement/table_schema.tex}

\section{Scaling arguments underlying the moment-partitioning model}
\label{sup:theory}

This section develops the results asserted in Section~5.2 of the manuscript.
Section~\ref{sup:scaling} establishes the scaling of seismic moment and average slip with rupture area.
Section~\ref{sup:subadditivity} shows that the scaling relation is not additive under subdivision of a rupture, from which it follows that the total moment must be obtained from the total area rather than by summation over components.
Section~\ref{sup:invariance} shows that an exponent of unity is the unique choice for which the moment apportioned within a segment yields an average slip that is invariant to the delineation of components within that segment.

\subsection{Scaling of seismic moment and average slip with rupture area}
\label{sup:scaling}

Empirical magnitude-area scaling relations are conventionally expressed in the form
\begin{equation}
	\label{eq:sup-ma}
	M_w = a\log_{10}A + b,
\end{equation}
where $A$ is the rupture area in km$^2$ and $a$ and $b$ are empirically determined.
The relation is said to be self-similar when $a=1$.
The relation of \citet{leonard_selfconsistent_2014} adopted in the manuscript is of this form with $a=1$.
For consistency the average of the strike-slip and dip-slip relations in Table 1 of \citet{leonard_selfconsistent_2014} is used, i.e. $A = 10^{M_w-3.995}$.

Moment magnitude is related to seismic moment by \citep{hanks_moment_1979}
\begin{equation}
	\label{eq:sup-hk}
	M_w = \frac{2}{3}\left(\log_{10}M_o - 9.05\right),
\end{equation}
with $M_o$ in N\,m, so that $\log_{10}M_o = \frac{3}{2}M_w + 9.05$.
Substituting Equation~(\ref{eq:sup-ma}) into this expression gives
\begin{equation}
	\label{eq:sup-moa}
	\log_{10}M_o = \frac{3}{2}a\log_{10}A + \frac{3}{2}b + 9.05,
	\qquad\text{i.e.}\qquad
	M_o = \kappa A^{p}, \quad p \equiv \tfrac{3}{2}a,
\end{equation}
where $\kappa = 10^{(3b/2)+9.05}$.
For a self-similar relation, $a=1$ and hence $p=3/2$.

The seismic moment of a rupture is related to its average slip displacement $\bar{D}$ by
\begin{equation}
	\label{eq:sup-mo-slip}
	M_o = \mu \bar{D} A,
\end{equation}
where $\mu$ is the shear modulus, taken here as constant.
Combining Equations~(\ref{eq:sup-moa}) and~(\ref{eq:sup-mo-slip}),
\begin{equation}
	\label{eq:sup-slip-scaling}
	\bar{D} = \frac{\kappa}{\mu}A^{p-1},
\end{equation}
so that for a self-similar relation $\bar{D}\propto A^{1/2}$.
Average slip therefore increases with rupture size, and seismic moment is not proportional to rupture area.
It is this departure from proportionality, and only this departure, that gives rise to the two results that follow; had $p=1$, both would be trivial.

\subsection{Non-additivity of the scaling relation under subdivision}
\label{sup:subadditivity}

Consider a rupture that occurs upon a single, continuous, planar fault surface of total area $A_T$ --- that is, a rupture which is in every physical respect a single-segment rupture --- and suppose that this surface is artificially subdivided into $N$ components of area $A_i$, with $\sum_{i=1}^{N}A_i = A_T$.
The components so obtained are a discretization of one continuous structure rather than a set of physically distinct structures, and the subdivision is therefore a modeling choice which leaves the earthquake itself unchanged.
Any quantity that the generator assigns to the rupture as a whole must accordingly be unaffected by it.

Suppose now that the scaling relation were applied to each such component independently, and the resulting moments summed.
Writing $f(A) = \kappa A^{p}$, the total moment so obtained is $\sum_i f(A_i) = \kappa\sum_i A_i^{p}$, whereas application of the relation to the rupture as a whole gives $f(A_T) = \kappa A_T^{p}$.

For $p>1$ the function $f$ is superadditive in the sense that
\begin{equation}
	\label{eq:sup-superadd}
	\sum_{i=1}^{N} A_i^{p} \;\leq\; \left(\sum_{i=1}^{N}A_i\right)^{p} = A_T^{p},
\end{equation}
with equality if and only if $N=1$, or all but one of the $A_i$ vanish.
Component-wise application of the scaling relation therefore always under-estimates the total moment, and does so by an amount that depends upon how the rupture has been subdivided.

\begin{figure}[htb!]
	\centering
	\begin{tikzpicture}
		\begin{loglogaxis}[
				width=0.78\linewidth, height=0.5\linewidth,
				xlabel={Number of equal-area components, $n$, into which a single continuous rupture is subdivided},
				ylabel={$\textstyle\sum_i f(A_i)\,/\,f(A_T)$},
				xlabel style={font=\small}, ylabel style={font=\small},
				xmin=1, xmax=110, ymin=0.07, ymax=1.5,
				xtick={1,2,4,10,20,50,100},
				xticklabels={1,2,4,10,20,50,100},
				ytick={0.1,0.2,0.5,1.0},
				yticklabels={0.1,0.2,0.5,1.0},
				grid=both,
				grid style={line width=0.2pt, draw=gray!25},
				major grid style={line width=0.3pt, draw=gray!45},
				tick align=outside,
				legend style={at={(0.03,0.06)}, anchor=south west, font=\small, draw=gray!60},
				legend cell align=left,
			]
			\addplot[domain=1:110, samples=2, very thick, black] {1};
			\addlegendentry{$p=1$: moment proportional to area}
			\addplot[domain=1:110, samples=140, very thick, red, dashed] {x^(-0.5)};
			\addlegendentry{$p=3/2$: self-similar scaling}
			\addplot[only marks, mark=*, mark size=1.8pt, red] coordinates {(4,0.5) (100,0.1)};
			\node[font=\small, anchor=south west] at (axis cs:4.4,0.52) {$n=4$: $0.50$};
			\node[font=\small, anchor=north east] at (axis cs:95,0.094) {$n=100$: $0.10$};
		\end{loglogaxis}
	\end{tikzpicture}
	\caption{Ratio of the total seismic moment obtained by applying a magnitude-area scaling relation to each component independently and summing, $\sum_i f(A_i)$, to that obtained by applying the same relation to the rupture as a whole, $f(A_T)$, as a function of the number of equal-area components $n$ into which a single continuous rupture surface is artificially subdivided.
		The ratio is $n^{1-p}$, per Equation~(\ref{eq:sup-nfold}).
		For a self-similar scaling relation, $p=3/2$, and the total moment obtained by summation falls as $n^{-1/2}$, tending to zero as the subdivision is refined; only for $p=1$ is the result invariant to the subdivision.}
	\label{fig:sup-subdivision}
\end{figure}
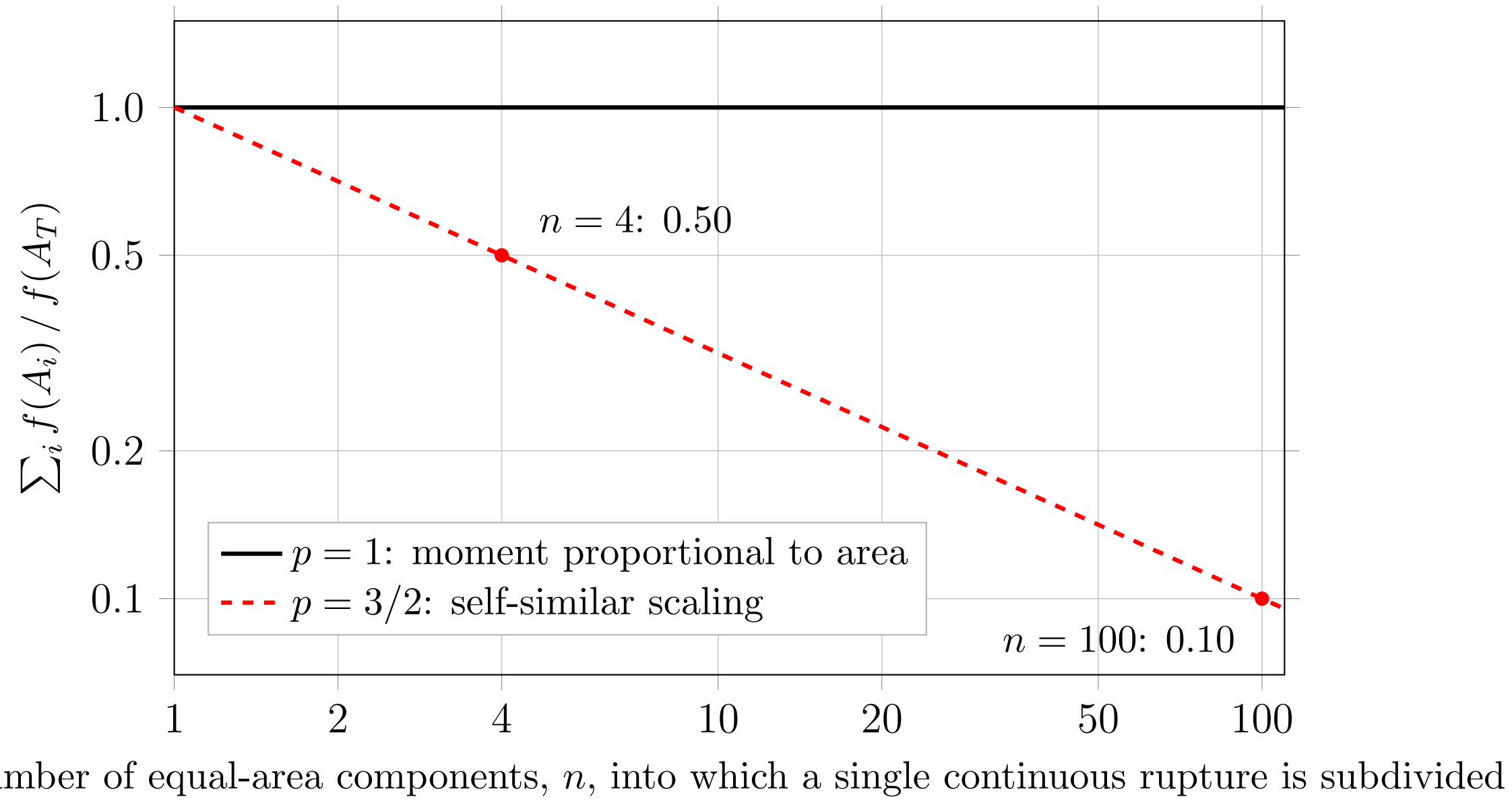

The magnitude of the discrepancy is most readily seen for a subdivision into $N=n$ components of equal area $A_T/n$:
\begin{equation}
	\label{eq:sup-nfold}
	\frac{\sum_i f(A_i)}{f(A_T)}
	= \frac{n\left(A_T/n\right)^{p}}{A_T^{p}}
	= n^{1-p},
\end{equation}
which for $p=3/2$ gives $n^{-1/2}$.
This behavior is illustrated in Figure~\ref{fig:sup-subdivision}.
Subdividing a single continuous rupture plane into four nominal components would thus halve its total moment, and into one hundred would reduce it tenfold; in the limit $n\to\infty$, corresponding to the discretization of a continuous rupture surface into arbitrarily small elements, the total moment tends to zero.
Only for $p=1$, i.e., for moment proportional to area, is the summation invariant to the subdivision.

Since the delineation of a single continuous rupture plane into components is a modeling choice and not a property of the earthquake, a procedure whose result depends upon that choice in the manner of Equation~(\ref{eq:sup-nfold}) is inadmissible.
The total moment is accordingly obtained from the total area, $M_{o,T}=f(A_T)$, and subsequently apportioned among the participating segments.

\subsection{Invariance of the within-segment apportionment}
\label{sup:invariance}

Let a segment $S_k$ be assigned moment $M_k$, and be represented as components of area $A_i$ with $\sum_i A_i = A_k$.
Consider apportioning $M_k$ among those components according to a general power rule
\begin{equation}
	\label{eq:sup-genrule}
	M_{o,i} = M_k\frac{A_i^{q}}{\sum_{j \in S_k} A_j^{q}},
\end{equation}
for some exponent $q$.
The average slip on component $i$ follows from Equation~(\ref{eq:sup-mo-slip}) as
\begin{equation}
	\label{eq:sup-genslip}
	\bar{D}_i = \frac{M_{o,i}}{\mu A_i} = \frac{M_k}{\mu}\cdot\frac{A_i^{q-1}}{\sum_{j \in S_k} A_j^{q}}.
\end{equation}

The components comprising a segment do not correspond to physically distinct structures, since a segment is defined as a set of components sufficiently close to one another to represent a single continuous fault, the internal boundaries arising from the manner in which the source model parameterizes changes in strike or dip.
The apportionment within a segment must therefore be invariant to that internal delineation.
Two conditions follow.

First, the average slip must not depend upon which component is considered.
From Equation~(\ref{eq:sup-genslip}), $\bar{D}_i \propto A_i^{q-1}$, so that for any two components of the same segment
\begin{equation}
	\label{eq:sup-sliprat}
	\frac{\bar{D}_i}{\bar{D}_j} = \left(\frac{A_i}{A_j}\right)^{q-1}.
\end{equation}
This ratio is unity for all $i,j$ if and only if $q=1$.
For $q=3/2$ it equals $(A_i/A_j)^{1/2}$, which is unbounded as the delineation is made increasingly uneven: two adjacent portions of one continuous structure would be assigned average slips differing by a factor of three were their areas to differ by an order of magnitude, purely as a consequence of where the boundary between them had been drawn.

Second, the average slip must not depend upon the partition itself.
Setting $q=1$ in Equation~(\ref{eq:sup-genslip}) gives
\begin{equation}
	\label{eq:sup-invariant}
	\bar{D}_i = \frac{M_k}{\mu\sum_{j \in S_k} A_j} = \frac{M_k}{\mu A_k},
\end{equation}
which depends only upon the segment moment and the segment area, and is therefore identical for every admissible partition of that segment into components.
For $q\neq 1$ no such invariance obtains, except in the degenerate case of a partition into components of equal area, for which symmetry alone yields a uniform slip.

The exponent of unity adopted within segments in Equation~(2) of the manuscript is thus not a convenient approximation but the unique choice consistent with the requirement that the generated rupture be insensitive to an arbitrary feature of the source model parameterization; this is requirement R2 of manuscript Section~2.
Between segments, by contrast, the relative areas correspond to physically distinct structures whose relative sizes are a property of the rupture rather than of its parameterization, and the exponent $p=3/2$ implied by the scaling relation is retained.

\subsection{Components and segments of differing rigidity}
\label{sup:rigidity}

Sections~\ref{sup:scaling}--\ref{sup:invariance} have taken the shear modulus $\mu$ to be constant over the participating components.
Where the components occupy differing depth ranges this is not adequate, since the one-dimensional velocity profile adopted in rupture generation typically exhibits reduced shear-wave velocity, and hence reduced rigidity, over the uppermost few kilometers.
A component confined to shallow depth therefore possesses a lower average rigidity, $\mu_i$, than one extending to seismogenic depth.

Let $\mu_i$ denote the average shear modulus of component $i$, and consider the within-segment apportionment in proportion to the \emph{rigidity-weighted} area,
\begin{equation}
	\label{eq:sup-rigidity-in}
	M_{o,i} = M_k\frac{\mu_i A_i}{\sum_{j \in S_k} \mu_j A_j}.
\end{equation}
The average slip upon component $i$ then follows from Equation~(\ref{eq:sup-mo-slip}) as
\begin{equation}
	\label{eq:sup-rigidity-slip}
	\bar{D}_i = \frac{M_{o,i}}{\mu_i A_i} = \frac{M_k}{\sum_{j \in S_k} \mu_j A_j},
\end{equation}
which is independent both of $i$ and of the partition of the segment into components.
Equation~(\ref{eq:sup-rigidity-in}) is therefore the generalization of Equation~(\ref{eq:sup-genrule}) with $q=1$ to the case of non-uniform rigidity, and it recovers Equation~(\ref{eq:sup-invariant}) when $\mu_i$ is the same for every component.
The invariance result of Section~\ref{sup:invariance} is thus not restricted to uniform rigidity; what is required is that moment be apportioned in proportion to the rigidity-weighted area, and it is only when the rigidity is uniform that apportionment by area alone suffices.

Apportionment by area alone where the rigidity is not uniform assigns to a low-rigidity component a moment appropriate to a rigidity it does not possess, and hence an average slip inflated by the factor $\bar{\mu}/\mu_i$, where $\bar{\mu}$ is the rigidity-weighted mean over the segment.
For a shallow component this factor may be appreciable, and the resulting slip is correspondingly unstable with respect to the depth extent assigned to that component by the source model.

The corresponding between-segment relation follows from the requirement that the scaling of average slip with segment area be preserved.
From Equation~(\ref{eq:sup-slip-scaling}), $\bar{D}_k \propto A_k^{p-1}$, and since $M_k = \mu_k\bar{D}_k A_k$,
\begin{equation}
	\label{eq:sup-rigidity-between}
	M_k = M_{o,T}\frac{\mu_k A_k^{p}}{\sum_{k'} \mu_{k'} A_{k'}^{p}},
\end{equation}
with $p=3/2$ for a self-similar scaling relation, so that the segment area is weighted by the segment rigidity rather than raised to a modified power.

\subsection{Calibration of the apportionment over the pooled population}
\label{sup:pooled}

Manuscript Section~5.4 summarizes the calibration of the apportionment model against the SRCMOD ruptures that report more than one component, fitting the exponent separately to the single-segment and multi-segment populations identified in Section~\ref{sup:curved}.
This section records the fitting procedure.

The residual is taken in natural logarithm, $\ln\left(M_{o,i}/M_{o,T}\right) - \ln\left(A_i^{p}/\sum_j A_j^{p}\right)$.
The fit is found by least squares regression on $p$ with the standard deviation estimated by cluster bootstrapping which accounts for the dependence between moment fractions within each cluster.
Between-segment bootstrapping clusters by rupture, and within-segment bootstrapping clusters by segment.

Over the $\jThreeLinearN{}$ components of the pooled (single- and multi-segment) population, an exponent of unity gives a mean residual of $\jThreeLinearMean{}$ with a standard deviation of $\jThreeLinearSd{}$, and an exponent of $3/2$ gives a mean residual of $\jThreeThreeHalvesMean{}$ with a standard deviation of $\jThreeThreeHalvesSd{}$.
These two statistics do not favor either model due to the considerable standard deviation in the residuals.
Fitting the exponent as a free parameter over the pooled population returns $\jThreeFittedExponent{} \pm \jThreeFittedStandardError{}$.

\begin{figure}[htb!]
	\centering
	\includegraphics[width=\linewidth]{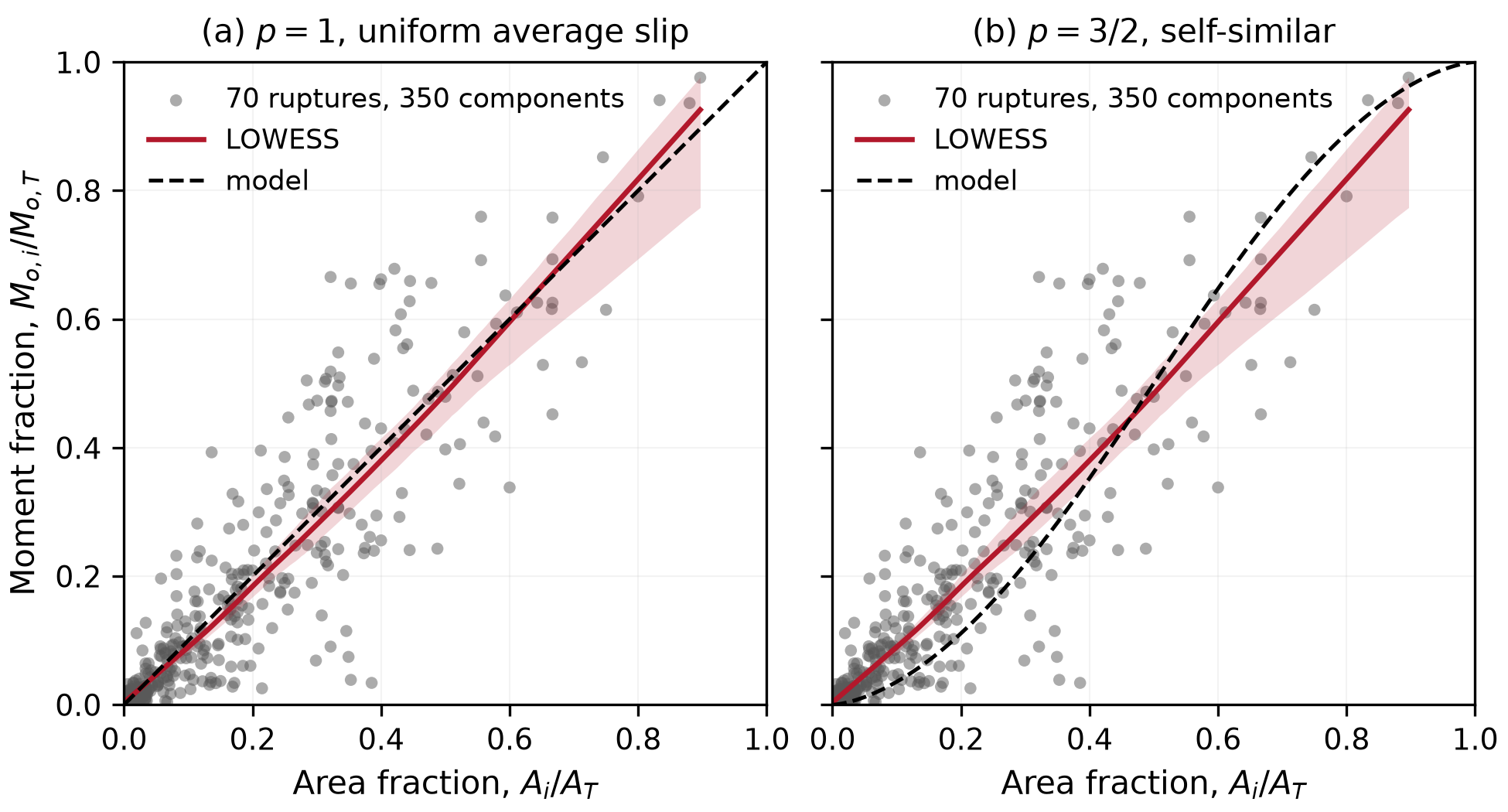}
	\caption{Component moment fraction against component area fraction for the components of all \dataReportedMulti{} SRCMOD crustal ruptures that report more than one component, pooled without regard to the number of segments, with the apportionment model (dashed) adopting (a) an exponent of unity and (b) an exponent of $3/2$.
		The solid line is a locally weighted regression of the data, with a bootstrap confidence band.}
	\label{fig:sup-pooled}
\end{figure}

Figure~\ref{fig:sup-pooled} shows the pooled population against each candidate exponent, with a locally weighted regression of the data.
The regression departs from the exponent of unity at large area fractions and from the exponent of $3/2$ at small ones, which is the pattern expected of a mixture of the two regimes.

\subsection{Stochastic perturbation of the apportionment}
\label{sup:momentPerturbation}

When the apportionment is evaluated stochastically, it is perturbed about the value given by Equations~(1) and~(2) of the manuscript.
The form of the perturbation follows from the structure of the scatter, which is examined first.
Writing $\varepsilon_i = M_{o,i}/M_{o,T} - A_i/A_T$ for the departure of a component's moment fraction from its area fraction, $\varepsilon_i$ is a difference of two proportions and is therefore bounded on $(-1,1)$.
An additive perturbation would accordingly be defined on the Fisher $z$-transformation, $z_i = \operatorname{artanh}\varepsilon_i$, which maps that interval onto the real line.
Over the $\jFourN{}$ components of the multi-segment ruptures, $z$ has a standard deviation of $\sigma_z = \jFourSigmaM{}$.

A homoscedastic additive model is not, however, sufficient.
Regressing $z^2$ on the area fraction gives $p \jFourHeteroscedasticityP{}$, so that the scatter depends strongly on component size, the departure growing with the area fraction.
A single constant $\sigma_z$ therefore over-perturbs small components and under-perturbs large ones.
Taking $\sigma_z(f) = s\,f^{\,b}$ in the area fraction $f = A_i/A_T$, which recovers the constant model at $b = 0$, and fitting by maximum likelihood returns $s = \cSixMomentScale{}$ and $b = \cSixMomentExponent{}$, with a $95\%$ profile-likelihood interval for $b$ of $\cSixMomentExponentLow{}$ to $\cSixMomentExponentHigh{}$, and the Akaike information criterion falls by $\cSixMomentDeltaAic{}$.
For a component holding one twentieth of the rupture area, the fitted model gives $\sigma_z = \cSixMomentAtTwentieth{}$, compared with the constant value of $\jFourSigmaM{}$.
An additive perturbation of constant scale is therefore too wide for the components of a rupture of many components, and it produces negative moment fractions for such ruptures.

Since $b$ is close to unity, the departure is approximately proportional to the area fraction, so that the error in the apportioned moment is relative, and the perturbation adopted is therefore multiplicative.
It is applied to the segment moments,
\begin{equation}
	\label{eq:sup-multiplicative}
	M_k = M_{o,T}\,\frac{\hat{M}_k\,e^{\epsilon_k}}{\sum_{k'} \hat{M}_{k'}\,e^{\epsilon_{k'}}},
	\qquad \epsilon_k \sim N\!\left(0,\sigma_M^2\right), \quad \left|\epsilon_k\right| \le 2\sigma_M,
\end{equation}
in which $\hat{M}_k$ is the segment moment given by manuscript Equation~(1), while the apportionment within each segment is left as manuscript Equation~(2).
The perturbation is drawn from the normal distribution truncated at $\pm 2\sigma_M$, since an untruncated draw occasionally assigns to a small segment a share of the moment, and hence an average slip, well beyond that of any segment in the calibration data.

The standard deviation $\sigma_M$ is calibrated from the residuals $\eta_k = \ln\left(M_k/M_{o,T}\right) - \ln\left(A_k^{3/2}/\sum A_{k'}^{3/2}\right)$ over the segments of the \kFourRuptures{} multi-segment SRCMOD ruptures.
The rescaling of Equation~(\ref{eq:sup-multiplicative}) removes a factor common to the segments of a rupture, so that only the departures of $\eta_k$ from their mean within each rupture are informative.
For a rupture of $K$ segments, these centered residuals have variance $\sigma_M^2 (K-1)/K$, and it follows that
\begin{equation}
	\label{eq:sup-sigmaM}
	\hat{\sigma}_M^2 = \frac{\sum_{\text{ruptures}} \sum_{k} \left(\eta_k - \bar{\eta}\right)^2}{\sum_{\text{ruptures}} \left(K - 1\right)},
\end{equation}
which gives $\sigma_M = \kFourSigmaM{}$ over \kFourComponents{} segments with \kFourDof{} degrees of freedom.

Regressing the squared centered residuals, scaled by $K/(K-1)$, on the segment area fraction gives $p = \kFourHeteroscedasticityP{}$ (manuscript Figure~6).
The log residuals are therefore homoscedastic, in contrast with the additive departures, and a single $\sigma_M$ suffices.

\section{Rupture-path model: supporting detail}
\label{sup:rupturePath}

\subsection{Sensitivity of the maximum-likelihood rupture tree to the characteristic jumping distance}
\label{sup:r0}

The characteristic jumping distance $r_0 = 3$~km of manuscript Equation~(3) was obtained by \citet{shaw_probabilities_2007} from Californian seismicity, and its transfer to the New Zealand setting therefore requires justification.
Without enough data to constrain $r_{0}$, a sensitivity study was instead conducted to see if different values would meaningfully change the results.
Maximum-likelihood tree selection and probabilities were analyzed for the \jFiveGeometries{} reference geometries of manuscript Figure~8 for seven values of $r_0$ spanning \jFiveRzeroLow{}--\jFiveRzeroHigh{}~km.
The maximum-likelihood tree was found to be the same for all values of $r_0$; this result is not a coincidence of the data.
The maximum-likelihood tree is found by Kruskal's algorithm, which sorts edges by weight and adds each in order, only discarding an edge where its addition would create a cycle.
Because of this, the maximum-likelihood tree found only depends on the relative ordering of the edges (sorted in code first by probability, then by separation), which is invariant to changes in $r_{0}$.
The probability of the maximum-likelihood tree is sensitive to $r_0$, decreasing from \jFiveKaikouraModalLow{} to \jFiveKaikouraModalHigh{} for Kaik\=oura and from \jFiveDarfieldModalLow{} to \jFiveDarfieldModalHigh{} for Darfield as $r_0$ increases from \jFiveRzeroLow{} to \jFiveRzeroHigh{}~km.
Given the lack of sensitivity in the maximum-likelihood path to $r_{0}$, and the lack of New Zealand data constraining it, the choice was made to keep $r_{0} = 3$~km with calibration as future work.

\subsection{Weighted spanning-tree distribution and its implementation}
\label{sup:spanningTree}

As shown in manuscript Equation~(5), the distribution over rupture trees is a weighted spanning-tree distribution with edge weights $w_{ij} = P_{ij}/(1-P_{ij})$, which is what permits the algorithm of \citet{wilson_generating_1996} to be applied to it directly.
In the implementation, the maximum-likelihood tree is obtained as the maximum spanning tree under the weights $\log w_{ij} = \log P_{ij} - \log\left(1 - P_{ij}\right)$, and the sampling routine likewise rescales to $w_{ij}$ before applying the algorithm of \citet{wilson_generating_1996}.

Because the leading product $\prod (1 - P_{ij})$ in manuscript Equation~(4) is common to every tree, the normalizing constant of the distribution is the total weighted spanning-tree count, which Kirchhoff's matrix-tree theorem evaluates in closed form as a determinant.
The probability of any individual rupture tree is therefore available without enumerating the trees, which is what makes the analysis of manuscript Section~6.3 tractable over a source model containing hundreds of thousands of ruptures.
Taken with unit weights, the same theorem gives the number of admissible rupture trees reported for the historical geometries in manuscript Section~6.2.

\subsection{Jump probabilities for the 2010 Darfield geometry}
\label{sup:darfield}

Table~\ref{tab:sup-darfield} lists the \jSixDarfieldEdges{} admissible edges between the \jSixDarfieldFaults{} components of the \citet{pajaro2026Darfield} geometry of the 2010 Darfield earthquake, with their separations and the jumping probabilities of manuscript Equation~(3), and identifies the edges of the maximum-likelihood rupture tree.
The leading alternative at the junction on the north-west/south-east section of the Greendale Fault is a connection between the central section and either the Charing Cross Fault or the north-east structure, both of which are also at the probability cap.

\begin{table}[htb!]
	\centering
	\caption{Admissible edges between the components of the 2010 Darfield geometry of \citet{pajaro2026Darfield}, with the separation $r$, the jumping probability $P$ of manuscript Equation~(3), and the edges of the maximum-likelihood rupture tree ($\bullet$).}
	\label{tab:sup-darfield}
	\footnotesize
	\input{analysis/out/supplement/table_darfield_edges.tex}
\end{table}

\subsection{Ambiguity of the rupture path over the 2022 NZ NSHM rupture set}
\label{sup:nshmAmbiguity}

\begin{figure}[htb!]
	\centering
	\includegraphics[width=\linewidth]{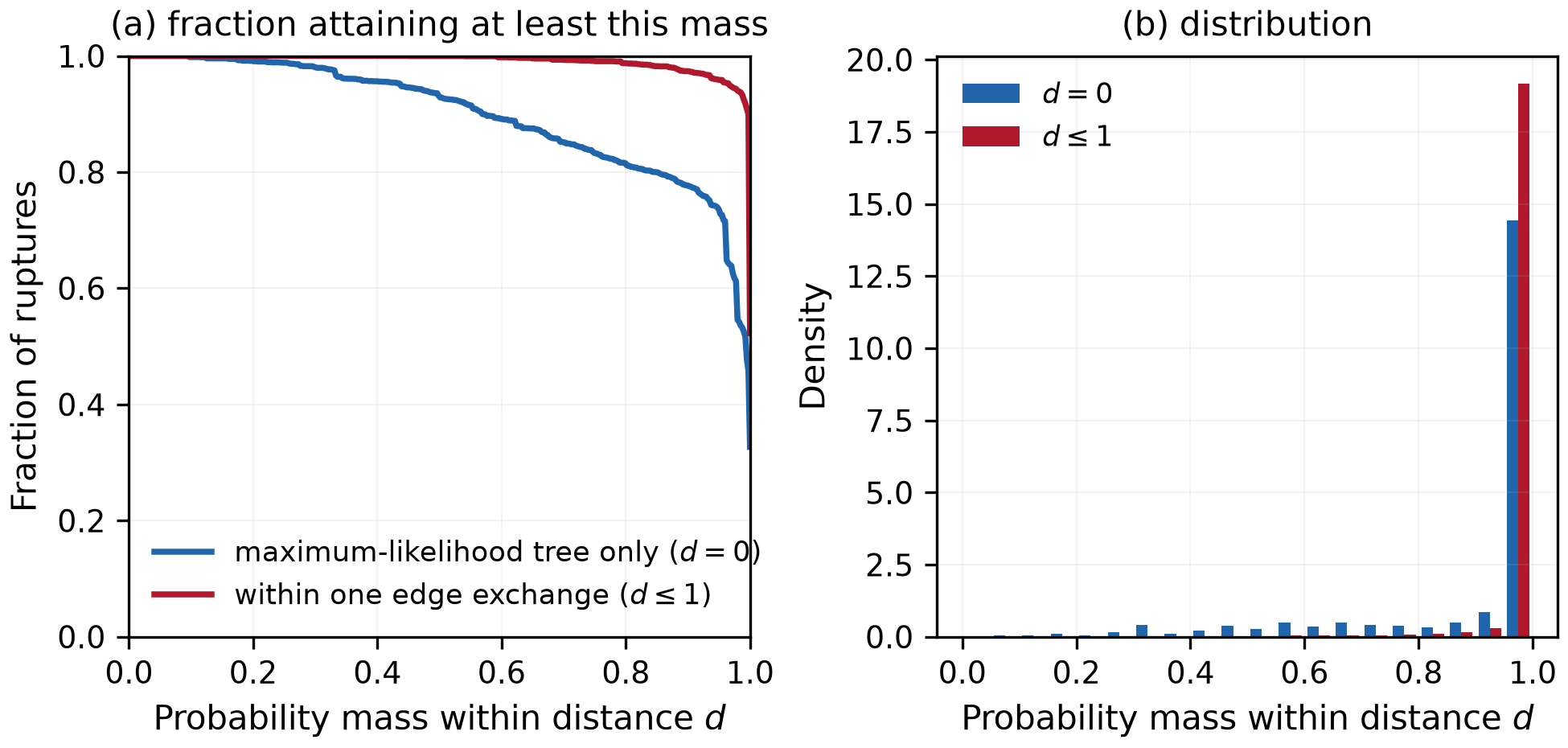}
	\caption{Ambiguity of the rupture path over the \jSevenDistanceN{} multi-segment ruptures of the 2022 NZ NSHM crustal fault model: (a) the portion of the population attaining at least a given probability mass, for the maximum-likelihood path alone and for paths within one edge exchange of it; and (b) the corresponding distributions.}
	\label{fig:sup-nshmTreeDistance}
\end{figure}

Figure~\ref{fig:sup-nshmTreeDistance} shows the ambiguity of the rupture path over the \jSevenDistanceN{} multi-segment ruptures of the 2022 NZ NSHM crustal fault model considered in Section~6.3 of the manuscript, expressed as the probability mass within a given number of edge exchanges of the maximum-likelihood path.
The rupture path was computed for the \jSevenDistanceN{} multi-segment ruptures of the 2022 NZ NSHM that are generated in manuscript Section~9.2, and Figure~\ref{fig:sup-nshmTreeDistance} shows the distribution of the resulting ambiguity of the rupture path.
Nearly all (\jSevenModalIsLinePct{}\%) of the maximum-likelihood rupture paths are unbranched.
Where a choice does arise, the alternatives are close to the maximum-likelihood tree.
Admitting the paths that differ from the maximum-likelihood tree by a single edge exchange raises the probability accounted for from \jSevenMassModalMean{} to \jSevenMassWithinOneMean{} in the mean.
The remaining ambiguity results from small step-over components in the NSHM introducing non-consecutive jumps with low probability.

\section{Jump location and time delay: supporting detail}
\label{sup:jumping}

\subsection{Modified jump point selection for the 2022 NZ NSHM benchmark}
\label{sup:nshm2022-jump-algorithm}
As mentioned in manuscript Section~7.1, reproduction of the 2022 NZ NSHM jump point and rupture generation success statistics is not easily done if one must generate a complete geometric discretization for the large geometries.
To this end, the analysis of manuscript Section~9.2 obtains the jump points by an optimization over the two component surfaces $C_i$ and $C_j$, as follows:

\begin{enumerate}
	\item A pair of local coordinate systems $(s,d)$ is established on the two components, in which $(s,d)\in[0,1]\times[0,1]$ denote the normalized along-strike and down-dip positions, respectively, each being accompanied by an embedding function, $F(s,d)$ and $F^{'}(s^{'},d^{'})$, into a common Cartesian coordinate system (here, NZTM2000).
	\item An objective function is defined as the Euclidean distance between a point on each of the two components, $L(s,d,s^{'},d^{'}) = \left|F(s,d) - F^{'}(s^{'},d^{'})\right|$.
	\item The jump point and conjugate point are obtained by minimizing $L$, subject to the constraint that both points lie at a depth of at least 5~km.
	      Where a component does not extend to that depth, the constraint is relaxed to the deepest point available on it.
\end{enumerate}

For a pair of planar components the minimization returns the global minimum of $L$, since $L$ is then a convex function over a rectangular domain.
Where a component comprises several planes convexity is not guaranteed, but repeating the minimization for the \jOptJumps{} jumps of manuscript Section~9.1 from 81 starting points improved no separation by more than \jOptMaxImprovement{}~m.

\subsection{Sensitivity of the jump location to the depth constraint}
\label{sup:depth}

The minimum depth of 5~km imposed on the jump-location optimization of manuscript Section~7.1 is a chosen value, and the computed jump locations are sensitive to it.
Recomputing every jump of the rupture of manuscript Section~9.1 for thresholds from $\jDepthSweepLow{}$ to $\jDepthSweepHigh{}$\,km moves the jump point by $\jDepthShiftMeanLow{}$ to $\jDepthShiftMeanHigh{}$\,km in the mean, and by as much as $\jDepthShiftMax{}$\,km in the worst case, corresponding to approximately $\jDepthShiftFractionMean{}\%$ of the along-strike extent of the triggered component in the mean and $\jDepthShiftFractionMax{}\%$ at most.
The jumping distance varies much less, from $\jDepthSeparationLow{}$ to $\jDepthSeparationHigh{}$\,km in the mean across the same range, and increases monotonically with the threshold, since constraining both points to lie deeper restricts the pair of surfaces over which the minimum is taken.
The consequence for the rupture time is therefore modest even where the jump location moves appreciably, since it is the separation, and not the position, that enters manuscript Equation~(6).
No jump requires relaxation of the constraint at $\jDepthSweepLow{}$\,km, whereas \jDepthRelaxedCount{} do at \jDepthRelaxedFromKm{}~km and above.

\subsection{Separation between the inferred and computed jump locations}
\label{sup:jumpLocationError}

\begin{figure}[htb!]
	\centering
	\includegraphics[width=\linewidth]{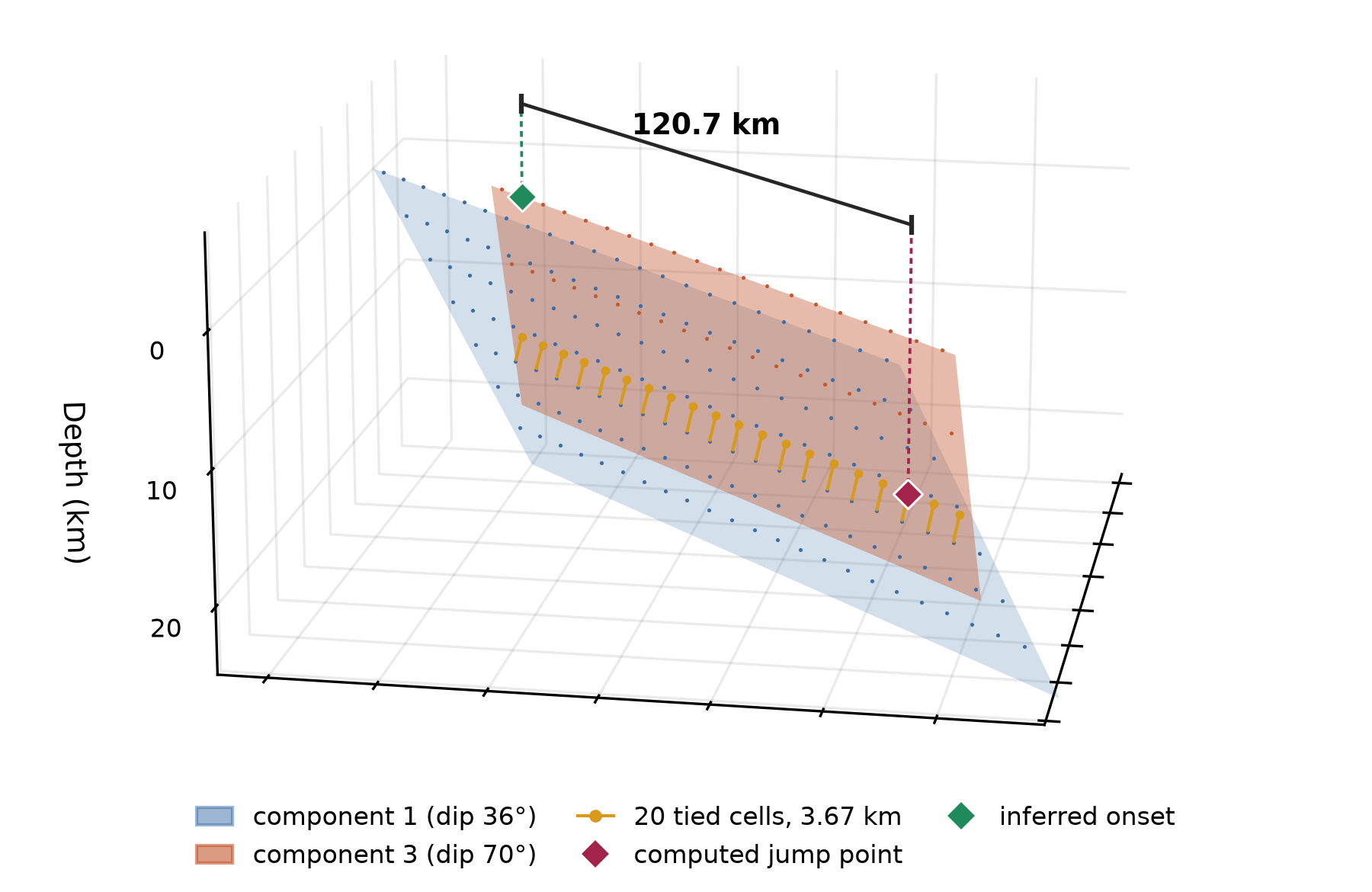}
	\caption{Close-up view of the \citet{hayes-ruptures-2017} inversion of the Kaik\=oura earthquake, seen from the north-east and oblique to strike, at the jump with the largest separation between the inferred and computed jump locations.
		Component~1 (blue, dip 36$^\circ$) triggers component~3 (orange, dip 70$^\circ$), and the dots mark the top-centers of their sub-faults.
		Because the two components share a strike, the 20 sub-faults in the lowest row of component~3 (gold) are all 3.67~km from component~1, and the gold tie lines join each to its nearest sub-fault on component~1.
		The green diamond marks the inferred onset and the red diamond marks the computed jump point.
	}
	\label{fig:kaikouraJumpPoints}
\end{figure}

Manuscript Figure~9 compares the jump locations inferred from the SRCMOD inversions with those computed by the closest-point construction of manuscript Section~7.1.
The separation between the jump location inferred from the inversion and that computed by the closest-point construction has a mean of $\jNineMeanKm{}$~km and a standard deviation of $\jNineSdKm{}$~km over the $\jNineN{}$ jump pairs for which both are defined, or $\jNineMeanFraction{} \pm \jNineSdFraction{}$ as a fraction of the along-strike extent of the triggered component.
Given the sample size, both statistics are heavily affected by a small number of outliers.
The largest separation is in the \citet{hayes-ruptures-2017} inversion of the Kaik\=oura earthquake (manuscript Figure~9(b) and~(c)).
Here, two overlapping planes are strike-parallel so that several closest-point pairs lie along a common line at the bottom of the steeper-dipping plane.
Because the jump algorithm is informed by geometry only, the closest point pair that it selects is, in this case, not unique.
This example represents a limitation of a purely geometric jump-point algorithm, and motivates its further development.
An along-strike view of this jump is shown in Figure~\ref{fig:kaikouraJumpPoints}.

\subsection{Sensitivity of the time-delay calibration}
\label{sup:delaySensitivity}

Under a constant (homoscedastic) scatter, regressing the observed delay on the jumping distance through the origin gives a gradient of $\jElevenGradient{} \pm \jElevenGradientSe{}$\,s\,km$^{-1}$ over the \jTenJumpPairs{} crustal pairs of manuscript Section~7.3, and hence an apparent velocity of $\jThirteenApparentVelocity{}$\,km\,s$^{-1}$ with a residual standard deviation of $\jTwelveSigmaDelta{}$\,s.
This result is sensitive to the range of jumping distances admitted.
Restricting the regression to jumps shorter than 4~km gives $\jThirteenVelocityUnderFour{}$\,km\,s$^{-1}$ with $\sigma_\delta = \jTwelveSigmaDeltaUnderFour{}$\,s, and shorter than 6~km gives $\jThirteenVelocityUnderSix{}$\,km\,s$^{-1}$ and $\jTwelveSigmaDeltaUnderSix{}$\,s.
The residual scatter therefore grows as long jumps are admitted, and the apparent velocity is unstable.
A test of the residuals for dependence on jumping distance is marginal over the full sample ($p = \jTwelveHeteroscedasticityP{}$), and the likelihood comparison which follows quantifies the dependence.

Allowing the scatter to vary as $\sigma_\delta(r) = s\,r^{\,g}$, which recovers the constant model at $g = 0$, and fitting $s$, $g$ and the gradient jointly by maximum likelihood (jointly, since a variance function estimated from the residuals about a mean fitted under the wrong variance assumption biases both) returns $g = \cSixDelayExponent{}$, with a $95\%$ profile-likelihood interval of $\cSixDelayExponentLow{}$ to $\cSixDelayExponentHigh{}$, and $s = \cSixDelayScale{}$\,s\,km$^{-g}$.
The Akaike information criterion falls by $\cSixDelayDeltaAic{}$ relative to the constant model, and the fitted velocity is $\cSixDelayVelocity{}$\,km\,s$^{-1}$, since the long jumps, which pull the constant-scatter estimate upward, no longer carry the same weight as the short ones.
This is Model~A of manuscript Equation~(6).

\begin{figure}[htb!]
	\centering
	\includegraphics[width=\linewidth]{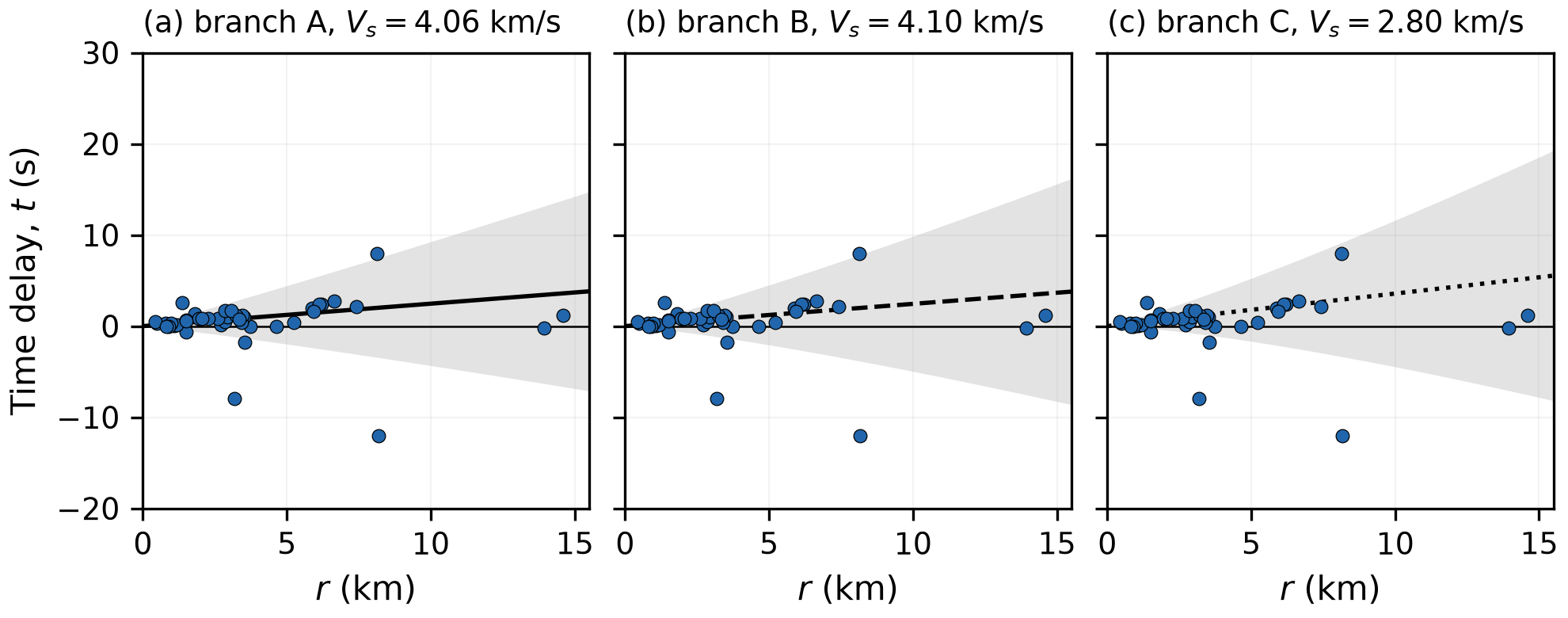}
	\caption{The three models of manuscript Equation~(6), each with its mean and $\pm 1\sigma_\delta(r)$ band, against the crustal jump pairs to which they are fitted.}
	\label{fig:sup-delayBranches}
\end{figure}

Table~\ref{tab:sup-delayBranches} lists the three models of manuscript Equation~(6), fitted to the \tFortyOnePairs{} crustal pairs on which the delay model is calibrated.
Figure~\ref{fig:sup-delayBranches} shows the $\pm 1\sigma_\delta(r)$ band of each model against the data.
Adding a term independent of the jumping distance (Model~B) raises the information criterion by $\tFortyOneBDeltaAic{}$, leaves the predicted share of negative delays at \tFortyOneBNegativePct{}\%, against \tFortyOneObservedNegativePct{}\% observed, and changes the apparent velocity little, to $\tFortyOneBVelocity{}$\,km\,s$^{-1}$.
Fixing the velocity at the within-segment rupture velocity (Model~C) changes the scatter little and saves one parameter relative to Model~B.
It follows that the apparent velocity is sensitive to the form of the scatter, which is a further indication that it reflects the uncertainty in the jump pair more than the propagation of rupture.
The exponent of the three models lies between $\tFortyOneAExponent{}$ and $\tFortyOneCExponent{}$, and implies a scatter of approximately \tFortyOneASigmaFifteen{} to \tFortyOneCSigmaFifteen{}~s at the 15~km limit.
The differences in AIC between the models are less than two, and the ranking is sensitive to individual ruptures: omitting the inversion of the 1948 Fukui earthquake makes Model~B the lowest, and omitting that of the 2013 Balochistan earthquake makes Model~C the lowest.
The two pairs of the 2016 Kaik\=oura inversion at \jTenKaikouraHighDistanceKm{} and \jTenKaikouraLowDistanceKm{}~km, with delays of \jTenKaikouraHighDelayS{} and \jTenKaikouraLowDelayS{}~s, are among the most influential on the fitted scatter.
No pair is longer than \jTenMaxDistanceKm{}~km, so that the scatter between there and the 15~km limit is an extrapolation under each model.

\begin{table}[htb!]
	\centering
	\caption{The three models of manuscript Equation~(6), fitted by maximum likelihood to the crustal jump pairs. $\Delta$AIC is relative to the lowest value; $k$ is the number of free parameters. The predicted share of negative delays is evaluated at the jumping distances of the pairs fitted; the truncated share is that of the causal bound over the \tFortyOneNshmJumps{} jumps of the rated NSHM ruptures of manuscript Section~9.2.}
	\label{tab:sup-delayBranches}
	\small
	\begin{tabular}{llccccccccc}
		\hline
		Pairs                       & Model & $V_s$ (km\,s$^{-1}$)          & $\sigma_0$ (s)         & $s$ (s\,km$^{-g}$)    & $g$                   & $k$ & AIC              & $\Delta$AIC           & Negative (\%)            & Truncated (\%)            \\
		\hline
		Crustal (\tFortyOnePairs{}) & A     & \tFortyOneAVelocity{}         & --                     & \tFortyOneAScaleSci{} & \tFortyOneAExponent{} & 3   & \tFortyOneAAic{} & \tFortyOneADeltaAic{} & \tFortyOneANegativePct{} & \tFortyOneATruncatedPct{} \\
		                            & B     & \tFortyOneBVelocity{}         & \tFortyOneBSigmaZero{} & \tFortyOneBScaleSci{} & \tFortyOneBExponent{} & 4   & \tFortyOneBAic{} & \tFortyOneBDeltaAic{} & \tFortyOneBNegativePct{} & \tFortyOneBTruncatedPct{} \\
		                            & C     & \tFortyOneCVelocity{} (fixed) & \tFortyOneCSigmaZero{} & \tFortyOneCScaleSci{} & \tFortyOneCExponent{} & 3   & \tFortyOneCAic{} & \tFortyOneCDeltaAic{} & \tFortyOneCNegativePct{} & \tFortyOneCTruncatedPct{} \\
		\hline
	\end{tabular}
\end{table}

The causal bound of manuscript Section~7.2 requires that $t_j \geq -\left[t_{init}(h_j^{'}) - t_{init}(h_i)\right]$, the bracket being the time taken for rupture to travel from the initiation point of the triggering component to the conjugate point.
A draw of $t_j$ that violates this bound is replaced by a draw from the normal distribution truncated at it.
Over the rated NSHM ruptures, nucleated at the center of the southernmost component and following the maximum-likelihood rupture tree, that travel time has a median of \tFortyOneTravelMedianS{}~s (interquartile range \tFortyOneTravelLowerS{}--\tFortyOneTravelUpperS{}~s), so that the bound is reached only where a component is triggered close to the point at which its triggering component was itself entered.
The share of draws truncated is computed exactly, as the mean over all jumps of the probability that an untruncated draw falls below the bound.

The conjugate is taken as the nearest sub-fault on any other component, irrespective of its onset time.
A rule that admits only conjugate points that ruptured before the entry point selects the conjugate using the onset that is being modeled, and so cannot observe a negative delay; it is therefore not used.
Of the \tFortyOnePairs{} calibration pairs, \jTenNegativePairs{} have a negative delay under the adopted rule, the largest being \jTenKaikouraLowDelayS{}~s at a jumping distance of \jTenKaikouraLowDistanceKm{}~km in the inversion of the 2016 Kaik\=oura earthquake.

\subsection{Classification of the jump pairs}
\label{sup:pairClasses}

Each SRCMOD inversion is assigned a tectonic setting from the earthquake it represents (column \texttt{setting} of \texttt{jump\_pairs.csv} in the released dataset), and only crustal inversions are used for calibration (manuscript Section~4).
Each jump pair is further classed by the arrangement of its two components.
Both components are projected onto the strike of the triggering component, and a pair whose projections overlap by more than 2~km, about one sub-fault, is classed as overlapping.
A pair that does not overlap but whose strike or dip differs by more than 30$^\circ$ is classed as oblique, and the remainder as end to end.
The classification is made per pair and not per earthquake, so that the 2016 Kumamoto inversions contribute pairs to both classes.

\section{Stochastic realizations of a generated rupture}
\label{sup:stochastic}

\begin{figure}[htb!]
	\centering
	\includegraphics[width=\linewidth]{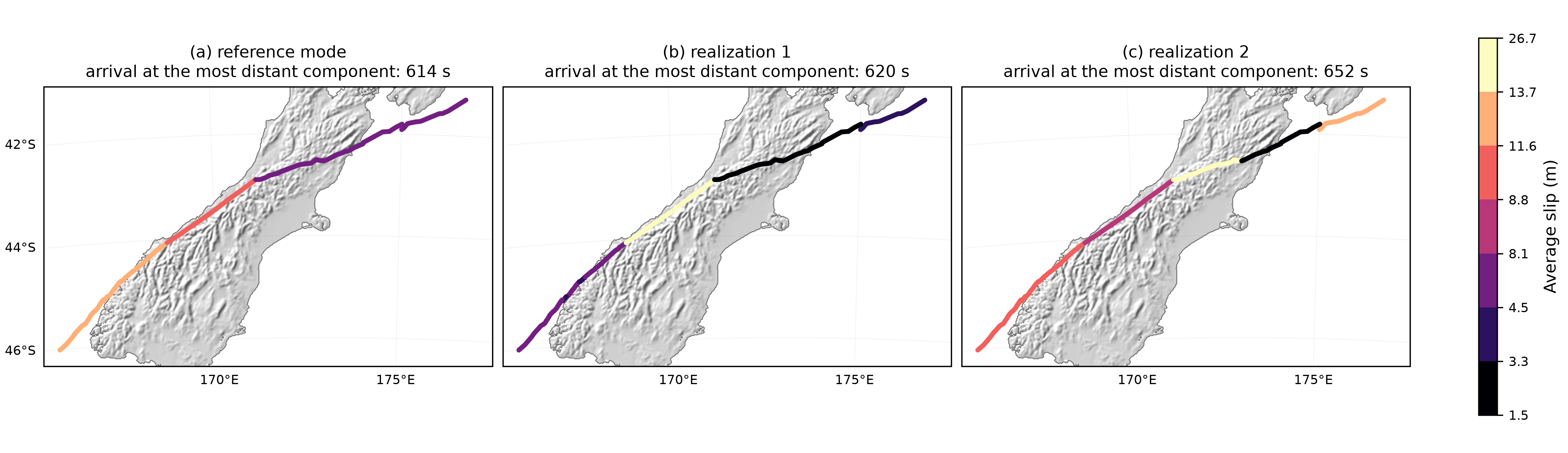}
	\caption{The rupture of manuscript Figure~12 in the reference mode (a) and in two stochastic realizations (b), (c), with the hypocenter held fixed.
		Average slip is on a common scale across the three panels, and the arrival of rupture at the most distant component is given above each.}
	\label{fig:sup-stochastic}
\end{figure}

Figure~\ref{fig:sup-stochastic} shows the rupture of manuscript Figure~12 in the reference mode and in two stochastic realizations, with the hypocenter held fixed so that the variability shown is that of the multi-segment framework alone, namely the rupture tree, the moment apportionment and the jump delays.
The reference apportionment varies smoothly along the rupture, following segment area, whereas in the realizations neighboring segments differ by a factor of several.
Over \kFiveRealizations{} realizations, the arrival of rupture at the most distant component lies between \kFiveArrivalLow{} and \kFiveArrivalHigh{}~s (5th and 95th percentiles), against \kFiveArrivalReference{}~s in the reference mode, and \kFiveNegativePct{}\% of the sampled delays are negative.
Average slip ranges from \kFiveMinSlipLow{} to \kFiveMinSlipHigh{}~m on the least-slipping component and from \kFiveMaxSlipLow{} to \kFiveMaxSlipHigh{}~m on the most-slipping one, against \kFiveRefSlipLow{} to \kFiveRefSlipHigh{}~m in the reference mode.

\bibliographystyle{abbrvnat}
\bibliography{paper}

%% file: analysis/out/numbers.tex
\newcommand{\jSixteenRuptureId}{71072}
\newcommand{\jSixteenMagnitude}{8.51}
\newcommand{\jSixteenArea}{20446}
\newcommand{\jSixteenSegments}{20}
\newcommand{\jSixteenComponents}{5}

\newcommand{\jSixteenJumps}{19}

\newcommand{\jSixteenRelaxedJumps}{3}

\newcommand{\dataTotalModels}{472}

\newcommand{\dataCandidateModels}{454}
\newcommand{\dataNonCrustalModels}{236}
\newcommand{\dataMagnitudeAreaModels}{218}

\newcommand{\dataComponentSeparation}{5}
\newcommand{\dataComponentDipDelta}{20}

\newcommand{\jOneSingleN}{197}
\newcommand{\jOneSingleMean}{0.036}
\newcommand{\jOneSingleSd}{0.275}
\newcommand{\jOneMultiN}{21}
\newcommand{\jOneMultiMean}{-0.015}
\newcommand{\jOneMultiSd}{0.217}

\newcommand{\jThreeWithinExponent}{1.080}
\newcommand{\jThreeWithinStandardError}{0.050}
\newcommand{\jThreeWithinModels}{67}
\newcommand{\jThreeWithinGroups}{71}
\newcommand{\jThreeWithinSegments}{320}

\newcommand{\jThreeBetweenExponent}{1.402}
\newcommand{\jThreeBetweenStandardError}{0.116}

\newcommand{\jThreeBetweenSegments}{52}

\newcommand{\nshmCrustalRated}{3,884}

\newcommand{\dataReportedMulti}{70}
\newcommand{\dataCurvedSingle}{49}
\newcommand{\dataTrueMulti}{21}

\newcommand{\kFourSigmaM}{0.79}
\newcommand{\kFourComponents}{52}
\newcommand{\kFourRuptures}{21}

\newcommand{\kFourHeteroscedasticityP}{0.98}

\newcommand{\jTenJumpPairs}{46}

\newcommand{\jNineN}{59}

\newcommand{\jTenPoolPairs}{46}
\newcommand{\jTenPoolEvents}{21}

\newcommand{\jTenEvents}{21}
\newcommand{\jTenTriggered}{72}
\newcommand{\jTenBeyondLimit}{7}
\newcommand{\jTenShallow}{19}

\newcommand{\jSevenDistanceN}{3,506}
\newcommand{\jSevenModalIsLinePct}{95.1}

\newcommand{\jSevenMassWithinOneMean}{0.992}

\newcommand{\jFiveGeometries}{3}

\newcommand{\jFourteenGeometricNegative}{6}
\newcommand{\jFourteenGeometricNegativePct}{13}

\newcommand{\tFortyOneAVelocity}{4.06}

\newcommand{\tFortyOneAExponent}{1.08}

\newcommand{\tFortyOneADeltaAic}{0.0}
\newcommand{\tFortyOneANegativePct}{34}
\newcommand{\tFortyOneBVelocity}{4.10}

\newcommand{\tFortyOneBDeltaAic}{1.6}
\newcommand{\tFortyOneBNegativePct}{34}
\newcommand{\tFortyOneCVelocity}{2.80}

\newcommand{\tFortyOneCExponent}{1.22}

\newcommand{\tFortyOneCDeltaAic}{1.0}
\newcommand{\tFortyOneCNegativePct}{28}

\newcommand{\kOneRated}{3,546}
\newcommand{\kOneCompleted}{3,506}

\newcommand{\kOneGeometryFailed}{40}

\newcommand{\kOneRuntimeMedian}{2.1}

\newcommand{\kOneRuntimeTotalHours}{3.9}
\newcommand{\kOneJumps}{18,824}
\newcommand{\kOneRelaxedJumps}{1,203}
\newcommand{\kOneRelaxedJumpsPct}{6.4}
\newcommand{\kOneRelaxedRuptures}{599}

\newcommand{\kFiveRealizations}{200}
\newcommand{\kFiveArrivalLow}{591}
\newcommand{\kFiveArrivalHigh}{636}
\newcommand{\kFiveArrivalReference}{614}
\newcommand{\kFiveNegativePct}{42}
\newcommand{\kFiveTruncatedPct}{0.7}
\newcommand{\kFiveDelayBranch}{B}

\newcommand{\kFiveMaxSlipLow}{11.6}
\newcommand{\kFiveMaxSlipHigh}{36.5}

\newcommand{\kFiveRefSlipHigh}{11.8}

\newcommand{\jSixDarfieldFaults}{7}
\newcommand{\jSixDarfieldEdges}{18}
\newcommand{\jSixDarfieldModalProbability}{0.108}
\newcommand{\jSixDarfieldTrees}{5{,}292}
\newcommand{\jSixKaikouraFaults}{12}
\newcommand{\jSixKaikouraEdges}{28}
\newcommand{\jSixKaikouraModalProbability}{0.087}
\newcommand{\jSixKaikouraTrees}{465{,}360}
\newcommand{\jSixAlpineHopeFaults}{20}
\newcommand{\jSixAlpineHopeEdges}{24}
\newcommand{\jSixAlpineHopeModalProbability}{0.972}
\newcommand{\jSixAlpineHopeTrees}{216}

\newcommand{\jTenKaikouraLowDelayS}{$-$12}
\newcommand{\jTenKaikouraLowDistanceKm}{8.2}
\newcommand{\jTenKaikouraHighDelayS}{+8}
\newcommand{\jTenKaikouraHighDistanceKm}{8.1}

\newcommand{\jSixDarfieldCappedEdges}{7}
\newcommand{\jSixDarfieldContactEdges}{4}
\newcommand{\jSixDarfieldOtherEdges}{11}
\newcommand{\jSixDarfieldNextProbability}{0.887}
\newcommand{\jSixDarfieldRemainingEdges}{10}
\newcommand{\jSixDarfieldRemainingMaxProbability}{0.287}
\newcommand{\jSixDarfieldWithinOne}{0.542}
\newcommand{\jSixDarfieldWithinTwo}{0.988}
\newcommand{\jSixKaikouraWithinOne}{0.430}
\newcommand{\jSixKaikouraWithinTwo}{0.836}

%% file: sections/00-abstract.tex
\begin{abstract}
	Contemporary seismic source models routinely define ruptures upon multiple, discontiguous fault structures, yet kinematic rupture generators for physics-based ground-motion simulation are formulated for single planar structures.
	This paper presents a generalization of kinematic rupture generation to arbitrarily complex multi-segment fault geometries.
	Three aspects are addressed: (i) the partitioning of seismic moment between the participating segments; (ii) the path of rupture over them from a prescribed hypocenter; and (iii) the location and timing of each jump between segments.
	A model is proposed for each, and calibrated against crustal multi-component ruptures in the SRCMOD database of finite-fault inversions.
	Moment is partitioned hierarchically, first between segments, which are physically distinct structures, and then between the components within each, so that the average slip does not depend on how a segment is divided into components.
	The rupture path model is a spanning tree of the components, weighted by a jumping probability that decreases with separation, with each jump between the closest points of the two components.
	The time delay in jumping between segments is modeled as proportional to the jumping distance, with a timing perturbation that reflects both real rupture jumping delays and uncertainty in locating where rupture crosses between segments.
	The developed multi-segment rupture generator requires no information beyond the participating fault geometries, the total magnitude, and a hypocenter, and is therefore applicable without modification to contemporary seismic source models used, with the 2022 New Zealand National Seismic Hazard Model adopted as an illustrative application.
\end{abstract}

\noindent\textbf{Keywords:} kinematic rupture generator; multi-segment rupture; physics-based ground-motion simulation; probabilistic seismic hazard analysis; earthquake rupture forecast

%% file: sections/01-introduction.tex
\section{Introduction}
\label{sec:introduction}

Multi-segment ruptures, the rupture of two or more fault structures approximated as distinct and discontiguous structures, such as the 2016 Kaik\=oura \citep{hamling_complex_2017,bradley_groundmotion_2017} and 2023 T\"urkiye \citep{melgar_sub_2023,jia_complex_2023} earthquakes, are increasingly being observed as the quality and quantity of post-earthquake observational data increase.
In parallel, the increasing sophistication of seismic source models now allows for complex ruptures that include multiple segments, motivated in part by the move to generalize away from the segmentation and characteristic earthquake paradigms \citep{field_estimating_2011,field_usgs_2023,gerstenberger_seismicity_2024}.

Multi-segment rupture geometries present a challenge for the modeling of earthquake-induced ground motions using both empirical and physics-based ground-motion models.
For example, in the NGA-West2 database, upon which the NGA-West2 ground-motion models (the most commonly used ground-motion models where multi-segment ruptures are of concern) are based \citep{gregor_comparison_2014}, multi-segment geometries are retained for a small number of earthquakes and source-to-site distances are measured to the closest segment, but the source parameters used as ground-motion model predictors (e.g., dip, rake and rupture width) are reduced to a single value for each event, and some multi-segment ruptures, including the 2010 Darfield earthquake, were simplified to a single plane \citep{ancheta_peer_2013,ancheta_nga-west2_2014}.
As a result, the use of empirical ground-motion models for multi-segment ruptures can be problematic, since such models adopt the overall rupture magnitude and the distance to the nearest point on the rupture ($R_{rup}$ or $R_{jb}$), irrespective of whether that point lies on the main rupture plane or on a small splay fault that contributes in a minor way to the total moment release \citep[e.g.,][]{thompson_case_2018,bradley_groundmotion_2017}.

For physics-based ground-motion simulation, the propagation of the seismic waves that result from a complex rupture is relatively tractable, as a result of the time-domain solution of the wave propagation problem.
However, the development of the multi-segment rupture description itself has not been addressed in a generalized form suitable for simulations that seek to provide an alternative to empirical ground-motion models in probabilistic seismic hazard analysis \citep[e.g.,][]{graves_cybershake_2011,rojas_earthquake_2024,bradley_cybershake_2026}.
In this regard, a generalized solution refers to a set of rules that allows a kinematic rupture to be generated over any multi-segment fault geometry.
Simulations of specific historical multi-segment earthquakes have instead relied upon manual heuristics, in which the partitioning of moment between segments, the sequence in which the segments rupture, and the inter-segment time delays are each prescribed subjectively on an event-by-event basis \citep[e.g.,][]{asano_estimation_2005,kuriyama_examination_2011,bradley_groundmotion_2017}.

Dynamic rupture modeling provides a natural computational setting in which to examine the propagation of rupture across a multi-segment fault structure, since whether rupture arrests at, or propagates through, a discontinuity between fault structures emerges from the physics of the problem.
However, a dynamic rupture calculation requires a fault friction law and its parameters, the initial stress state over the participating fault surfaces, and a nucleation procedure \citep{ramos_working_2022}, none of which are specified by a conventional seismic source model that defines only the fault geometry, magnitude and hypocenter location.
These quantities are also not generally resolvable from the observations that are typically available \citep[e.g.,][]{guatteri_what_2000,ripperger_earthquake_2007}.
For these reasons, kinematic rupture generators, which reproduce the well-established features of dynamic rupture models and are sometimes referred to as pseudo-dynamic rupture generators, are most commonly used in ground-motion simulation for seismic hazard and engineering applications \citep{baker_seismic_2021}.

The objective of this study is to extend kinematic rupture generation from a single planar structure (or a non-planar but contiguous structure that can be mapped onto a plane) to a complex multi-segment rupture geometry.
Section~\ref{sec:usecase} outlines the target application for such a generator, and hence the design requirements that determine the approach taken.
Section~\ref{sec:framework} introduces the three aspects that must be addressed to extend single-plane kinematic generators to the multi-segment case, and Sections~\ref{sec:momentPartitioning} to~\ref{sec:jumping} propose a model for each and calibrate it against the SRCMOD database of finite-fault rupture models (Section~\ref{sec:srcmod}).
Section~\ref{sec:procedure} sets out the resulting generation procedure, and Section~\ref{sec:application} demonstrates its application for forward modeling.
Finally, Section~\ref{sec:conclusions} summarizes the conclusions of this first attempt at a multi-segment kinematic generator, together with its limitations.

%% file: sections/02-usecase.tex
\section{Target use-case and design requirements}
\label{sec:usecase}

\begin{figure}[htb!]
	\centering
	\includegraphics[width=\linewidth]{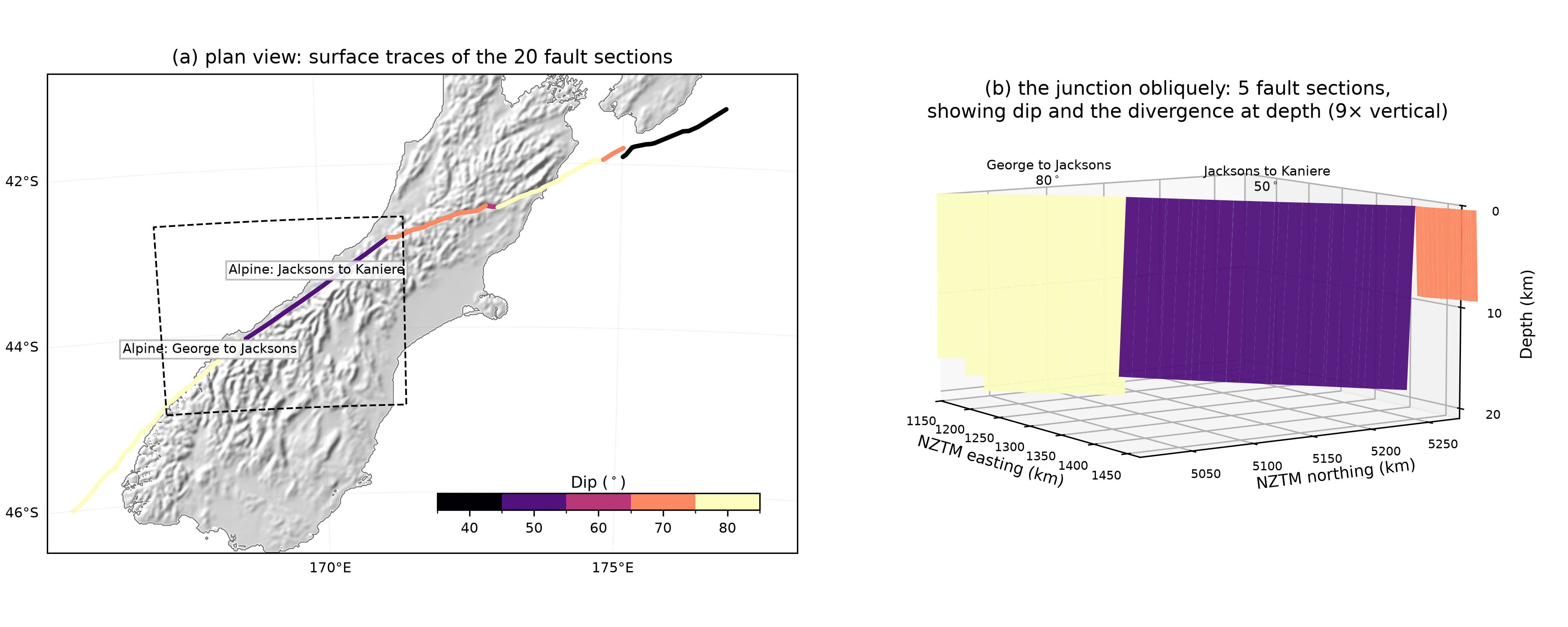}
	\caption{Illustration of fault surfaces as represented in a contemporary source model, for the \jSixteenSegments{} fault sections (components) participating in one Alpine--Marlborough rupture of the 2022 NZ NSHM \citep{gerstenberger_seismicity_2024}: (a) plan view of the surface traces, colored by dip, with the broken rectangle marking the region of panel (b); and (b) the same fault sections near one junction, viewed obliquely and close to along-strike so that the dip of each surface can be read directly.}
	\label{fig:sourceGeometryUseCase}
\end{figure}

The target application considered here is the use of the generator for probabilistic seismic hazard analysis (PSHA) and, in particular, national seismic hazard mapping.
In such applications the fault structures are defined as planar polygons, which are regionally accurate but often do not possess a high level of geometric complexity.
There may also be incompatibility at the boundaries of different fault components, since each component has a fixed dip angle, so that two adjoining faults with a connected surface trace but differing dip angles exhibit a, sometimes significant, discontinuity at depth.
Figure~\ref{fig:sourceGeometryUseCase} illustrates both features for the faults participating in one Alpine--Marlborough rupture (NSHM rupture ID 71072) of the 2022 New Zealand National Seismic Hazard Model (NZ NSHM).
In plan the fault geometries are very nearly lines, and the surface traces of the George to Jacksons and Jacksons to Kaniere sections of the Alpine Fault join, yet the two fault surfaces diverge by $30^\circ$ with depth.
It is therefore desirable that the rupture generator is not overly sensitive to such incompatibilities.
These incompatibilities will ideally be addressed in the development of improved seismic source models over time, but it is preferable that changes to a source model fault geometry are generally not required in order to generate realistic ruptures for ground-motion simulation.

In this problem setting, a specific fault geometry is prescribed by the earthquake rupture forecast of the seismic source model, i.e., the set of ruptures, together with their rates of occurrence, that the seismic source model defines.
A hypocenter location is also prescribed or, alternatively, randomly sampled.
This is the setting in which CyberShake applications have previously been undertaken, with the distinction that prior CyberShake studies have adopted seismic source models comprising single planar structures \citep{graves_cybershake_2011,rojas_earthquake_2024,callaghan_cybershake_2026,bradley_cybershake_2026}.
As a result, CyberShake programs have arguably not been able to adopt the most recent seismic source models in their respective regions, and have instead adopted source models of an earlier generation.
In California, for example, two US national seismic hazard models have adopted UCERF3-style multi-segment rupture forecasts \citep{field_usgs_2023}, and in New Zealand the 2022 NZ NSHM also adopted this methodology \citep{gerstenberger_seismicity_2024}.

A further characteristic of the target application, which is easily overlooked, is one of scale.
Inversion-based earthquake rupture forecasts contain very large numbers of candidate ruptures.
For example, UCERF3 solves for the rates of approximately 250,000 fault-based ruptures, an increase from approximately 8,000 in its predecessor \citep{field_uniform_2014}, and the crustal fault rupture set of the 2022 NZ NSHM comprises~\nshmCrustalRated{} unique non-zero rate ruptures, obtained by applying a maximum jump distance of 15~km together with the plausibility filters of \citet{milner_enumerating_2022} to the fault geometries of the deformation model \citep{gerstenberger_seismicity_2024}.
A ground-motion simulation-based hazard calculation requires a kinematic rupture description for each rupture considered, and in general for multiple realizations of each.
It follows that the generator must operate without expert intervention on any individual rupture, in contrast with the ability to adopt heuristic decisions for a single event of interest.

Taken together, these considerations lead to four requirements that the multi-segment kinematic rupture generator developed herein is designed to satisfy:

\begin{enumerate}
	\item[\textbf{R1}] \emph{Operate on the source model geometry as given.}
	      The generator should accept the fault surfaces defined by the earthquake rupture forecast without modification, so that the adoption of physics-based ground-motion simulation does not require a bespoke revision of the seismic source model.
	\item[\textbf{R2}] \emph{Tolerate geometric incompatibility between adjoining components.}
	      The resulting rupture should not be unduly sensitive to discontinuities that arise from the independent parameterization of adjoining fault structures, in particular differences in dip beneath a connected surface trace.
	\item[\textbf{R3}] \emph{Require no fault-specific information beyond that which the source model provides.}
	      The fault-specific inputs available are the participating fault geometries, the total magnitude, and a hypocenter location that is either prescribed or sampled.
	      As with single-segment kinematic generators, the generator may adopt default parameters that vary regionally (e.g., the crustal velocity structure), but it should not require fault-specific quantities, such as the initial stress state or frictional parameters, that the earthquake rupture forecast does not specify.
	\item[\textbf{R4}] \emph{Admit both deterministic and stochastic execution, at scale.}
	      The generator should be automatable over the full rupture set of an earthquake rupture forecast, and should permit each of its aspects to be evaluated either as a single best-estimate reference case or as a randomly sampled realization.
\end{enumerate}

%% file: sections/03-framework.tex
\section{Framework for generalizing from single- to multi-segment kinematic rupture generation}
\label{sec:framework}

Figure~\ref{fig:MultiSegFrameworkDiag} schematically illustrates a multi-segment source geometry over which rupture is to be determined, and this section introduces the three aspects that are required to extend a single-segment kinematic rupture generator to such a geometry.
In order to develop the mathematical approach, the following terms are defined:
$S_k$ is the 2D surface of segment $k$, of area $A_k$, which may be planar or non-planar;
a segment may comprise several components, the pieces by which the source model parameterizes it (each reported plane in SRCMOD, and each fault section in the NZ NSHM, which may comprise several planes joined end to end), and $C_i$ is the surface of component $i$;
$A_i$ is the component area;
$M_{o,i}$ is the seismic moment of component $C_i$;
$h_i$ is the `hypocenter' of component $i$, i.e., the point at which rupture initiates on that component (only $h_1$ is a true hypocenter, but the remainder serve the same function in a modeling context);
$h_i^{'}$ is the conjugate point of $h_i$, i.e., the location on the triggering component that acts as the reference point for the subsequent initiation of rupture on $C_i$ at $h_i$;
$r_i = \left|h_i - h_i^{'}\right|$ is the jumping distance;
and $t_i = t_{init}(h_i) - t_{init}(h_i^{'})$ is the time delay between slip initiation at $h_i^{'}$ and at $h_i$.

\begin{figure}[htb!]
	\centering
	\resizebox{\linewidth}{!}{%
		\begin{tikzpicture}[every node/.style={font=\large}]
			\tikzset{
			segline/.style={line width=0.9pt, black},
			front/.style={dashed, red, line width=0.9pt},
			pt/.style={circle, fill=black!65, inner sep=0pt, minimum size=7pt},
			jump/.style={{Latex[length=2.2mm]}-{Latex[length=2.2mm]}, blue!50!black, line width=0.8pt},
			}
			\coordinate (1a) at (0.9,4.00);  \coordinate (1b) at (4.1,4.62);
			\coordinate (1c) at (4.1,1.42);  \coordinate (1d) at (0.9,0.80);
			\coordinate (1e) at (7.3,4.92);  \coordinate (1f) at (7.3,1.72);
			\coordinate (2a) at (7.95,4.92); \coordinate (2b) at (14.35,4.32);
			\coordinate (2c) at (14.35,1.10); \coordinate (2d) at (7.95,1.72);
			\coordinate (ia) at (13.6,4.57); \coordinate (ib) at (20.0,5.67);
			\coordinate (ic) at (20.0,2.45); \coordinate (id) at (13.6,1.36);
			\coordinate (h1) at (1.75,2.85);
			\coordinate (h2p) at (6.58,3.94); \coordinate (h2) at (8.25,3.95);
			\coordinate (hip) at (12.95,3.30); \coordinate (hi) at (14.25,3.80);

			\begin{scope}
				\clip (ia) -- (ib) -- (ic) -- (id) -- cycle;
				\foreach \radius in {1.6, 3.9, 6.1}
				\draw[front] (hi) circle (\radius);
			\end{scope}
			\draw[segline] (ia) -- (ib) -- (ic) -- (id);

			\fill[white] (2a) -- (2b) -- (2c) -- (2d) -- cycle;
			\begin{scope}
				\clip (2a) -- (2b) -- (2c) -- (2d) -- cycle;
				\foreach \radius in {1.0, 3.0, 5.6}
				\draw[front] (h2) circle (\radius);
			\end{scope}
			\draw[segline] (2a) -- (2b) -- (2c) -- (2d) -- cycle;
			\draw[black, dotted, line width=0.9pt] (ia) -- (id);
			\draw[black, dotted, line width=0.9pt] (id) -- ($(id)!0.11!(ic)$);

			\begin{scope}
				\clip (1a) -- (1b) -- (1e) -- (1f) -- (1c) -- (1d) -- cycle;
				\foreach \radius in {1.4, 2.7, 4.1, 5.5}
				\draw[front] (h1) circle (\radius);
			\end{scope}
			\draw[segline] (1a) -- (1b) -- (1e) -- (1f) -- (1c) -- (1d) -- cycle;
			\draw[black, dotted, line width=0.9pt] (1b) -- (1c);

			\node[star, star points=5, star point ratio=2.2, fill=white, draw=black,
				inner sep=2.4pt] at (h1) {};
			\node[anchor=west] at ($(h1)+(0.25,-0.05)$) {$h_1$};
			\node[pt] at (h2p) {}; \node[anchor=north] at ($(h2p)+(0,-0.15)$) {$h_2^{'}$};
			\node[pt] at (h2) {};  \node[anchor=north west] at ($(h2)+(0.05,-0.1)$) {$h_2$};
			\draw[jump] ($(h2p)+(0.15,0)$) -- ($(h2)+(-0.15,0)$)
			node[midway, above=3pt, blue!50!black] {$r_2$};
			\node[pt] at (hip) {}; \node[anchor=north] at ($(hip)+(0,-0.15)$) {$h_i^{'}$};
			\node[pt] at (hi) {};  \node[anchor=south west] at ($(hi)+(0.1,0.1)$) {$h_i$};
			\draw[jump] ($(hip)!0.12!(hi)$) -- ($(hip)!0.88!(hi)$)
			node[midway, above left=1pt, blue!50!black] {$r_i$};

			\node[rotate=5] at (3.9,5.05) {Segment 1};
			\node[rotate=-5] at (11.2,5.0) {Segment 2};
			\node[rotate=10] at (16.9,5.55) {Segment $i$};
			\node[align=left, anchor=north west] at (3.9,1.30) {Area, $A_1$\\Moment, $M_{o,1}$};
			\node[anchor=north] at (10.9,1.25) {$A_2$, $M_{o,2}$};
			\node[anchor=north] at (17.0,1.80) {$A_i$, $M_{o,i}$};
			\node[red, anchor=north] (rf) at (1.4,0.45) {Rupture front};
			\draw[red, -{Latex[length=2mm]}] (rf.north) -- ($(h1)+(-48:1.4)$);
		\end{tikzpicture}%
	}
	\caption{Schematic illustration of a multi-segment rupture geometry and the notation of Section~\ref{sec:framework}.
		Segment~1 changes strike along its length (dotted line), and is therefore non-planar but a single segment, here parameterized as a single component of two planes.}
	\label{fig:MultiSegFrameworkDiag}
\end{figure}

It is important to distinguish between a rupture segment and a planar rupture geometry.
Kinematic rupture generators based on 2D spatial random field theory have been applied to non-planar single-segment geometries, which principally comprise two variations, namely: (1) a change in strike along the length of the rupture; and/or (2) a change in dip down the width of the rupture.
In both cases, these variations are handled by a mapping between the non-planar geometry and a corresponding planar geometry upon which the random field is generated \citep[e.g.,][]{graves_broadband_2010,graves_kinematic_2016,pitarka_kinematic_2020}.
In the multi-segment extension here, the same terminology is retained, so that the focus is on discontinuities in the fault rupture that are sufficiently large that a simple geometric mapping is insufficient.
These include not only the separation distances between fault structures but also differences in fault characteristics, such as the sense of slip (e.g., strike-slip and dip-slip) and the maximum depth of the participating structures (e.g., splay faults).

With this distinction established, the three aspects that must be prescribed in order to generate a kinematic rupture over the geometry of Figure~\ref{fig:MultiSegFrameworkDiag} can be stated.
First is the partitioning of moment.
Since the total moment of the rupture, $M_{o,T}$, is obtained from the total rupture area, $A_T$, by way of a magnitude-area scaling relation, the moment $M_{o,i}$ of each participating component must be determined as some function of the segment and component properties, and in particular of the component area $A_i$.
For example, the partitioning adopted determines whether the average slip displacement is equal over all segments or varies between them.
Second is the macro-scale path of the rupture, i.e., which component triggers the rupture of each component, and hence the sequence in which the segments rupture (the grouping of components into segments is defined in Section~\ref{sec:apportionmentModel}).
For a geometry as simple as that of Figure~\ref{fig:MultiSegFrameworkDiag} the path is self-evident, but general rules are required for the arbitrarily complex geometries that a contemporary earthquake rupture forecast contains.
Third is the specification of the locations at which jumping occurs and of the associated timing, i.e., the location $h_i$ at which rupture initiates on each triggered component, its conjugate point $h_i^{'}$ on the triggering component, and the time delay $t_i$ between slip initiation at the two locations.

%% file: sections/04-srcmod-data.tex
\section{Calibration data: the SRCMOD database}
\label{sec:srcmod}

The models proposed for the three aspects in Sections~\ref{sec:momentPartitioning}, \ref{sec:rupturePath} and~\ref{sec:jumping}, respectively, are calibrated against the SRCMOD database of finite-fault rupture models \citep{mai_srcmod_2014}, using the distribution of the database available as of 2~December~2024 (see Data and Resources).
Each aspect requires different quantities to be reported in the source inversion, and the calibration of the moment partitioning and of the jumping also requires that the rupture comprise multiple components; a different subset of the database is used for each aspect as summarized in Table~\ref{tab:srcmod-datasets}.
Models in which every sub-fault patch is reported as a separate component are excluded throughout, since such models do not distinguish the segmentation of the source from its discretization, leaving \dataCandidateModels{} of the \dataTotalModels{} models with usable data.
Only inversions of crustal earthquakes are used since the generator's primary application is to the crustal fault model of the 2022 NZ NSHM.
Excluding the \dataNonCrustalModels{} subduction-interface, intraslab and oceanic models leaves \dataMagnitudeAreaModels{} models, all of which are used for the magnitude-area relationship (Section~\ref{sec:maScaling}).
All \dataReportedMulti{} of the remaining models that report more than one component are used for the moment partitioning (Section~\ref{sec:apportionmentCalib}), of which \dataTrueMulti{} contain a discontinuity between segments, and \jTenEvents{} report rupture-onset times and jump pairs, supplying \jTenJumpPairs{} jump pairs for the calibration of the jumping location and time delay (Section~\ref{sec:jumpCalibration}).
Rupture areas are trimmed by the rule of \citet{somerville_characterizing_1999} throughout.
Section~S1 of the electronic supplement documents the extraction of the dataset, the rules that define each subset, and the event lists.

\begin{table}[htb!]
	\centering
	\begin{threeparttable}
		\caption{Subsets of the SRCMOD database utilized to calibrate each aspect of the framework.}
		\label{tab:srcmod-datasets}
		\small
		\begin{tabular}{@{}l c p{5.4cm} r r@{}}
			\toprule
			Aspect                          & Section                      & Information required                                    & Models                     & Jump pairs       \\
			\midrule
			Magnitude--area scaling         & \ref{sec:maScaling}          & Moment magnitude and rupture area              & \dataMagnitudeAreaModels{} & ---              \\
			\addlinespace
			Moment partitioning             & \ref{sec:apportionmentCalib} & Multiple components, forming one or more segments, with slip and geometric dimensions & \dataReportedMulti{}       & ---              \\
			\quad Single segment   &                              &                                                         & \dataCurvedSingle{}        & ---              \\
			\quad Two or more segments      &                              &                                                         & \dataTrueMulti{}           & ---              \\
			\addlinespace
			Jumping location and time delay & \ref{sec:jumpCalibration}    & Two or more components, with rupture-onset times          & \jTenEvents{}              & \jTenJumpPairs{} \\
			\bottomrule
		\end{tabular}
	\end{threeparttable}
\end{table}

%% file: sections/05-moment-partitioning.tex
\section{Moment partitioning}
\label{sec:momentPartitioning}

The generation of a kinematic rupture over a multi-segment geometry requires that the total seismic moment of the rupture be distributed over the participating segments.
This section first examines whether the magnitude-area scaling relations from which the total moment is obtained are applicable to ruptures that are geometrically multi-segment (Section~\ref{sec:maScaling}).
Section~\ref{sec:apportionmentModel} then develops the model adopted for apportioning the total moment, which is illustrated by means of a simple example in Section~\ref{sec:apportionmentExample} and calibrated against the crustal ruptures of the SRCMOD database in Section~\ref{sec:apportionmentCalib}.

\subsection{Magnitude-area scaling of multi-segment ruptures}
\label{sec:maScaling}

The total seismic moment of a rupture, $M_{o,T}$, is obtained from its total area, $A_T$, by way of an empirical magnitude-area scaling relation, for which the average-rake form of the relation of \citet{leonard_selfconsistent_2014}, $A = 10^{M_w - 3.995}$, is adopted herein (the coefficient being the mean of the strike-slip and dip-slip values).
Such relations are derived predominantly from ruptures that have been represented as single planar structures, although some of the larger historical ruptures may have been multi-segment even where they were idealized as a single plane.
Their applicability to ruptures that are geometrically multi-segment is therefore not self-evident.

\begin{figure}[htb!]
	\centering
	\includegraphics[width=0.65\linewidth]{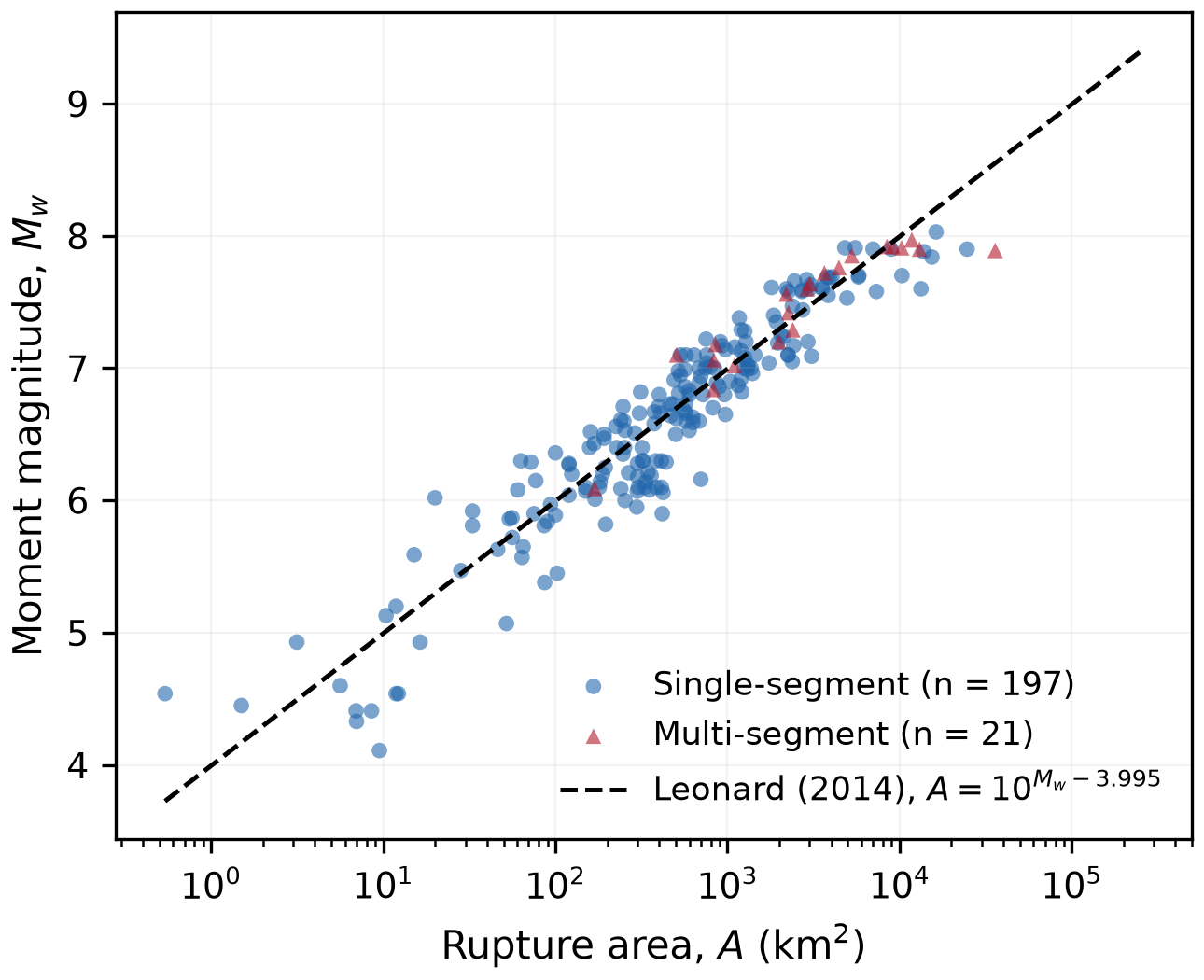}
	\caption{Comparison of single- and multi-segment SRCMOD crustal ruptures with the \citet{leonard_selfconsistent_2014} magnitude-area scaling relation, illustrating that multi-segment ruptures exhibit consistent $M_w$--$A$ scaling.}
	\label{fig:mw_area}
\end{figure}

Figure~\ref{fig:mw_area} compares the moment magnitude and rupture area of the \dataMagnitudeAreaModels{} SRCMOD crustal rupture models with the adopted \citet{leonard_selfconsistent_2014} relation, distinguishing the \jOneSingleN{} single-segment models from the \jOneMultiN{} multi-segment models.
Segments are identified by the grouping rule subsequently discussed in Section~\ref{sec:apportionmentModel}.
It is evident that the multi-segment ruptures are no less consistent with the relation than the single-segment ruptures, the residual in the natural logarithm of area having a mean of \lnof{\jOneMultiMean} and a standard deviation of \lnof{\jOneMultiSd} for the multi-segment ruptures, compared with \lnof{\jOneSingleMean} and \lnof{\jOneSingleSd} for the single-segment ruptures.
On this basis, $M_{o,T} = f(A_T)$ is adopted for multi-segment ruptures throughout.

\subsection{Apportionment of moment between segments}
\label{sec:apportionmentModel}

Magnitude-area scaling relations are approximately self-similar, having the general form $M_w = \log_{10}A_T + c$.
Since moment magnitude is related to seismic moment by $M_w = \frac{2}{3}\left(\log_{10}M_o - 9.05\right)$, such a relation implies that $M_o \propto A^{3/2}$.
Furthermore, since $M_o = \mu \bar{D} A$, where $\bar{D}$ is the average slip displacement and the shear modulus $\mu$ may be taken as constant, it follows that $\bar{D} \propto A^{1/2}$, so that average slip increases with rupture size and seismic moment is not proportional to rupture area.
As a result, the moment of a multi-segment rupture cannot be obtained by applying the scaling relation to each segment independently and summing.
As an example, for a rupture of total area $A_T$ subdivided into $n$ components of equal area, $\sum_{i} f(A_i) = n\left(A_T/n\right)^{3/2} = A_T^{3/2}n^{-1/2}$, which is smaller than $f(A_T) = A_T^{3/2}$ by a factor of $\sqrt{n}$.
Such a total moment would depend upon the number of components into which the rupture happens to have been divided, which for a single continuous rupture plane is an arbitrary choice.
It follows that the total moment must be obtained from the total area, $M_{o,T} = f(A_T)$, and subsequently apportioned among the segments.

The same consideration governs how the apportionment is constructed, and motivates a distinction between what are referred to herein as rupture \emph{segments} and the \emph{components} within a segment.
A segment is a set of components that are sufficiently close to one another to represent a single physical structure.
Within a segment, the modeled components arise from the manner in which the source model parameterizes a continuous fault (e.g., a change in strike along its length or in dip down its width), and not from a physical discontinuity.
Because this internal delineation is an artifact of the source model, the apportionment within a segment must be invariant to it.
Assigning moment in proportion to component area has this property, since it implies an average slip $\bar{D}_i = M_k / \left(\mu \sum_{j \in S_k} A_j\right)$ that is the same for every component of the segment and independent of how the segment has been subdivided.
In contrast, an exponent of $3/2$ applied within a segment does not, since the average slip of a component would then depend upon where the component boundaries had been placed, in the ratio $\bar{D}_i/\bar{D}_j = \left(A_i/A_j\right)^{1/2}$, which grows without bound as the delineation becomes increasingly uneven (Section~S2.3 of the electronic supplement).
Between segments, the relative areas reflect physically distinct structures, and the self-similar exponent of $3/2$ is retained.
These concepts are subsequently shown to be consistent with empirical data (Section~\ref{sec:apportionmentCalib}).

Components are grouped into segments according to their separation and dip, two components belonging to the same segment when the distance between their closest points below a depth of 5~km (the depth constraint of Section~\ref{sec:jumpLocation}) is less than \dataComponentSeparation{}~km and their dips differ by less than $\dataComponentDipDelta{}^\circ$.
Following this grouping, the total moment $M_{o,T}$ is distributed between the segments, $S_k$, of areas $A_k$, in proportion to $A_k^{3/2}$,
\begin{equation}
	\label{eq:moment-distribution-between-component}
	M_{k} = M_{o,T}\frac{A_{k}^{3/2}}{\sum_{k}A_{k}^{3/2}},
\end{equation}
and a component $C_i$ within a segment of moment $M_k$ is apportioned moment in proportion to its area,
\begin{equation}
	\label{eq:moment-distribution-in-component}
	M_{o,i} = M_k\frac{A_{i}}{\sum_{j \in S_k}A_{j}},
\end{equation}
so that the total moment of the rupture is preserved.
Equations~(\ref{eq:moment-distribution-between-component}) and~(\ref{eq:moment-distribution-in-component}) assume that the shear modulus is uniform over the participating components.
Where components occupy differing depth ranges this is not the case, and the apportionment generalizes to a rigidity-weighted area, $\mu_i A_i$, which preserves the invariance established above (Section~S2.4 of the electronic supplement).

\subsection{Illustrative example}
\label{sec:apportionmentExample}

Figure~\ref{fig:simple-rupture} illustrates the moment apportionment for a simple multi-segment geometry that comprises three components.
Components $C_1$ and $C_2$, of areas 300 and 150~km$^2$, are separated by 0.5~km and are therefore treated as belonging to the same segment, $S_1$, while component $C_3$, of area 300~km$^2$, is separated from $C_2$ by 8~km and constitutes a segment of its own, $S_2$.
For the total rupture area of 750~km$^2$, the \citet{leonard_selfconsistent_2014} relation gives $M_w = 6.87$, and hence $M_{o,T} \approx 2.26 \times 10^{19}$\,N\,m.
Applying Equation~(\ref{eq:moment-distribution-between-component}) to the two segments, of areas 450 and 300~km$^2$, gives $M_1 = 450^{3/2}/(450^{3/2} + 300^{3/2})\,M_{o,T} \approx 0.65M_{o,T}$, and hence $M_2 \approx 0.35M_{o,T}$.
Equation~(\ref{eq:moment-distribution-in-component}) then apportions $M_1$ between $C_1$ and $C_2$ in proportion to their areas, so that they receive $43\%$ and $22\%$ of the total moment, respectively, while $C_3$, as the only component of $S_2$, receives the whole of $M_2$.
It is instructive to observe that $C_1$ and $C_3$ are of equal area, yet $C_1$ is apportioned appreciably greater moment, and hence greater average slip, as a result of belonging to the larger of the two segments.

\begin{figure}[htb!]
	\centering
	\begin{tikzpicture}

		\def\segHeight{1}
		\def\segWidth{3}
		\def\smallSegWidth{1.5}

		\def\yBase{0}
		\def\labelOffset{0.4}  %
		\def\boxPad{0.2}       %

		\def\xA{0}     %
		\def\xB{4.5}   %
		\def\xC{6.0}   %
		\def\xD{9.5}   %

		\draw[fill=gray!10] (\xA, \yBase) rectangle ++(\segWidth, \segHeight);
		\node[above] at (\xA + 0.5*\segWidth, \yBase + \segHeight) {\( C_1 \)};
		\node at (\xA + 0.5*\segWidth, \yBase + 0.5*\segHeight) {\( 300\,\text{km}^2 \)};

		\draw[fill=gray!10] (\xB, \yBase) rectangle ++(\smallSegWidth, \segHeight);
		\node[above] at (\xB + 0.5*\smallSegWidth, \yBase + \segHeight) {\( C_2 \)};
		\node at (\xB + 0.5*\smallSegWidth, \yBase + 0.5*\segHeight) {\( 150\,\text{km}^2 \)};

		\draw[fill=gray!10] (\xD, \yBase) rectangle ++(\segWidth, \segHeight);
		\node[above] at (\xD + 0.5*\segWidth, \yBase + \segHeight) {\( C_3 \)};
		\node at (\xD + 0.5*\segWidth, \yBase + 0.5*\segHeight) {\( 300\,\text{km}^2 \)};

		\draw[dashed] (\xA + \segWidth, \yBase + 0.5*\segHeight) -- (\xB, \yBase + 0.5*\segHeight);
		\node[above] at ({0.5*(\xA + \segWidth + \xB)}, \yBase + 0.5) {0.5 km};

		\draw[dashed] (\xC, \yBase + 0.5*\segHeight) -- (\xD, \yBase + 0.5*\segHeight);
		\node[above] at ({0.5*(\xC + \xD)}, \yBase + 0.5) {8 km};

		\draw[dashed, thick]
		(\xA - \boxPad, \yBase - \boxPad)
		rectangle
		(\xB + \smallSegWidth + \boxPad, \yBase + \segHeight +
		\labelOffset + \boxPad);

		\draw[dashed, thick]
		(\xD - \boxPad, \yBase - \boxPad)
		rectangle
		(\xD + \segWidth + \boxPad, \yBase + \segHeight + \labelOffset + \boxPad);

		\node[above] at
		({0.5*((\xA - \boxPad) + (\xB + \smallSegWidth + \boxPad))},
		{\yBase + \segHeight + \labelOffset + \boxPad}) {\( S_1 \)};
		\node[above] at
		({\xD + 0.5*\segWidth},
		{\yBase + \segHeight + \labelOffset + \boxPad}) {\( S_2 \)};

	\end{tikzpicture}

	\vspace{1mm}
	{\small (a)}

	\vspace{5mm}

	\begin{tikzpicture}[
			grow=right,
			level 1/.style={sibling distance=34mm, level distance=48mm},
			level 2/.style={sibling distance=26mm, level distance=48mm},
			every node/.style={draw=none, inner sep=4pt},
			edge from parent/.style={draw, ->, thick, shorten >=3pt, shorten <=3pt}
		]

		\node {\(M_{o,T}\)}
		child {
				node {\(M_2 = 0.35M_{o,T}\)}
				child {
						node {\(M_{o,3} = 0.35M_{o,T}\)}
						edge from parent node[above, midway, font=\small] {\(1.00\)}
					}
				edge from parent node[below, sloped, midway, font=\small] {\(0.35\)}
			}
		child {
				node {\(M_1 = 0.65M_{o,T}\)}
				child {
						node {\(M_{o,2} = 0.22M_{o,T}\)}
						edge from parent node[below, sloped, midway, font=\small] {\(1/3\)}
					}
				child {
						node {\(M_{o,1} = 0.43M_{o,T}\)}
						edge from parent node[above, sloped, midway, font=\small] {\(2/3\)}
					}
				edge from parent node[above, sloped, midway, font=\small] {\(0.65\)}
			};

		\node[font=\small\bfseries, anchor=south] at (0mm,38mm) {Total};
		\node[font=\small\bfseries, anchor=south] at (48mm,38mm) {Segments};
		\node[font=\small\bfseries, anchor=south] at (96mm,38mm) {Components};

	\end{tikzpicture}

	\vspace{1mm}
	{\small (b)}

	\caption{Illustrative apportionment of moment for a rupture of three components and two segments:
		(a) the rupture geometry, in which the components share a common dip, so that their grouping depends on separation distance alone;
		and (b) the two-stage apportionment by Equations~(\ref{eq:moment-distribution-between-component}) and~(\ref{eq:moment-distribution-in-component}), in which edge labels give the fraction of the parent moment apportioned along each branch and node labels the resulting fraction of the total moment.}
	\label{fig:simple-rupture}
\end{figure}

\subsection{Calibration against the SRCMOD database}
\label{sec:apportionmentCalib}

\begin{figure}[htb!]
	\centering
	\includegraphics[width=\linewidth]{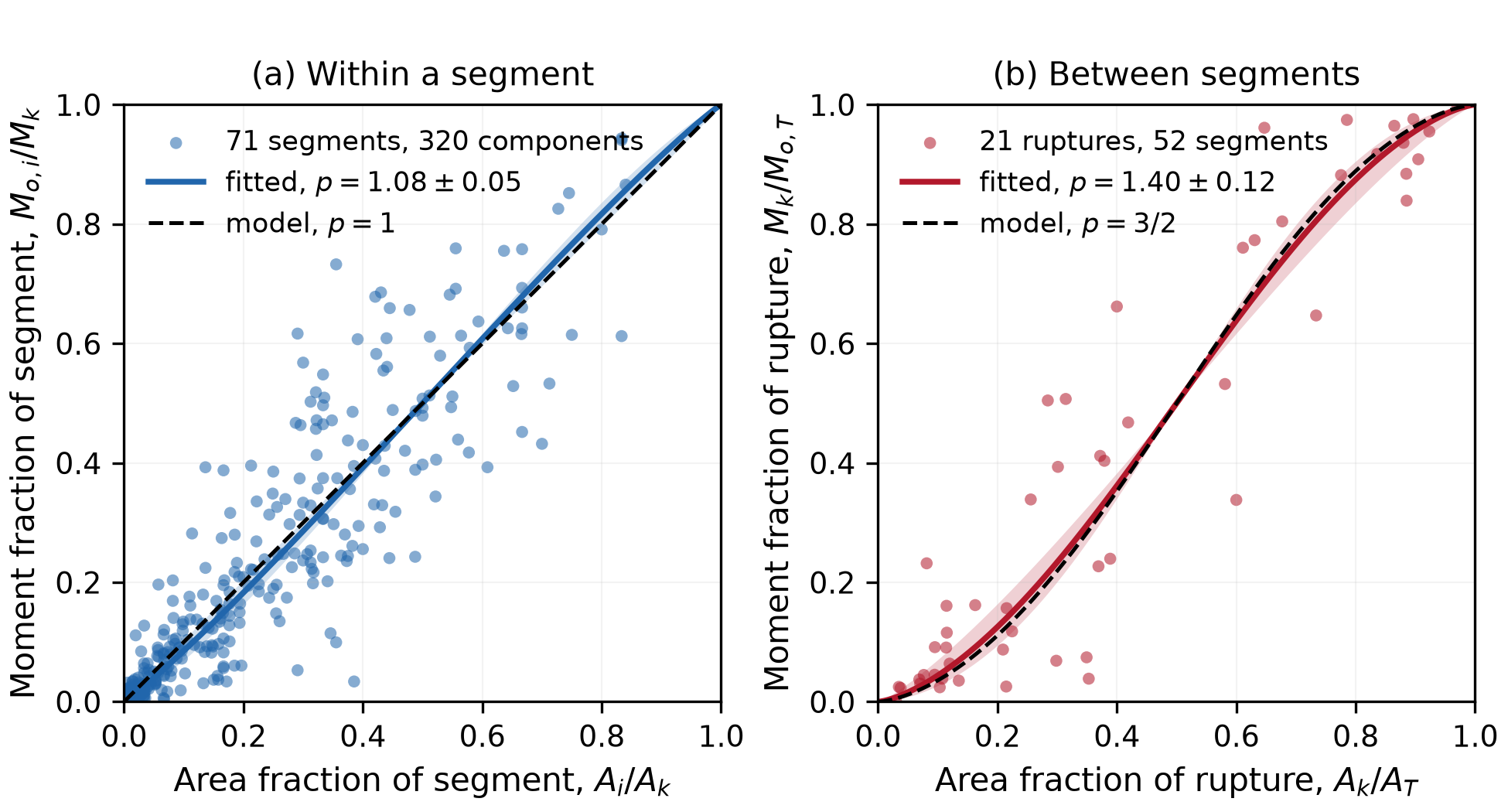}
	\caption{Moment fraction against area fraction for the multi-component SRCMOD crustal ruptures: (a) the \jThreeWithinGroups{} segments of two or more components, from \jThreeWithinModels{} ruptures, as fractions of the segment, against the within-segment exponent of unity; and (b) the segments of the \dataTrueMulti{} multi-segment ruptures, as fractions of the total rupture, against the between-segment exponent of $3/2$.
	In each panel the solid line is the exponent fitted to that population, with the $95\%$ confidence interval of the exponent shaded, and the dashed line is the model.
	Curves are drawn for two components or segments, for which one holding a fraction $f$ of the area holds $f^p/[f^p + (1-f)^p]$ of the moment.}
	\label{fig:mom_area_ratio}
\end{figure}

Figure~\ref{fig:mom_area_ratio} compares the apportionment of Equations~(\ref{eq:moment-distribution-between-component}) and~(\ref{eq:moment-distribution-in-component}) with that inferred from the multi-component SRCMOD crustal ruptures, in which the moment fraction of each component or segment is plotted against its area fraction.
In \dataCurvedSingle{} of the \dataReportedMulti{} multi-component crustal models the components are planar facets of a single curved structure, which form one segment under the rule of Section~\ref{sec:apportionmentModel}, and the remaining \dataTrueMulti{} models contain a discontinuity between segments (see Table~\ref{tab:srcmod-datasets}).
The within-segment rule (an exponent of unity) is tested on every segment of two or more components, using the moment and area fractions of each component within its own segment, $M_{o,i}/M_k$ and $A_i/A_k$.
Such segments number \jThreeWithinGroups{}, from \jThreeWithinModels{} ruptures, since several of the multi-segment ruptures also contain segments of more than one component.
The between-segment rule (an exponent of $3/2$) is tested on the segments of the \dataTrueMulti{} multi-segment ruptures, using the moment and area fractions of each segment within the rupture, $M_k/M_{o,T}$ and $A_k/A_T$.
Fitting the exponent separately to each population returns $\jThreeWithinExponent{} \pm \jThreeWithinStandardError{}$ over \jThreeWithinSegments{} components and $\jThreeBetweenExponent{} \pm \jThreeBetweenStandardError{}$ over \jThreeBetweenSegments{} segments, respectively, each being within two standard errors of the proposed model.
It follows that the hierarchical apportionment of Equations~(\ref{eq:moment-distribution-between-component}) and~(\ref{eq:moment-distribution-in-component}) is supported by the data directly, in addition to the invariance argument of Section~\ref{sec:apportionmentModel}.
Section~S2.5 of the electronic supplement gives the analysis of the two populations pooled, for which a single exponent fits neither regime, thus confirming the importance of the two-tiered segment--component hierarchy.

\begin{figure}[htb!]
	\centering
	\includegraphics[width=0.6\linewidth]{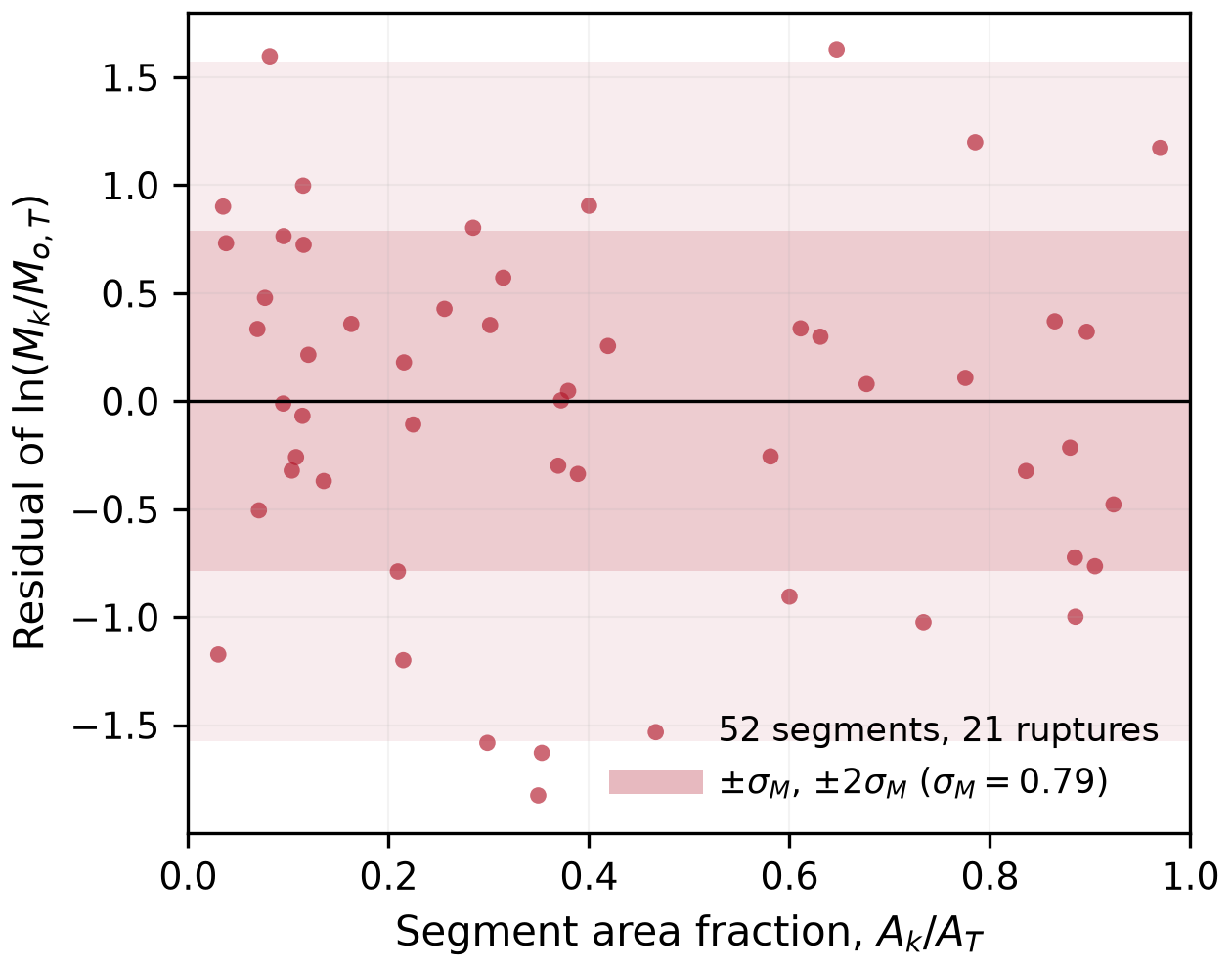}
	\caption{Residual of the logarithm of segment moment fraction about the between-segment apportionment of Equation~(\ref{eq:moment-distribution-between-component}), centered within each rupture and scaled by $\sqrt{K/(K-1)}$ for a rupture of $K$ segments, against segment area fraction, for the \kFourComponents{} segments of the \kFourRuptures{} multi-segment SRCMOD crustal ruptures.
		Shading marks $\pm\sigma_M$ and $\pm 2\sigma_M$.}
	\label{fig:momentScatter}
\end{figure}

The scatter about the model in Figure~\ref{fig:mom_area_ratio} provides the basis for the random perturbation applied when the apportionment is evaluated stochastically (requirement R4 in Section~\ref{sec:usecase}).
The departure of a component's moment fraction from its area fraction grows approximately in proportion to the area fraction (Section~S2.6 of the electronic supplement), so that the error in the apportioned moment is relative, and the perturbation is accordingly applied multiplicatively.
The moment of each segment is multiplied by a lognormally distributed factor, $M_k \propto \hat{M}_k\exp(\epsilon_k)$, with $\epsilon_k \sim N(0, \sigma_M^2)$ truncated at $\pm 2\sigma_M$ in order to avoid generating excessively large or small segment moments, in which $\hat{M}_k$ is the moment of Equation~(\ref{eq:moment-distribution-between-component}), and the segment moments are then rescaled so that their sum is $M_{o,T}$, while the apportionment within each segment remains that of Equation~(\ref{eq:moment-distribution-in-component}).
This form ensures that every perturbed moment is positive, and it preserves the invariance of the within-segment apportionment to the delineation of components in every realization.
The standard deviation is calibrated from the residuals of $\ln(M_k/M_{o,T})$ about Equation~(\ref{eq:moment-distribution-between-component}) over the segments of the \kFourRuptures{} multi-segment ruptures, and is $\sigma_M = \kFourSigmaM{}$.
Figure~\ref{fig:momentScatter} shows that these residuals do not depend on the segment area fraction ($p = \kFourHeteroscedasticityP{}$), so that a single value of $\sigma_M$ suffices.
Section~S2.6 of the electronic supplement gives further details on the development of this model.

%% file: sections/06-rupture-path.tex
\section{Rupture path}
\label{sec:rupturePath}
\label{sec:triggeringBasis}

This section addresses how the rupture evolves over the participating segments, given the set of their components, $\left\{C_i\right\}_{i=1}^{N}$, and the hypocenter location $h_1$.
Single-segment kinematic rupture generators resolve this question within a segment, by prescribing a rupture front that propagates outwards from the hypocenter at some fraction of the local shear-wave velocity.
What is additionally required for a multi-segment geometry is the macro-scale sequence in which the segments rupture.

An earthquake rupture forecast of the kind described in Section~\ref{sec:usecase} does not present an arbitrary collection of fault surfaces, since the rupture set has already been filtered for plausibility during its construction, for example, by rules that limit the maximum jump distance between participating segments and that exclude configurations judged to be mechanically implausible \citep{milner_enumerating_2022}.
As a result, the source model has already asserted that the segments comprising a given rupture do rupture together, and the rupture generator only needs to determine the order in which the segments break.
This permits a purely geometric treatment, in which the sequence is based upon the separation distances between segments, without recourse to the resolved Coulomb stress changes that would be required were the admissibility of the rupture itself in question.

\subsection{Jump probability and rupture-tree model}
\label{sec:rupturePathModel}

In forward simulation applications, the evolution of the rupture over time is not known beyond the specified initiating hypocenter position.
Thus, a simple model has been developed to obtain the rupture path from the component geometries and the initiating component.
The model is formulated over components, since these are the units in which a source model specifies a rupture (Section~\ref{sec:framework}), and the sequence in which the segments rupture follows from it.
This model has two parts, namely: (i) a model for the probability that rupture jumps between two components; and (ii) a model that weighs the likelihood of a set of jumps against all of the possible rupture paths.

For components separated by a distance of $r$~km, the probability that rupture jumps between them is estimated using the first-order model proposed by \citet{shaw_probabilities_2007}:
\begin{equation}
	\label{eq:shaw-model}
	P\left( {\text{Component}\ C_{i}\ \text{jumps}\ \text{to}\ C_{j}\ \text{at}\ \text{distance}\ {r\text{ km}}} \right) = \min\left(1,\ e^{- \frac{r - r_c}{r_{0}}}\right),
\end{equation}
with $r_0 = 3$~km and an offset $r_c = 1$~km, below which components are treated as certain to communicate.
Since the components of a segment are typically contiguous, rupture passes between them with near certainty, and the uncertainty in the rupture path therefore resides principally at the discontinuities between segments.
Probabilities are additionally capped at $0.99$, so that no edge is treated as a certainty in Equation~(\ref{eq:tree-probability}), since the rupture-tree distribution otherwise degenerates.
This local model does not consider direction, i.e., $P\left( C_i\ \text{jumps to}\ C_j \right) = P\left( C_j\ \text{jumps to}\ C_i \right)$.

A maximum separation of 15~km is adopted, beyond which a jump is not considered.
This value is the threshold used in the construction of the rupture set to which the generator is applied \citep{milner_enumerating_2022,gerstenberger_seismicity_2024}, since a smaller value would prevent some ruptures of the source model from being generated, and a larger value would admit connections that the source model has excluded; requirement R1 therefore determines it.
In contrast, the characteristic jumping distance $r_0$ is not determined in this manner, and its value of 3~km was obtained by \citet{shaw_probabilities_2007} from observations of rupture segmentation in California.
The maximum-likelihood trees do not change for seven values of $r_0$ spanning 2--5~km for the \jFiveGeometries{} geometries considered in Sections~\ref{sec:rupturePathCalib} and~\ref{sec:rupturePathProspective} (electronic supplement Section~S3.1).
A full sensitivity study is left as future work.

A probability graph is then constructed, with the components as nodes and with the edges between components weighted by their jumping probability, no edge being included between components separated by more than 15~km.
Each spanning tree of this graph is a rupture tree, which represents the path that the rupture takes from the initiating component.
A rupture tree $T$ is sampled according to
\begin{equation}
	\label{eq:tree-probability}
	P(T) \propto \prod_{(C_i,C_j) \in T}P\left( {C_i\ \text{triggers}\ {C_j}} \right)\prod_{(C_i,C_j) \notin T}\left( 1 - P\left( {C_i\ \text{triggers}\ {C_j}} \right) \right),
\end{equation}
in which each triggering pair is assumed to be independent.
The second product in Equation~(\ref{eq:tree-probability}) may be written as the product over all admissible edges, which is common to every tree, divided by the product over the edges contained in $T$, so that Equation~(\ref{eq:tree-probability}) is equivalent to
\begin{equation}
	\label{eq:tree-probability-weights}
	P(T) \propto \prod_{(C_i,C_j)\in T} w_{ij},
	\qquad
	w_{ij} = \frac{P\left(C_i\ \text{triggers}\ C_j\right)}{1 - P\left(C_i\ \text{triggers}\ C_j\right)}.
\end{equation}
This is the conditional-Bernoulli construction of \citet[Method 3]{chen_sampling_1994}, there conditioned on a fixed sample size and here on the edges forming a spanning tree.
The probability mass function over rupture trees is therefore a weighted spanning-tree distribution with edge weights $w_{ij}$, and two algorithms make use of it, namely: (i) a fair sampling method using the algorithm of \citet{wilson_generating_1996}, which samples rupture trees in proportion to $P(T)$; and (ii) a maximum-likelihood method, which returns the maximum spanning tree under the weights $\log w_{ij}$, and is useful when only a few instances of a single event are simulated.
Both algorithms extend without modification to asymmetric jumping probabilities, $P\left( C_i\ \text{jumps to}\ C_j \right) \ne P\left( C_j\ \text{jumps to}\ C_i \right)$.
Further details are given in Section~S3.2 of the electronic supplement.

\subsection{Calibration against historical ruptures}
\label{sec:rupturePathCalib}

In this section, the rupture-path model is evaluated against historical multi-segment ruptures for which a finite-fault inversion provides an independent estimate of the rupture geometry and, where rupture timing is reported, of the rupture sequence.
Most of the multi-segment ruptures in the SRCMOD database are geometrically simple, comprising nearly collinear segments for which the rupture path is trivial.
Thus, the evaluation rests upon the 2010 Darfield and 2016 Kaik\=oura earthquakes, whose geometries are complex, and for each the rupture-path model is applied to the inferred fault geometry in order to obtain a probability mass function over the admissible rupture trees.

In order to illustrate the rupture-path model, Figure~\ref{fig:rupPathSchematic} shows its application to the Darfield earthquake.
The Darfield geometry is that of \citet{pajaro2026Darfield}, based on a modification of the geometry of \citet{beavan_fault_2012}.
This comprises the three sections of the Greendale Fault (north-west/south-east, central and east), the Charing Cross Fault, the fault near Hororata, and two unnamed structures striking north-north-west and north-east.
Evaluating Equation~(\ref{eq:shaw-model}) for the actual separations gives \jSixDarfieldEdges{} admissible edges between these \jSixDarfieldFaults{} components.
Of these, \jSixDarfieldCappedEdges{} are capped at 0.99, since their components are separated by less than 1~km (and in \jSixDarfieldContactEdges{} cases are in contact).
Of the remaining \jSixDarfieldOtherEdges{} edges, one has a probability of $\jSixDarfieldNextProbability{}$ and the other \jSixDarfieldRemainingEdges{} have probabilities of $\jSixDarfieldRemainingMaxProbability{}$ or less (Table~S3 of the electronic supplement).
It follows that the rupture path is uncertain only within the two clusters of components that are effectively in contact, labeled J1 and J2 in Figure~\ref{fig:rupPathSchematic}.
At each junction the three possible configurations are equally probable, and the maximum-likelihood tree has approximately $1/9$ of the total probability as a result of these two junctions alone, compared with $\jSixDarfieldModalProbability{}$ over the full graph.
Because the components at each junction are in near-equidistant proximity, the practical implication of the exact serial order of triggering at these junctions is likely to be inconsequential.

\begin{figure}[htb!]
	\centering
	\begin{tikzpicture}[scale=0.82, every node/.style={font=\small}]

		\tikzset{
		seg/.style={line width=3.2pt, gray!70, line cap=round},
		segsel/.style={line width=3.2pt, black!75, line cap=round},
		rup/.style={-{Latex[length=2.2mm]}, thick, black},
		junc/.style={dashed, draw=red!75, line width=0.8pt},
		}

		\useasboundingbox (-0.2,-0.95) rectangle (13.6,4.45);

		\coordinate (HO1) at (0.3,0.6);   \coordinate (HO2) at (1.9,2.1);   %
		\coordinate (GW1) at (2.8,2.4);   \coordinate (GW2) at (5.8,0.7);   %
		\coordinate (GC1) at (5.8,0.7);   \coordinate (GC2) at (10.0,0.85); %
		\coordinate (GE1) at (10.0,0.85); \coordinate (GE2) at (13.3,1.15); %
		\coordinate (NE1) at (10.0,0.85); \coordinate (NE2) at (11.9,3.4);  %
		\coordinate (LK1) at (6.2,0.35);  \coordinate (LK2) at (7.6,2.1);   %
		\coordinate (HY1) at (6.9,1.6);   \coordinate (HY2) at (8.2,4.1);   %
		\foreach \a/\b in {HO1/HO2, GW1/GW2, GC1/GC2, GE1/GE2, NE1/NE2, LK1/LK2, HY1/HY2}
		\draw[seg] (\a) -- (\b);

		\coordinate (H) at ($(HY1)!0.5!(HY2)$);
		\node[star, star points=5, star point ratio=2.2, fill=white, draw=black,
			inner sep=1.6pt] at (H) {};
		\node[anchor=west] at ($(H)+(0.3,-0.15)$) {$h_1$};

		\newcommand{\alongtrace}[5]{%
			\coordinate (p) at ($(#1)!#3!(#2)$);
			\coordinate (q) at ($(#1)!#4!(#2)$);
			\draw[rup] ($(p)!#5!90:(#2)$) -- ($(q)!#5!-90:(#1)$);}
		\def\off{0.22cm}
		\alongtrace{HY1}{HY2}{0.58}{0.92}{\off}
		\alongtrace{HY2}{HY1}{0.58}{0.92}{-\off}
		\alongtrace{LK2}{LK1}{0.12}{0.85}{\off}
		\alongtrace{GW2}{GW1}{0.10}{0.88}{\off}
		\alongtrace{GC1}{GC2}{0.10}{0.88}{\off}
		\draw[rup] ($(GW1)!0.15cm!(HO2)$) -- ($(HO2)!0.15cm!(GW1)$);
		\alongtrace{HO2}{HO1}{0.12}{0.78}{-\off}
		\alongtrace{GE1}{GE2}{0.10}{0.88}{\off}
		\alongtrace{NE1}{NE2}{0.12}{0.88}{\off}

		\draw[junc] (6.15,0.85) circle (0.95);
		\node[red!75, anchor=north] at (6.15,-0.15) {\textbf{J1}};
		\draw[junc] (10.10,1.15) circle (0.95);
		\node[red!75, anchor=south] at (10.10,2.15) {\textbf{J2}};

		\node[anchor=north east, font=\scriptsize] at (1.35,1.05) {Hororata};
		\node[anchor=south west, font=\scriptsize] at (2.75,2.5) {Greendale NW--SE};
		\node[anchor=north, font=\scriptsize] at (8.6,0.68) {Greendale central};
		\node[anchor=north, font=\scriptsize] at (12.4,0.98) {Greendale east};
		\node[anchor=south east, font=\scriptsize] at (11.75,3.45) {NE structure};
		\node[anchor=east, font=\scriptsize, fill=white, inner sep=1pt] at (6.3,1.95) {Charing Cross};
		\draw[gray!60!black, thin] (6.32,1.90) -- (6.93,1.30);
		\node[anchor=south east, font=\scriptsize] at (7.35,3.75) {NNW structure};

		\node[anchor=north west, align=left, font=\scriptsize, gray!60!black]
		at (0.0,-0.15) {N $\uparrow$\ \ (schematic; not to scale)};

	\end{tikzpicture}

	\vspace{1mm}
	{\small (a)}

	\vspace{4mm}

	\begin{tikzpicture}[scale=0.82, every node/.style={font=\small}]

		\tikzset{
		sn/.style={rounded corners=2pt, draw=black, fill=white, minimum width=15mm,
				minimum height=7.5mm, inner sep=1pt, align=center, font=\scriptsize},
		det/.style={-{Latex[length=2.2mm]}, thick, black},
		jed/.style={-{Latex[length=2.2mm]}, thick, red!80},
		pl/.style={font=\scriptsize, inner sep=1.5pt, fill=white},
		plr/.style={font=\scriptsize, inner sep=1.5pt, fill=white, text=red},
		jbox/.style={dashed, draw=red!70, line width=0.8pt, rounded corners=7pt},
		}

		\useasboundingbox (-0.2,-1.25) rectangle (13.6,5.55);

		\draw[jbox] (2.9,0.55) -- (6.0,4.25) -- (9.7,0.55) -- cycle;
		\node[red!80, anchor=east, font=\small] at (4.0,3.5) {\textbf{J1}};
		\draw[jbox] (6.9,0.45) -- (11.9,4.7) -- (13.4,0.0) -- cycle;
		\node[red!80, anchor=south, font=\small] at (10.4,4.4) {\textbf{J2}};

		\node[sn] (S7) at (7.3,5.1) {NNW\\structure};
		\node[sn] (S6) at (6.0,3.4) {Charing\\Cross};
		\node[sn] (S1) at (0.9,1.4) {Hororata};
		\node[sn] (S2) at (4.3,1.4) {Greendale\\NW--SE};
		\node[sn] (S3) at (8.3,1.4) {Greendale\\central};
		\node[sn] (S5) at (11.7,3.8) {NE\\structure};
		\node[sn] (S4) at (11.9,0.7) {Greendale\\east};

		\draw[det] (S7) -- (S6) node[pl, pos=0.55, right] {$0.99$};
		\draw[det] (S2) -- (S1) node[pl, midway, above] {$0.89$};

		\draw[jed] (S6) -- (S2) node[plr, midway, above left] {$0.99$};
		\draw[jed] (S6) -- (S3) node[plr, midway, above right] {$0.99$};
		\draw[jed] (S2) to[bend left=20] node[plr, midway, above] {$0.99$} (S3);
		\draw[jed] (S3) to[bend left=20] (S2);

		\draw[jed] (S3) -- (S5) node[plr, midway, above left] {$0.99$};
		\draw[jed] (S3) -- (S4) node[plr, midway, below] {$0.99$};
		\draw[jed] (S5) to[bend left=20] node[plr, midway, right] {$0.99$} (S4);
		\draw[jed] (S4) to[bend left=20] (S5);

		\node[anchor=west, font=\scriptsize, gray!60!black, align=left] at (0.0,-0.35)
		{Black: edges for which the triggering component is unambiguous. Red: edges at a
			three-way junction, for which\\ more than one rupture tree is admissible.
			Opposing arrow pairs denote that either direction of triggering is possible.};

	\end{tikzpicture}

	\vspace{1mm}
	{\small (b)}

	\caption{Illustration of the rupture-path model applied to the multi-segment geometry of the 2010 Darfield earthquake \citep{pajaro2026Darfield}:
		(a) schematic map view of the participating components, the hypocenter $h_1$, and the direction of rupture propagation for one sampled rupture tree;
		and (b) the corresponding graph, in which the components are nodes and the edges are weighted by the jumping probability of Equation~(\ref{eq:shaw-model}) for the actual separations between components.
		Only the eight edges of appreciable probability are drawn.}
	\label{fig:rupPathSchematic}
\end{figure}

\begin{figure}[htb!]
	\centering
	\includegraphics[width=\linewidth]{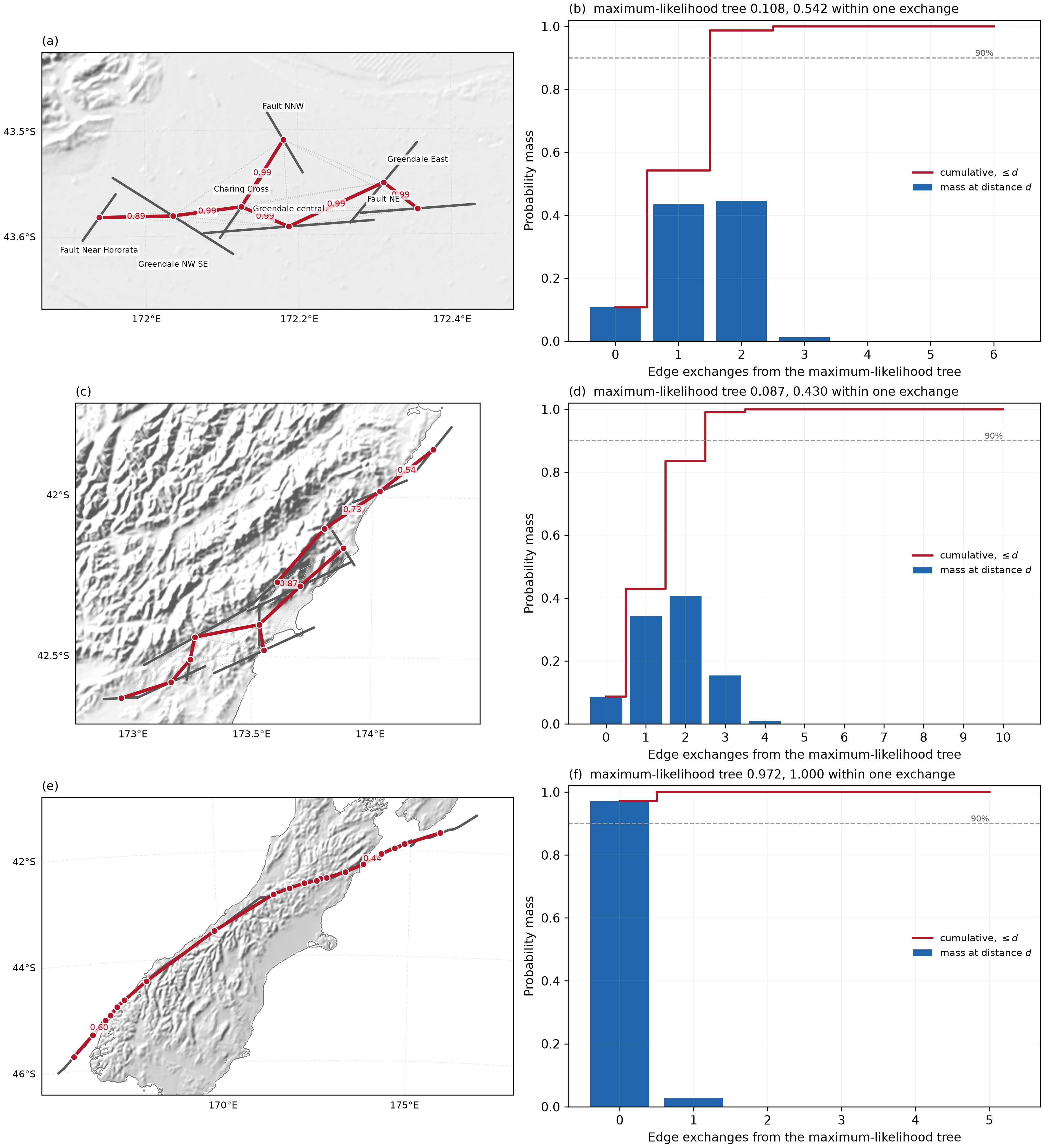}
	\caption{Concentration of probability about the maximum-likelihood rupture tree, for: (a), (b) the 2010 Darfield earthquake, using the source geometry of \citet{pajaro2026Darfield}; (c), (d) the 2016 Kaik\=oura earthquake, using the source geometry of \citet{bradley_groundmotion_2017}; and (e), (f) a prospective multi-segment Alpine--Marlborough fault rupture from the 2022 NZ NSHM \citep{gerstenberger_seismicity_2024}.
		The left column shows the fault traces, with the maximum-likelihood tree in red labeled with jump probabilities, drawn between the along-strike centers of the components (filled circles), and the admissible connections not selected as dotted lines.
		Unlabeled edges are at the 0.99 cap.
		The right column shows the probability mass at each edge exchange distance from the maximum-likelihood tree together with the cumulative mass.}
	\label{fig:rupPathExamples}
\end{figure}

Figure~\ref{fig:rupPathExamples} shows the maximum-likelihood rupture trees for the Darfield and Kaik\=oura geometries, together with the distribution of tree probabilities plotted against edge-exchange distance.
The Darfield geometry comprises $\jSixDarfieldFaults{}$ components connected by $\jSixDarfieldEdges{}$ admissible edges, and the Kaik\=oura geometry $\jSixKaikouraFaults{}$ components connected by $\jSixKaikouraEdges{}$ edges, whereas a spanning tree requires only $n-1$ edges.
As a result, there are $\jSixDarfieldTrees{}$ admissible rupture trees for Darfield and $\jSixKaikouraTrees{}$ for Kaik\=oura, and the maximum-likelihood tree carries only $\jSixDarfieldModalProbability{}$ and $\jSixKaikouraModalProbability{}$ of the total probability, respectively.
Taken alone, these probabilities would suggest that the rupture path of either event is largely undetermined.
However, almost all of the remaining probability lies on trees that differ from the maximum-likelihood tree by one or two edges.
The difference between two rupture trees is measured here by the number of edge exchanges required to transform one into the other, which is their distance in the tree graph of \citet{cummins_hamilton_1966}.
On this measure, the probability within one exchange of the maximum-likelihood tree is $\jSixDarfieldWithinOne{}$ for Darfield and $\jSixKaikouraWithinOne{}$ for Kaik\=oura, and within two exchanges it is $\jSixDarfieldWithinTwo{}$ and $\jSixKaikouraWithinTwo{}$, respectively.
This implies that the competing paths are, in effect, the maximum-likelihood tree with a single jump rerouted, and not distinct rupture sequences.

\subsection{Behavior on prospective source-model geometries}
\label{sec:rupturePathProspective}

The behavior of the rupture-path model over the multi-segment ruptures of the 2022 NZ NSHM crustal fault model follows from the comparison of historical and prospective geometries in Figure~\ref{fig:rupPathExamples}.
The rupture geometries inferred from finite-fault inversions of the Darfield and Kaik\=oura earthquakes are appreciably more complex than the multi-segment ruptures of contemporary source models.
The Kaik\=oura rupture of Figure~\ref{fig:rupPathExamples}(c) involves a dozen mapped and unmapped faults with numerous three-way junctions, whereas the prospective Alpine--Marlborough rupture of Figure~\ref{fig:rupPathExamples}(e) is a near-linear structure.
It comprises $\jSixAlpineHopeFaults{}$ components but only $\jSixAlpineHopeEdges{}$ edges, admits only $\jSixAlpineHopeTrees{}$ rupture trees, and its maximum-likelihood tree carries $\jSixAlpineHopeModalProbability{}$ of the total probability.
Thus, the rules developed here are calibrated upon geometries that are more demanding than those to which they are subsequently applied, giving confidence in their extrapolation.
Figure~S3 of the electronic supplement shows that \jSevenModalIsLinePct{}\% of the \jSevenDistanceN{} multi-segment ruptures of the 2022 NZ NSHM have maximum-likelihood rupture paths that are unbranched, similar to that of Figure~\ref{fig:rupPathExamples}(e) (see Section~S3.4 of the electronic supplement for further details).

%% file: sections/07-jumping-and-delay.tex
\section{Segment jumping location and time delay}
\label{sec:jumping}

The rupture path determined in Section~\ref{sec:rupturePath} establishes which pairs of components trigger one another, but it does not specify where and when the rupture initiates on each component.
This section therefore addresses the third aspect of the framework, namely the determination of the jump point $h_j$ at which rupture initiates on the triggered component, its conjugate point $h_j^{'}$ on the triggering component, and the time delay $t_j$ between slip initiation at the two locations, as defined in Section~\ref{sec:framework}.

\subsection{Jump location and conjugate point}
\label{sec:jumpLocation}

A rupture tree over $N$ components contains $N-1$ ordered edges, each of which constitutes a \emph{jump pair}.
For each pair $(C_i,C_j)$, in which $C_i$ triggers $C_j$, the jump point and its conjugate are obtained as the closest pair of points between the discretized geometries of $C_i$ and $C_j$, subject to the constraint that both points are below a depth of 5~km.
The depth constraint is imposed in order to avoid the artificial rupture delays that would otherwise result wherever two components approach one another most closely near the ground surface, where the rupture velocity is low, and is relaxed for a component that does not extend beyond 5~km in depth.

The minimizing pair defines the jump point $h_j$ on the triggered component $C_j$ and its conjugate point $h_j^{'}$ on the triggering component $C_i$, separated by the jumping distance $r_j = \left|h_j - h_j^{'}\right|$.
The hypocenter of the initiating component, $h_1$, is either prescribed or randomly sampled (Section~\ref{sec:usecase}).

For SRCMOD and generated ruptures, using the discretized geometry makes the jump points contingent on the resolution of the fault geometry.
For generated ruptures the resolution is often 100~m or less, so that this artifact is small, at the cost of large rupture files.
For ease of external reproducibility in generating the statistics for the complete 2022 NZ NSHM rupture set, the validation adopts an optimization method to find the jump pairs which does not require discretization.
This optimization algorithm is summarized in Section~S4.1 of the electronic supplement.

\subsection{Rupture time delay}
\label{sec:timeDelay}

For a jump pair $(C_i,C_j)$ for which the initiation times on $C_i$ have already been determined, the initiation time at the jump point $h_j$ is obtained from that at the conjugate point $h_j^{'}$ by the addition of the time delay $t_j$:
\begin{equation}
	\label{eq:rupture-time}
	t_{init}(h_j) = t_{init}(h_j^{'}) + t_j,
	\qquad
	t_j = \frac{r_j}{V_s} + \delta_j,
	\qquad
	\delta_j \sim N\!\left(0, \sigma_\delta^2(r_j)\right),
	\qquad
	\sigma_\delta^2(r) = \sigma_0^2 + \left(s\,r^{\,g}\right)^2,
\end{equation}
in which $V_s$ is the apparent velocity of propagation across the jump, and $\delta_j$ is a normally distributed random perturbation whose standard deviation comprises a term $\sigma_0$ independent of the jumping distance and a term that grows with it.
Since the time delay is the least constrained part of the framework, owing to the smaller size of the calibration data (Table~\ref{tab:srcmod-datasets}), three versions of Equation~(\ref{eq:rupture-time}) are considered for use in a logic tree (Section~\ref{sec:procedure}):
\begin{description}
	\item[Model A:] $V_s$, $s$ and $g$ are fitted, with $\sigma_0 = 0$;
	\item[Model B:] $V_s$, $\sigma_0$, $s$ and $g$ are all fitted;
	\item[Model C:] $\sigma_0$, $s$ and $g$ are fitted, with $V_s$ fixed at a within-segment average rupture velocity of 2.8~km\,s$^{-1}$ ($0.8 \times 3.5$~km\,s$^{-1}$).
\end{description}
Model~B contains the other two as special cases.
Negative delays may be drawn under each model, for the reason given in Section~\ref{sec:jumpCalibration}, but no component may initiate before its triggering component begins to rupture, $t_{init}(h_j) \geq t_{init}(h_i)$ (Section~S4.4 of the electronic supplement).

\subsection{Calibration against the SRCMOD database}
\label{sec:jumpCalibration}

\begin{figure}[htb!]
	\centering
	\includegraphics[width=\linewidth]{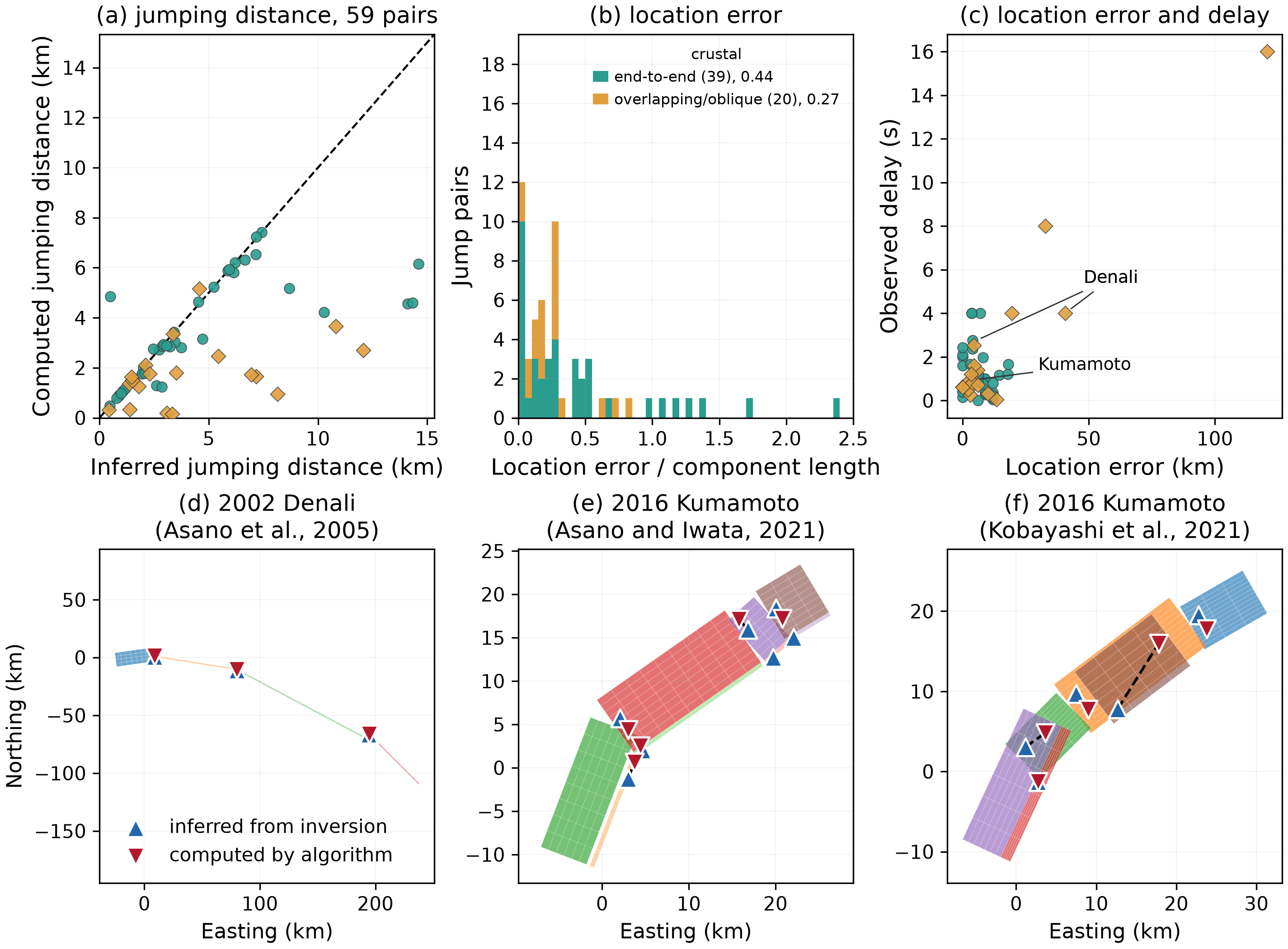}
	\caption{Comparison of the jump location inferred from the source inversion with that computed by the algorithm of Section~\ref{sec:jumpLocation}, over the \jNineN{} jump pairs for which both are defined, distinguishing pairs arranged end to end from pairs that overlap or are oblique:
		(a) computed against inferred jumping distance;
		(b) the separation between the two locations, as a fraction of the along-strike extent of the triggered component;
		(c) separation against the observed delay, with the pairs of the Denali and Kumamoto earthquakes indicated;
		and (d)--(f) a comparison between computed and inferred jump points for three chosen models.}
	\label{fig:jumpPoints}
\end{figure}

\begin{figure}[htb!]
	\centering
	\includegraphics[width=\linewidth]{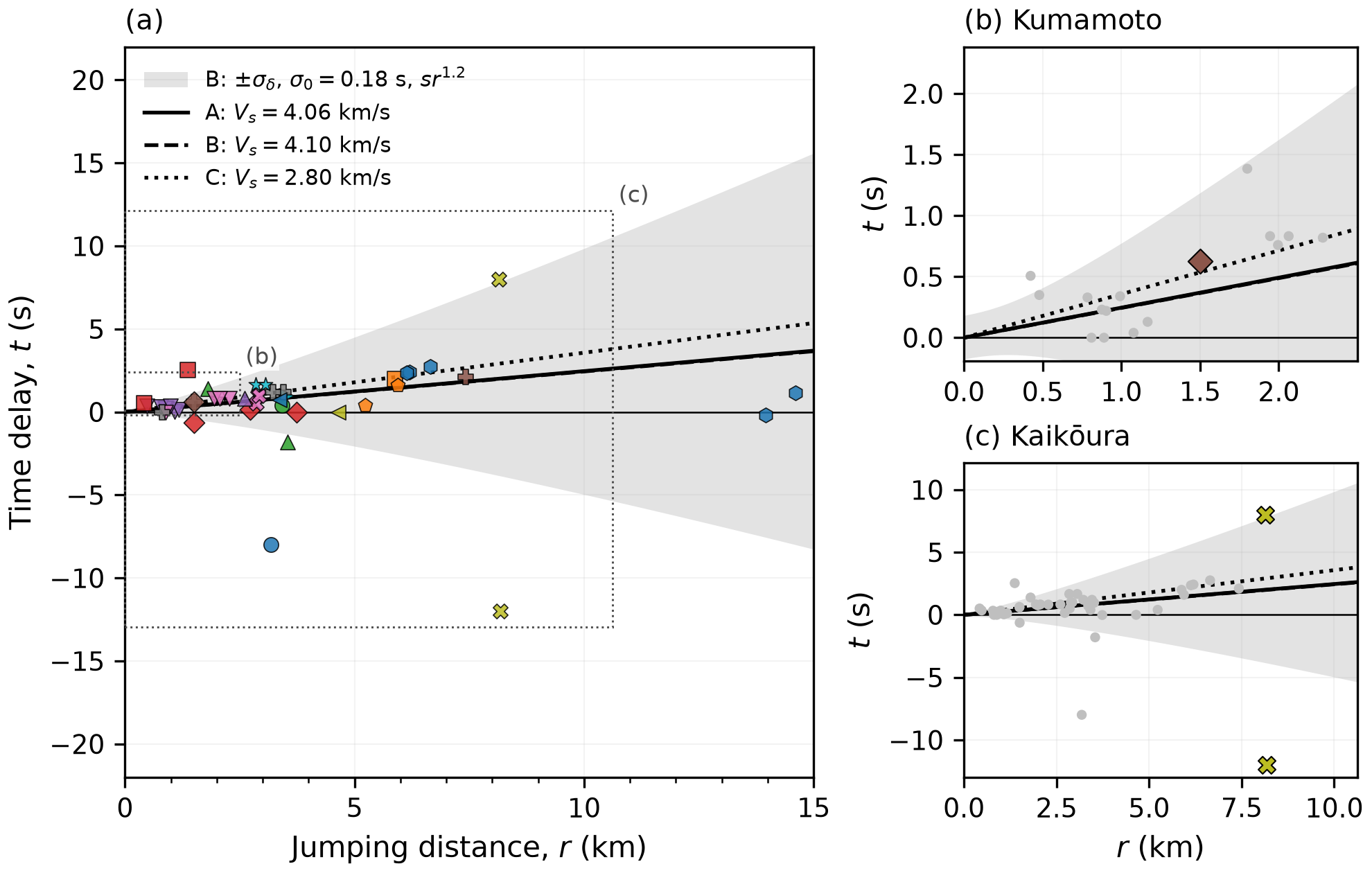}
	\caption{Inferred time delay of rupture between the components of multi-segment ruptures in the SRCMOD database, against the jumping distance computed following Section~\ref{sec:jumpLocation}, with symbols distinguishing the jump pairs of different ruptures.
		The lines are the means of Models A (solid), B (dashed) and C (dotted) of Equation~(\ref{eq:rupture-time}), and the shading is the $\pm 1\sigma_\delta(r)$ band of Model~B; the bands of all three models are shown in Figure~S5 of the electronic supplement.
		Panels (b) and (c) highlight the 2016 Kumamoto (Kobayashi inversion of the Mw~6.1 foreshock) and Kaik\=oura earthquakes, respectively, with the pairs of other ruptures in gray.}
	\label{fig:rupTimeDelay}
\end{figure}

Figure~\ref{fig:jumpPoints} compares the jump locations inferred from SRCMOD inversions with those computed by the construction of Section~\ref{sec:jumpLocation} (the separation between the two is examined in Section~S4.3 of the electronic supplement), and Figure~\ref{fig:rupTimeDelay} shows the corresponding time delays as a function of the jumping distance.
The calibration uses the SRCMOD ruptures that report rupture-onset times and comprise more than one component, whether those components form one segment or several.
The conjugate is taken as the nearest sub-fault on any other component, irrespective of its onset.
Each sub-fault is located at its top-center, as the SRCMOD files report it, so that the jumping distances and the 5~km depth requirement are evaluated at that point.
Pairs joining two components of the same segment are retained, since the generator applies Equation~(\ref{eq:rupture-time}) at every edge of the rupture tree, including those within a segment.
Where two components are each the nearest to the entry point of the other, the contact is recorded once, as the pair whose delay is not negative.
Pairs whose onset times coincide are retained, since the cause of the coincidence cannot be determined from the inversion.
Of the \jTenTriggered{} triggered components, \jTenBeyondLimit{} are excluded by the limit on jumping distance and a further \jTenShallow{} by the depth requirement, leaving \jTenPoolPairs{} jump pairs from \jTenPoolEvents{} ruptures.
Because the conjugate point is chosen without reference to its onset, the delay may be negative, and \jFourteenGeometricNegative{} of the \jTenJumpPairs{} crustal pairs ($\jFourteenGeometricNegativePct{}\%$) have a negative delay.

Fitting each model by maximum likelihood, with the mean and the scatter estimated jointly, gives $V_s = \tFortyOneAVelocity{}$, $\tFortyOneBVelocity{}$ and $\tFortyOneCVelocity{}$\,km\,s$^{-1}$ for Models A, B and C.
Relative to the lowest value, the Akaike information criterion (AIC) is larger by $\tFortyOneADeltaAic{}$, $\tFortyOneBDeltaAic{}$ and $\tFortyOneCDeltaAic{}$, respectively.
The fitted parameters and the AIC of each model are listed in Table~S4 of the electronic supplement.
No weights are prescribed here, and the AIC is reported in order that users of the framework may derive their own (e.g., as Akaike weights).
The differences in AIC are less than two, so that the data do not meaningfully discriminate between the three models.
The apparent velocities of Models~A and~B exceed the crustal shear-wave velocity of 3.5--3.6~km\,s$^{-1}$ commonly adopted in ground-motion simulation \citep[e.g.,][]{boore_simulation_2003}, and are therefore not interpreted as velocities of rupture propagation; they reflect the uncertainty in the inferred location of the jump points and the coarse discretization of the inverted models.
Rupture within each component continues to propagate at the rupture velocity of the single-segment generator.
The scatter of all three models grows with the jumping distance, approximately as $r^{\,\tFortyOneAExponent{}}$ to $r^{\,\tFortyOneCExponent{}}$, and is influenced appreciably by a small number of ruptures, including the two Kaik\=oura pairs at \jTenKaikouraHighDistanceKm{} and \jTenKaikouraLowDistanceKm{}~km, with delays of \jTenKaikouraHighDelayS{} and \jTenKaikouraLowDelayS{}~s (Section~S4.4 of the electronic supplement).

Since the jump point and its conjugate are approximations of where rupture crossed, negative delays are admitted in the generated ruptures as an indirect representation of the uncertainty in jumping location.
At the jumping distances of the calibration, Models A, B and C give a negative delay in approximately \tFortyOneANegativePct{}\%, \tFortyOneBNegativePct{}\% and \tFortyOneCNegativePct{}\% of random realizations, against the $\jFourteenGeometricNegativePct{}\%$ ``observed'' in the SRCMOD data.

%% file: sections/08-procedure-and-uncertainty.tex
\section{Rupture generation procedure and treatment of uncertainty}
\label{sec:procedure}

Together, the three aspects developed in Sections~\ref{sec:momentPartitioning}--\ref{sec:jumping} constitute a complete procedure for generating a kinematic rupture over a prescribed multi-segment geometry.
Given the components of the participating segments, $\left\{C_i\right\}_{i=1}^{N}$, the total rupture area $A_T$, and a hypocenter $h_1$ upon a nominated initiating component, the procedure is as follows.

\begin{enumerate}
	\item Obtain the total seismic moment of the rupture from the total area, $M_{o,T} = f(A_T)$, using the adopted magnitude-area scaling relation (Section~\ref{sec:maScaling}).
	\item Group the participating components into segments according to their separation and dip, and apportion $M_{o,T}$ between the segments by Equation~(\ref{eq:moment-distribution-between-component}) and within each segment by Equation~(\ref{eq:moment-distribution-in-component}) (Section~\ref{sec:apportionmentModel}).
	\item Construct the probability graph over the components, assigning to each admissible pair the jumping probability of Equation~(\ref{eq:shaw-model}), and omitting pairs separated by more than 15~km (Section~\ref{sec:rupturePathModel}).
	\item Obtain the rupture tree $T$, rooted at the initiating component, either as the maximum spanning tree under the weights of Equation~(\ref{eq:tree-probability-weights}) or by sampling from the distribution that those weights define.
	\item For each of the $N-1$ jump pairs of $T$, obtain the jump point $h_j$ and its conjugate $h_j^{'}$ by the algorithm of Section~\ref{sec:jumpLocation}, and hence the jumping distance $r_j$.
	\item Generate the slip and rupture initiation times upon each component independently, using a single-segment kinematic rupture generator with the component moment $M_{o,i}$ from step~2 and the hypocenter $h_j$ from step~5.
	\item Reference the components to one another in time by a breadth-first traversal of $T$, applying Equation~(\ref{eq:rupture-time}) at each jump pair under the selected branch, subject to the bound that no component initiates before its triggering component (Section~\ref{sec:timeDelay}).
\end{enumerate}

Decisions in these stages are of two distinct kinds.
At some stages a functional form must be selected --- how the moment is partitioned, how the jumping probability depends upon separation distance, how the jump location is determined, and how the time delay is related to the jumping distance.
These are choices between alternative models, and represent epistemic uncertainty.
At the remaining stages a value must be drawn from a distribution that the selected model itself prescribes --- the moment apportioned to a segment, the rupture tree, the jump location, and the time delay.
These are realizations of the apparent aleatory variability that the calibration of the preceding sections quantifies.

Figure~\ref{fig:logicTree} presents these stages as a logic tree, grouped under the three aspects of the framework.
A complete traversal of the tree, from left to right, constitutes one realization of the generator.
Two branches, drawn with dashed lines, denote models that the framework admits but that are not implemented herein, and which are identified in order to indicate where the framework may subsequently be extended; their weights in the present work are accordingly zero, $p_{4,2}=p_{6,2}=0$.
Stage~(viii) carries the three branches of the time-delay model of Section~\ref{sec:timeDelay}.
The weights $p_{i,j}$ are left symbolic, since their values depend on the application; for stage~(viii), the information criteria of Section~\ref{sec:jumpCalibration} provide one basis for them.
Since it addresses only the multi-segment aspects of the generator, the logic tree is intended, in application, to be coupled with the corresponding logic tree for the single-segment rupture generator, which describes the choice of generator and the uncertainty in its parameters.

\begin{figure}[p]
	\centering
	\rotatebox{90}{\resizebox{0.86\textheight}{!}{%
			\begin{tikzpicture}[
					x=1cm, y=1cm,
					sel/.style={black, line width=1.1pt},
					alt/.style={black, line width=1.1pt},
					unimp/.style={black, line width=1.1pt, dashed},
					ray/.style={gray!45, line width=0.5pt},
					sl/.style={font=\scriptsize, align=left, text width=1.5cm, inner sep=1.0pt},
					al/.style={font=\scriptsize, align=left, text width=1.5cm, inner sep=1.0pt},
					st/.style={font=\scriptsize\itshape, align=center, text width=2.0cm,
							text=gray!50!black, inner sep=1.0pt},
					hd/.style={font=\bfseries, align=center},
					lg/.style={font=\scriptsize, align=left, inner sep=1.0pt, text=gray!45!black},
					sep/.style={dashed, gray!60, line width=0.6pt},
				]

				\draw[sep] (5.75,-4.20) -- (5.75,5.2);
				\draw[sep] (10.35,-4.20) -- (10.35,5.2);
				\draw[gray!60, line width=0.6pt] (-2.7,4.60) -- (21.5,4.60);
				\node[hd] at (1.5,4.95) {Moment partitioning};
				\node[hd] at (8.05,4.95) {Rupture path};
				\node[hd] at (15.9,4.95) {Jumping location and time delay};

				\node[sl, anchor=east, text width=1.3cm, align=right] at (-1.4,0) {$\{C_i\}$, $h_1$};
				\node[sl, anchor=west, text width=1.6cm] at (20.0,1.6) {Multi-segment kinematic rupture};

				\draw[sel] (-1.3,0) -- (0,0);
				\draw[sel] (0,0) -- (0.55,1.6) -- (2.3,1.6);
				\draw[sel] (2.3,1.6) -- (2.85,0) -- (4.6,0);
				\draw[sel] (4.6,0) -- (6.9,0);
				\draw[sel] (6.9,0) -- (7.45,1.6) -- (9.2,1.6);
				\draw[sel] (9.2,1.6) -- (11.5,1.6);
				\draw[sel] (11.5,1.6) -- (12.05,0) -- (13.8,0);
				\draw[sel] (13.8,0) -- (16.1,0);
				\draw[sel] (16.1,0) -- (16.65,1.6) -- (18.4,1.6);
				\draw[sel] (18.4,1.6) -- (19.9,1.6);

				\draw[alt] (0,0) -- (0.55,-1.6) -- (1.95,-1.6);
				\draw[alt] (2.3,1.6) -- (2.85,3.2) -- (4.25,3.2);
				\draw[unimp] (6.9,0) -- (7.45,-1.6) -- (8.85,-1.6);
				\draw[unimp] (11.5,1.6) -- (12.05,3.2) -- (13.45,3.2);
				\draw[alt] (16.1,0) -- (18.05,0);
				\draw[alt] (16.1,0) -- (16.65,-1.6) -- (18.05,-1.6);

				\node[sl, anchor=south west] at (0.65,1.70) {Self-similar, $M_k \propto A_k^{3/2}$};
				\node[sl, anchor=north west] at (0.65,1.50) {(Eq.~\ref{eq:moment-distribution-between-component}); ($p_{1,1}$)};
				\node[al, anchor=south west] at (0.65,-1.50) {Uniform average slip, $M_k \propto A_k$};
				\node[al, anchor=north west] at (0.65,-1.70) {($p_{1,2}$)};
				\node[st, anchor=north] at (1.1,-3.60) {(i) partitioning model};

				\node[sl, anchor=south west] at (2.95,0.10) {Separation threshold $d_c$};
				\node[sl, anchor=north west] at (2.95,-0.10) {(Sec.~\ref{sec:apportionmentModel}); ($p_{2,1}$)};
				\node[al, anchor=south west] at (2.95,3.30) {Strike criterion also};
				\node[al, anchor=north west] at (2.95,3.10) {($p_{2,2}$)};
				\node[st, anchor=north] at (3.4,-3.60) {(ii) segment delineation};

				\fill[gray!14, draw=gray!45, line width=0.4pt]
				(4.6,0) -- ++(20:1.2) arc[start angle=20, end angle=-20, radius=1.2] -- cycle;
				\draw[ray] (4.6,0) -- ++(10:1.2);
				\draw[ray] (4.6,0) -- ++(-10:1.2);
				\node[sl, anchor=south west] at (4.75,0.55) {$M_k \propto \hat{M}_k\,e^{\epsilon_k}$, $\epsilon_k \sim N\!\left(0, \sigma_M^2\right)$, $|\epsilon_k| \le 2\sigma_M$};
				\node[st, anchor=north] at (5.7,-3.60) {(iii) moment realization};

				\node[sl, anchor=south west] at (7.55,1.70) {$P = e^{-(r - r_c)/r_0}$};
				\node[sl, anchor=north west] at (7.55,1.50) {(Eq.~\ref{eq:shaw-model}); ($p_{4,1}$)};
				\node[al, anchor=south west] at (7.55,-1.50) {Coulomb-informed $P$};
				\node[al, anchor=north west] at (7.55,-1.70) {($p_{4,2}$)};
				\node[st, anchor=north] at (8.0,-3.60) {(iv) jump probability model};

				\draw[ray] (9.2,1.6) -- ++(20:1.2);
				\draw[ray] (9.2,1.6) -- ++(10:1.2);
				\draw[ray] (9.2,1.6) -- ++(-10:1.2);
				\draw[ray] (9.2,1.6) -- ++(-20:1.2);
				\node[sl, anchor=south west] at (9.35,2.15) {$T \sim P(T)$ (Eq.~\ref{eq:tree-probability-weights})};
				\node[st, anchor=north] at (10.3,-3.60) {(v) rupture path realization};

				\node[sl, anchor=south west] at (12.15,0.10) {Closest point, $\min L$};
				\node[sl, anchor=north west] at (12.15,-0.10) {(Sec.~\ref{sec:jumpLocation}); ($p_{6,1}$)};
				\node[al, anchor=south west] at (12.15,3.30) {Hypocenter-weighted objective};
				\node[al, anchor=north west] at (12.15,3.10) {($p_{6,2}$)};
				\node[st, anchor=north] at (12.6,-3.60) {(vi) jump-point objective};

				\draw[gray!55, dashed, line width=0.5pt]
				(13.8,0) -- ++(20:1.2) arc[start angle=20, end angle=-20, radius=1.2] -- cycle;
				\node[al, anchor=south west] at (13.95,0.55) {$h_j$ perturbed, $\sigma_h$};
				\node[st, anchor=north] at (14.9,-3.60) {(vii) jump location realization};

				\node[sl, anchor=south west] at (16.75,1.70) {A: $V_s$ fitted, $\sigma_0 = 0$};
				\node[sl, anchor=north west] at (16.75,1.50) {(Eq.~\ref{eq:rupture-time}); ($p_{8,1}$)};
				\node[al, anchor=south west] at (16.75,0.10) {B: $V_s$, $\sigma_0$ fitted};
				\node[al, anchor=north west] at (16.75,-0.10) {($p_{8,2}$)};
				\node[al, anchor=south west] at (16.75,-1.50) {C: $V_s = 2.8$\,km\,s$^{-1}$};
				\node[al, anchor=north west] at (16.75,-1.70) {($p_{8,3}$)};
				\node[st, anchor=north] at (17.2,-3.60) {(viii) time-delay model};

				\fill[gray!14, draw=gray!45, line width=0.4pt]
				(18.4,1.6) -- ++(20:1.2) arc[start angle=20, end angle=-20, radius=1.2] -- cycle;
				\draw[ray] (18.4,1.6) -- ++(10:1.2);
				\draw[ray] (18.4,1.6) -- ++(-10:1.2);
				\node[sl, anchor=south west, text width=1.4cm] at (18.5,2.15) {$\delta_j \sim N\!\left(0, \sigma_\delta^2(r)\right)$, causally bounded (Eq.~\ref{eq:rupture-time})};
				\node[st, anchor=north] at (19.5,-3.60) {(ix) time-delay realization};

				\draw[sel] (-2.7,-4.75) -- ++(0.75,0);
				\node[lg, anchor=west, text=black] at (-1.85,-4.75) {branch implemented herein, with weight $p_{i,j}$};
				\draw[unimp] (-2.7,-5.45) -- ++(0.75,0);
				\node[lg, anchor=west, text=black] at (-1.85,-5.45) {potential future branch extension};
				\fill[gray!14, draw=gray!45, line width=0.4pt]
				(-2.7,-6.15) -- ++(16:0.75) arc[start angle=16, end angle=-16, radius=0.75] -- cycle;
				\node[lg, anchor=west] at (-1.85,-6.15) {range from which the realization is drawn};

			\end{tikzpicture}}}
	\caption{Schematic logic tree for the generation of a multi-segment kinematic rupture, showing its nine stages grouped under the three aspects of the framework (Section~\ref{sec:procedure}).
		Solid branches are the models implemented herein, with weight $p_{i,j}$ on branch $j$ of stage $i$, and shaded ranges are the distributions from which each realization is drawn; dashed lines denote extensions that the framework admits but that are not implemented herein, including the jump-location realization of stage~(vii).}
	\label{fig:logicTree}
\end{figure}

Two modes of execution follow from the logic-tree framework of Figure~\ref{fig:logicTree}.
In the \emph{reference} mode, the highest-weighted model is adopted at each stage at which a model is selected, and the central estimate at each stage at which a value is drawn, so that a single, reproducible, best-estimate rupture is obtained for a given geometry and hypocenter.
This is the appropriate mode for scenario simulation, and for any comparison against a specific historical earthquake, in which the intent is to represent one event rather than a population.
In the \emph{stochastic} mode, the logic tree is randomly sampled, so that repeated application to the same geometry and hypocenter yields a population of ruptures whose variability reflects that of the calibration data.
This is the mode required for PSHA, in which the apparent aleatory variability and epistemic uncertainty of the source must be integrated over rather than fixed.

Simulation-based hazard calculations for single-segment sources already sample the hypocenter location and the slip realization of the rupture generator.
The multi-segment extension introduces three further sources of aleatory variability, namely the moment apportioned to each segment, the rupture tree and the time delays, and a fourth, the jump locations, once stage~(vii) is implemented.
For tractability, the nine stages are treated as independent.
Figure~\ref{fig:logicTree} also deliberately places the epistemic and aleatory stages within a single diagram.
This is a convenient representation of the complete generation procedure, but it should not be read as a logic tree in the sense conventionally employed in PSHA, in which the epistemic branches alone constitute the logic tree and the aleatory variability is integrated over within each branch rather than enumerated alongside it.

%% file: sections/09-demonstration.tex
\section{Demonstration}
\label{sec:application}

The preceding sections developed the three aspects of the generator and set out the procedure by which they are combined.
This section demonstrates the result, first for a single multi-segment rupture, and then for the complete multi-segment rupture set of a contemporary source model.
Its intent is to show what the generator produces, and how it behaves when applied at the scale for which it is intended; the ground motions that result from the generated ruptures are examined elsewhere \citep[e.g.,][]{bradley_groundmotion_2026}.

\subsection{A generated multi-segment rupture}
\label{sec:demoSingle}

\begin{figure}[htb!]
	\centering
	\includegraphics[width=\linewidth]{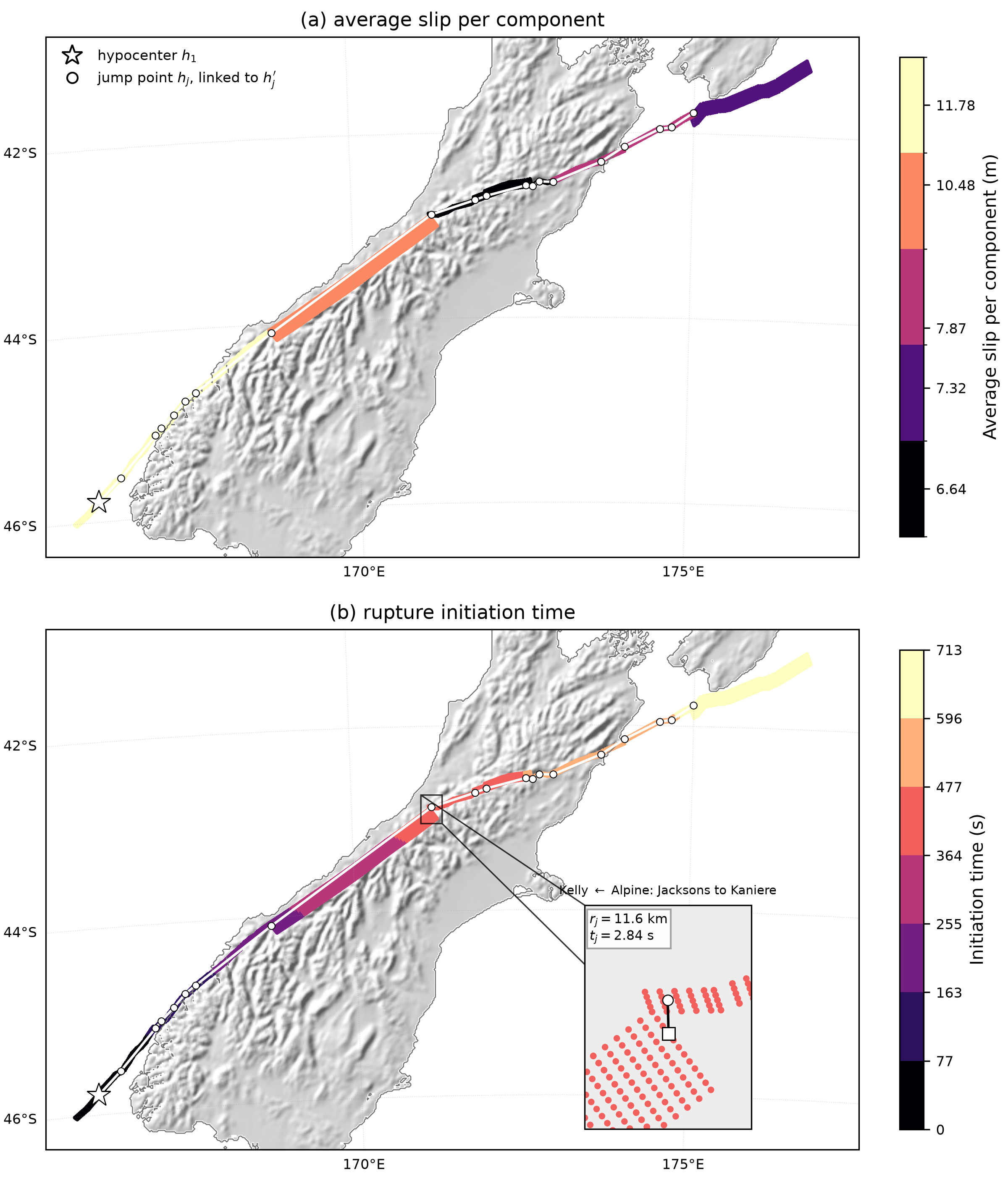}
	\caption{Multi-segment rupture generated for NSHM rupture \jSixteenRuptureId{} (Alpine--Marlborough), $M_w\,\jSixteenMagnitude{}$, total area $A_T = \jSixteenArea{}$\,km$^2$, comprising \jSixteenSegments{} components grouped into \jSixteenComponents{} segments under the rule of Section~\ref{sec:apportionmentModel}.
		(a) Average slip apportioned to each component by Equations~(\ref{eq:moment-distribution-between-component}) and~(\ref{eq:moment-distribution-in-component}).
		(b) Rupture initiation time, in which the star marks the hypocenter $h_1$ and circles mark the jump point $h_j$ on each triggered component, joined by a line to its conjugate point (squares) $h_j^{'}$ on the triggering component.}
	\label{fig:generatedRupture}
\end{figure}

Figure~\ref{fig:generatedRupture} presents the rupture generated for a single multi-segment Alpine Fault--Marlborough Fault Zone geometry from the 2022 NZ NSHM \citep{gerstenberger_seismicity_2024}, executed in the reference mode of Section~\ref{sec:procedure} with Model~\kFiveDelayBranch{} of Equation~(\ref{eq:rupture-time}).
Three features of the generated rupture warrant attention, each being the visible consequence of one of the three aspects of the framework:
\begin{enumerate}
	\item The average slip differs between segments, and is approximately uniform between the components of a single segment, which is the direct consequence of Equations~(\ref{eq:moment-distribution-between-component}) and~(\ref{eq:moment-distribution-in-component}).
	\item The rupture front does not propagate continuously across the discontinuity between segments.
	      Rather, the front reaches the conjugate point $h_j^{'}$ on the triggering component, and rupture is re-nucleated at the jump point $h_j$ on the triggered component.
	      A new front propagates outwards from the nucleation point, so that the rupture-time contours are discontinuous at the jump and approximately concentric about each $h_j$.
	\item The rupture time at $h_j$ exceeds that at $h_j^{'}$ by the delay $t_j$ of Equation~(\ref{eq:rupture-time}), and the offset between the contour values on either side of a jump is therefore a direct reading of the time-delay model.
\end{enumerate}

Executing the same geometry in the stochastic mode introduces variability in the rupture tree, the moment apportionment and the delays, and hence in the jump locations, which follow from the sampled tree, as illustrated in Figure~S6 of the electronic supplement.
Over \kFiveRealizations{} realizations of the rupture of Figure~\ref{fig:generatedRupture}, with the hypocenter held fixed, the arrival of rupture at the most distant component lies between \kFiveArrivalLow{} and \kFiveArrivalHigh{}~s (5th and 95th percentiles), against \kFiveArrivalReference{}~s in the reference mode.
The average slip of the most-slipping segment ranges from \kFiveMaxSlipLow{} to \kFiveMaxSlipHigh{}~m, against \kFiveRefSlipHigh{}~m in the reference mode, since the perturbation of Section~\ref{sec:apportionmentCalib} is applied to the segment as a whole, and a small segment can therefore receive a large share of the moment.
Under Model~\kFiveDelayBranch{} of Equation~(\ref{eq:rupture-time}), \kFiveNegativePct{}\% of the sampled delays are negative, and \kFiveTruncatedPct{}\% of draws are truncated by the causal bound.

\subsection{Behavior over a complete source model}
\label{sec:demoScale}

In order to demonstrate requirement R4 from Section~\ref{sec:usecase}, that the generator operates without expert intervention across the whole rupture set of an earthquake rupture forecast, it was applied to the multi-segment ruptures of the crustal fault model of the 2022 NZ NSHM.
The behavior of the multi-segment rupture framework is independent of the generator used to realize slip and rupture times within a connected sequence of contiguous fault planes; this generator is taken as given and the tests only exercise the novel framework.
Of the \kOneRated{} such ruptures assigned a non-zero mean annual rate of occurrence by the inversion \citep{gerstenberger_seismicity_2024}, \kOneGeometryFailed{} were removed because the sub-planes of a single participating fault cannot be assembled into one continuous surface, an inconsistency in the representation of that fault in the source model as distributed.
The generator processed all of the remaining \kOneCompleted{} ruptures to completion without intervention.

A total of \kOneRelaxedJumps{} of the \kOneJumps{} jumps ($\kOneRelaxedJumpsPct{}\%$), in \kOneRelaxedRuptures{} ruptures, have the depth constraint relaxed because one of the components does not extend below 5~km.
For example, in the rupture of Section~\ref{sec:demoSingle} it is relaxed for \jSixteenRelaxedJumps{} of the \jSixteenJumps{} jumps, each involving a component whose lower edge lies between 3 and 3.5~km.

Computational cost is not a constraint on the intended use.
Excluding the single-segment slip generation (step~6 of Section~\ref{sec:procedure}), which is unchanged from the existing generator, the time required per rupture has a median of \kOneRuntimeMedian{}\,s on a single processor core, for a total of \kOneRuntimeTotalHours{} hours over the whole population.
This distribution is strongly skewed, since the cost is dominated by the pairwise separation calculation and therefore grows with the square of the number of participating components.

Taken together with the preceding sections, this demonstrates that the four requirements of Section~\ref{sec:usecase} are satisfied, namely: the source model geometry is adopted as given (R1); the generated rupture is not unduly sensitive to the incompatibilities between independently parameterized adjoining components (R2); no fault-specific information is required beyond the participating geometries, the total magnitude and a hypocenter (R3); and the procedure is automatable over a complete rupture set and admits both deterministic and stochastic execution (R4).

%% file: sections/11-conclusions.tex
\section{Conclusions}
\label{sec:conclusions}

This manuscript has presented a generalization of kinematic rupture generation from single planar fault structures to arbitrarily complex multi-segment geometries, motivated by the incompatibility between the multi-segment rupture sets of contemporary seismic source models and the single-segment formulation of the rupture generators used in physics-based ground-motion simulation.
The approach was designed to satisfy four requirements (Section~\ref{sec:usecase}), namely that it: (R1) operate on the source model geometry as given; (R2) tolerate geometric incompatibility between adjoining components; (R3) require no fault-specific information beyond the participating fault geometries, the total magnitude and a hypocenter; and (R4) admit both deterministic and stochastic execution, at scale.
It is these requirements that determine the form of the models adopted.

A model was proposed and calibrated for each of the three aspects of the problem.
It was found that the total seismic moment must be partitioned hierarchically, between segments, which represent physically distinct structures, in proportion to $A_k^{3/2}$ as implied by a self-similar magnitude-area scaling relation, and within a segment in proportion to component area.
Apportionment in proportion to component area is the unique choice for which the average slip of a component is invariant to the delineation of components within the segment, and fitting the exponent separately to the corresponding populations of SRCMOD ruptures returns $\jThreeWithinExponent{} \pm \jThreeWithinStandardError{}$ within segments and $\jThreeBetweenExponent{} \pm \jThreeBetweenStandardError{}$ between them.
The rupture path was obtained as a weighted spanning tree of a graph whose edges carry the jumping probability of \citet{shaw_probabilities_2007}, with the maximum separation beyond which triggering is not considered inherited from the rupture-set construction of the source model.
For the multi-segment ruptures of the 2022 NZ NSHM, the maximum-likelihood path, together with the paths that differ from it by a single rerouted jump, accounts for \pctof{\jSevenMassWithinOneMean}\% of the total probability in the mean (Section~S3.4 of the electronic supplement).
Finally, the jumping time delay between segments was modeled as proportional to the jumping distance, with a scatter that grows with the jumping distance and admits negative delays.
Since the delay is the least constrained part of the framework, three alternative models were calibrated on crustal ruptures, two with fitted apparent velocities of $\tFortyOneAVelocity{}$ and $\tFortyOneBVelocity{}$\,km\,s$^{-1}$, and a third with the apparent velocity fixed at $\tFortyOneCVelocity{}$\,km\,s$^{-1}$.

The resulting multi-segment rupture generator satisfies the four design requirements set out in Section~\ref{sec:usecase}, and its application to the multi-segment rupture set of the 2022 NZ NSHM demonstrates that it can be executed without expert intervention at the scale of a national seismic hazard model, in both a deterministic reference mode and a stochastic mode.
It is also instructive to observe that the historical ruptures upon which the models have been calibrated are, as a class, geometrically more complex than the prospective ruptures to which they are applied, which is the principal basis for confidence in their extrapolation.
Nonetheless, the multi-segment rupture generator developed here, and its constituent models summarized in Figure~\ref{fig:logicTree}, can be considered a first-generation model, with limitations that arise from: (i) the sparsity of the data available for calibration, which will be relieved as the population of published multi-segment source inversions grows; (ii) modeling choices whose consequences have not been fully examined, which could be relieved now by sensitivity studies; and (iii) the decision to construct a kinematic model, which can be relieved only by extending the framework, as indicated by the potential extensions identified in Figure~\ref{fig:logicTree}.

%% file: sections/12-data-and-resources.tex
\section*{Data and Resources}
\label{sec:datacode}
\addcontentsline{toc}{section}{Data and Resources}

The finite-fault source inversions used in this study were obtained from the SRCMOD database \citep{mai_srcmod_2014}, in the MATLAB-format distribution of 2~December~2024 (\url{http://equake-rc.info/srcmod/}).
The fault geometries and rupture rates of the 2022 New Zealand National Seismic Hazard Model were obtained from \citet{gerstenberger_seismicity_2024}.
The shaded relief of the map figures is from the Copernicus GLO-30 digital elevation model (\url{https://doi.org/10.5270/ESA-c5d3d65}), as resampled onto a 30~m New Zealand Transverse Mercator grid by the New Zealand Community Velocity Model (\texttt{resources/dem.zarr}).
All code and data required to reproduce the results presented in this paper, including the distilled calibration dataset described in Section~S1 of the electronic supplement, are archived, together with the SRCMOD models in the form used, at \url{https://doi.org/10.5281/zenodo.24546050}.

%% file: sections/13-acknowledgements.tex
\section*{Acknowledgments}
\addcontentsline{toc}{section}{Acknowledgments}

This research was partially supported through QuakeCoRE: The New Zealand Centre for Earthquake Resilience, the New Zealand Natural Hazards Commission, and the New Zealand Marsden Fund.
The authors acknowledge the many investigators whose published finite-fault inversions are compiled in the SRCMOD database.
This is QuakeCoRE publication number 1296.

%% file: analysis/out/supplement/table_fields.tex
\begin{tabular}{@{}lp{0.28\linewidth}lp{0.36\linewidth}@{}}
\toprule
\textbf{Field} & \textbf{Description} & \textbf{Units} & \textbf{Read or derived} \\
\midrule
\texttt{evTAG} & Event identifier & --- & read \\
\texttt{event} & Event name & --- & read \\
\texttt{evDAT} & Event date & --- & read \\
\texttt{srcMwMoS} & Moment magnitude & --- & read (first element) \\
\texttt{invSEGM} & Number of reported components & --- & read \\
\texttt{invREF} & Inversion reference & --- & read \\
\texttt{invAUTH} & Inversion author & --- & read \\
\texttt{srcAStke, segNAStke} & Component strike & deg & read \\
\texttt{srcDipAn, segNDipAn} & Component dip & deg & read \\
\texttt{srcDimWL, segNDimWL} & Component width and length & km & read \\
\texttt{geoLAT, geoLON} & Latitude and longitude of the sub-fault top-center & deg & read \\
\texttt{geoX, geoY, geoZ} & Local coordinates of the sub-fault top-center & km & read \\
\texttt{srcSLIP, slipSPL} & Sub-fault slip & m & read \\
\texttt{srcTRUP, srcTIME} & Sub-fault rupture onset & s & read where reported \\
\texttt{srcTRISE, srcRISE} & Sub-fault rise time & s & read where reported \\
$dx$, $dz$ & Sub-fault dimensions & km & derived: $L/n_{\text{strike}}$, $W/n_{\text{dip}}$, or from adjacent node spacing \\
$A_i$ & Component area & km$^2$ & derived: $n_{\text{cells}}\,dx\,dz$ after trimming \\
$M_{o,i}$ & Component moment & N\,m & derived: $\mu A_i \bar{D}_i$ with $\mu = 3.3\times10^{10}$\,Pa \\
\bottomrule
\end{tabular}

%% file: analysis/out/supplement/table_events_jump_delay.tex
\begin{longtable}{@{}lp{0.15\linewidth}lllp{0.35\linewidth}@{}}
\caption{Models comprising the jump-position and delay subset, with the event name and date as recorded by SRCMOD and the number of components $n$ in each model.}\tabularnewline
\label{tab:sup-events-jump-delay}\tabularnewline
\toprule
\textbf{Model ID} & \textbf{Event} & \textbf{Date} & \textbf{$M_w$} & \textbf{$n$} & \textbf{Inversion reference} \tabularnewline
\midrule
\endfirsthead
\multicolumn{6}{@{}l}{\footnotesize\textit{Jump-position and delay subset --- continued}} \tabularnewline
\toprule
\textbf{Model ID} & \textbf{Event} & \textbf{Date} & \textbf{$M_w$} & \textbf{$n$} & \textbf{Inversion reference} \tabularnewline
\midrule
\endhead
\bottomrule
\endfoot
\texttt{s1948FUKUIJ01ICHI} & \raggedright Fukui (Japan) & 06/28/1948 & 6.65 & 2 & \raggedright Ichinose, G. A., P. Somerville, H. K. Thio, S. Matsushima, and T. Sato. 2005. Rupture process of the 1948 Fukui earthquake (M 7.1) from the joint inversion of seismic waveform and geodetic data. J. Geophys. Res., 110 (B5) B05301, doi:10.1029/2004JB003437. \tabularnewline
\texttt{s1999DUZCET01DELO} & \raggedright Duzce (Turkey) & 11/12/1999 & 7.18 & 3 & \raggedright Delouis, B., P. Lundgren, and D. Giardini. 2004. Slip distributions of the 1999 Düzce (Mw 7.2) and Izmit (Mw 7.6) earthquakes on the North Anatolian Fault (Turkey): A combined analysis. internal report, 2004. \tabularnewline
\texttt{s1999HECTOR01JIxx} & \raggedright Hector Mine (Calif.) & 10/16/1999 & 7.17 & 3 & \raggedright Ji, C., D. J. Wald, and D. V. Helmberger. 2002. Source description of the 1999 Hector Mine, California, earthquake, part II: Complexity of slip history. Bull. Seis. Soc. Am 92 (4):1208-1226. \tabularnewline
\texttt{s1999HECTOR01SALI} & \raggedright Hector Mine (Calif.) & 10/16/1999 & 7.14 & 4 & \raggedright Salichon, J., et al. 2004. Slip history of the 16 October 1999 Mw 7.1 Hector Mine earthquake (California) from the inversion of InSAR, GPS, and teleseismic data. Bull. Seis. Soc. Am 94 (6):2015-2027. doi:10.1785/0120030038. \tabularnewline
\texttt{s1999IZMITT01BOUC} & \raggedright Izmit (Turkey) & 08/17/1999 & 7.59 & 8 & \raggedright Bouchon, M., M. N. Toksoz, H. Karabulut, M. P. Bouin, M. Dietrich, M. Aktar, and M. Edie. 2002. Space and time evolution of rupture and faulting during the 1999 Izmit (Turkey) earthquake. Bull. Seis. Soc. Am 92 (1):256-266. \tabularnewline
\texttt{s1999IZMITT01DELO} & \raggedright Izmit (Turkey) & 08/17/1999 & 7.56 & 4 & \raggedright Delouis, B., D. Giardini, P. Lundgren, and J. Salichon. 2002. Joint inversion of InSAR, GPS, teleseismic, and strong-motion data for the spatial and temporal distribution of earthquake slip: Application to the 1999 Izmit mainshock. Bull. Seis. Soc. Am 92 (1):278-299. \tabularnewline
\texttt{s1999IZMITT01SEKI} & \raggedright Izmit (Turkey) & 08/17/1999 & 7.44 & 4 & \raggedright Sekiguchi, H., and T. Iwata. 2002. Rupture process of the 1999 Kocaeli, Turkey, earthquake estimated from strong-motion waveforms. Bull. Seis. Soc. Am 92 (1):300-311. \tabularnewline
\texttt{s2000TOTTOR01SEMM} & \raggedright Tottori (Japan) & 10/06/2000 & 6.73 & 2 & \raggedright Semmane, F., F. Cotton, and M. Campillo. 2005. The 2000 Tottori earthquake: A shallow earthquake with no surface rupture and slip properties controlled by depth. J. Geophys. Res. 110 (B3): B03306, doi:10.1029/2004JB003194. \tabularnewline
\texttt{s2002DENALI01HAYE} & \raggedright Denali (Alaska) & 11/03/2002 & 7.97 & 5 & \raggedright Hayes NEIC, Alaska Central 2015. Preliminary Finite Fault Results for the Nov 03, 2002 Mw 8.0 76 km ENE of Cantwell, Alaska Earthquake (Version 1), \url{http://earthquake.usgs.gov/earthquakes/eventpage/usp000bg0m#finite-fault} \tabularnewline
\texttt{s2011FUKUSH01TANA} & \raggedright Fukushima Hamadori, Japan & 04/11/2011 & 6.68 & 2 & \raggedright Tanaka, M., K. Asano, T. Iwata, and H. Kubo. 2014. Source rupture process of the 2011 Fukushima-ken Hamadori earthquake: how did the two subparallel faults rupture? Earth Planets Space 66:101. doi:10.1186/1880-5981-66-101. \tabularnewline
\texttt{s2013BALOCH01AVOU} & \raggedright Balochistan, Pakistan & 09/24/2013 & 7.70 & 7 & \raggedright Avouac, J.-P., F. Ayoub, S. Wei, J.-P. Ampuero, L. Meng, S. Leprince, R. Jolivet, Z. Duputel, and D. Helmberger. 2014. The 2013, Mw 7.7 Balochistan earthquake, energetic strike-slip reactivation of a thrust fault. Earth and Planetary Science Letters 391, 128-134. \tabularnewline
\texttt{s2013BALOCH01HAYE} & \raggedright Balochistan, Pakistan & 09/24/2013 & 7.72 & 5 & \raggedright Hayes, G.P., 2017. The finite, kinematic rupture properties of great-sized earthquakes since 1990. Earth and Planetary Science Letters, 468, pp.94-100. \tabularnewline
\texttt{s2015MURGHA01HAYE} & \raggedright Murghab, Tajikistan & 12/07/2015 & 7.17 & 3 & \raggedright Hayes, G.P., 2017. The finite, kinematic rupture properties of great-sized earthquakes since 1990. Earth and Planetary Science Letters, 468, pp.94-100. \tabularnewline
\texttt{s2016KUMAMO01KOBA} & \raggedright Kumamoto, Japan & 04/14/2016 & 6.19 & 3 & \raggedright Kobayashi, H., Koketsu, K., \& Miyake, H. (2017). Rupture processes of the 2016 Kumamoto earthquake sequence: Causes for extreme ground motions. Geophysical Research Letters, 44, 6002-6010. \url{https://doi.org/10.1002/2017GL073857} \tabularnewline
\texttt{s2016KUMAMO02ASAN} & \raggedright Kumamoto, Japan & 04/15/2016 & 7.04 & 2 & \raggedright Asano, K. and T. Iwata. 2016. Source rupture processes of the foreshock and mainshock in the 2016 Kumamoto earthquake sequence estimated from the kinematic waveform inversion of strong motion data. Earth Planets Space 68:147, doi:10.1186/s40623-016-0519-9. \tabularnewline
\texttt{s2016KUMAMO02KOBA} & \raggedright Kumamoto, Japan & 04/14/2016 & 6.09 & 2 & \raggedright Kobayashi, H., Koketsu, K., \& Miyake, H. (2017). Rupture processes of the 2016 Kumamoto earthquake sequence: Causes for extreme ground motions. Geophysical Research Letters, 44, 6002-6010. \url{https://doi.org/10.1002/2017GL073857} \tabularnewline
\texttt{s2016KUMAMO03ASAN} & \raggedright Kumamoto, Japan & 04/15/2016 & 7.06 & 11 & \raggedright Asano, K. and T. Iwata. 2021. Revisiting the source rupture process of the mainshock of the 2016 Kumamoto earthquake and implications for the generation of near-fault ground motions and forward‐directivity pulse. Bull. Seismol. Soc. Am. 111(5): 2426–2440. \tabularnewline
\texttt{s2016KUMAMO03KOBA} & \raggedright Kumamoto, Japan & 04/15/2016 & 7.02 & 6 & \raggedright Kobayashi, H., Koketsu, K., \& Miyake, H. (2017). Rupture processes of the 2016 Kumamoto earthquake sequence: Causes for extreme ground motions. Geophysical Research Letters, 44, 6002-6010. \url{https://doi.org/10.1002/2017GL073857} \tabularnewline
\texttt{s2016SOUTHI01HAYE} & \raggedright South Island, New Zealand & 11/13/2016 & 7.89 & 5 & \raggedright Hayes, G.P., 2017. The finite, kinematic rupture properties of great-sized earthquakes since 1990. Earth and Planetary Science Letters, 468, pp.94-100. \tabularnewline
\texttt{s2018IBURIH01ASAN} & \raggedright Iburi, Hokkaido, Japan & 09/05/2018 & 6.80 & 3 & \raggedright Asano, K. and T. Iwata. 2019. Source rupture process of the 2018 Hokkaido Eastern Iburi earthquake deduced from strong-motion data considering seismic wave propagation in three-dimensional velocity structure. Earth Planets Space 71:101, doi:10.1186/s40623-019-1080-0. \tabularnewline
\texttt{s2018IBURIH01KOBA} & \raggedright Iburi, Hokkaido, Japan & 09/05/2018 & 6.63 & 2 & \raggedright Kobayashi, H., Koketsu, K., \& Miyake, H. (2019). Rupture process of the 2018 Hokkaido Eastern Iburi earthquake derived from strong motion and geodetic data. Earth, Planets and Space, 71, 63. \url{https://doi.org/10.1186/s40623-019-1041-7} \tabularnewline
\end{longtable}

%% file: analysis/out/supplement/table_schema.tex
\paragraph{\texttt{ruptures.csv}}
One record per rupture; 472 records.

\begin{tabular}{lp{0.62\linewidth}}
\toprule
\textbf{Column} & \textbf{Meaning and units} \\
\midrule
\texttt{model} & SRCMOD model ID (event tag), e.g. s2016KUMAMO03ASAN \\
\texttt{event} & Event name, from the SRCMOD MATLAB field event (e.g. Hector Mine (Calif.)) \\
\texttt{date} & Event date, MM/DD/YYYY as recorded by SRCMOD \\
\texttt{magnitude} & Moment magnitude Mw, as reported by the inversion \\
\texttt{n\_segments} & Number of components (SRCMOD segments) in the model, after planes with no slip are deleted \\
\texttt{population} & single $\vert$ multi $\vert$ excluded, by component count, n\_segments (see the component/patch rule) \\
\texttt{area\_km2} & Total rupture area after trimming, km$^{2}$ \\
\texttt{moment\_nm} & Total seismic moment, N m, at mu = 3.3e10 Pa \\
\texttt{reference} & Inversion reference, as recorded by SRCMOD; the placeholder entries of s1999HECTOR01SALI and s2011LORCAx01LOPE are replaced by the published reference \\
\texttt{author} & Inversion author, as recorded by SRCMOD \\
\texttt{trim\_method} & Trimming heuristic applied (somerville) \\
\texttt{trim\_keep\_top} & 1 if the shallowest row was preserved \\
\bottomrule
\end{tabular}

\paragraph{\texttt{segments.csv}}
One record per component; 866 records.

\begin{tabular}{lp{0.62\linewidth}}
\toprule
\textbf{Column} & \textbf{Meaning and units} \\
\midrule
\texttt{model} & SRCMOD model ID \\
\texttt{segment} & Zero-based component (SRCMOD segment) index within the model \\
\texttt{area\_km2} & Component area after trimming, km$^{2}$ \\
\texttt{mean\_slip\_m} & Mean slip over retained cells, m \\
\texttt{moment\_nm} & Component moment, N m, at mu = 3.3e10 Pa \\
\texttt{area\_fraction} & Component area divided by total rupture area \\
\texttt{moment\_fraction} & Component moment divided by total rupture moment \\
\texttt{n\_cells} & Number of retained subfault cells \\
\texttt{dx\_km} & Along-strike cell dimension, km \\
\texttt{dz\_km} & Down-dip cell dimension, km \\
\texttt{trim\_status} & trimmed $\vert$ no-op $\vert$ failed $\vert$ skipped-* (planes with no slip are deleted) \\
\texttt{retained\_fraction} & Cells retained by trimming, as a fraction \\
\bottomrule
\end{tabular}

\paragraph{\texttt{jump\_pairs.csv}}
One record per jump pair and kind; 219 records.

\begin{tabular}{lp{0.62\linewidth}}
\toprule
\textbf{Column} & \textbf{Meaning and units} \\
\midrule
\texttt{model} & SRCMOD model ID \\
\texttt{to\_segment} & Index of the triggered component \\
\texttt{from\_segment} & Index of the triggering component \\
\texttt{distance\_km} & 3-D distance between jump point and conjugate point, km \\
\texttt{delay\_s} & Rupture delay across the jump, s \\
\texttt{to\_depth\_km} & Depth of the jump point, km \\
\texttt{from\_depth\_km} & Depth of the conjugate point, km \\
\texttt{to\_trup\_s} & Rupture onset at the jump point, s after nucleation \\
\texttt{from\_trup\_s} & Rupture onset at the conjugate point, s \\
\texttt{kind} & geometric (conjugate point is the nearest cell on any other component, irrespective of onset; the delay calibration) $\vert$ inferred (nearest cell on a component that ruptured earlier; the jump-location comparison only) $\vert$ algorithmic (the generator's closest-point construction for the inferred pair) \\
\texttt{n\_segments} & Number of components (SRCMOD segments) in the parent model, after planes with no slip are deleted \\
\texttt{to\_x} & Jump point, easting in the model's local frame, km \\
\texttt{to\_y} & Jump point, northing in the model's local frame, km \\
\texttt{to\_z} & Jump point depth, km \\
\texttt{from\_x} & Conjugate point, easting in the model's local frame, km \\
\texttt{from\_y} & Conjugate point, northing in the model's local frame, km \\
\texttt{from\_z} & Conjugate point depth, km \\
\texttt{to\_segment\_length\_km} & Along-strike extent of the triggered component, km \\
\texttt{overlap\_km} & Along-strike overlap of the two components on the triggering component's strike, km (negative = gap) \\
\texttt{strike\_difference\_deg} & Axial strike difference between the two components, degrees \\
\texttt{dip\_difference\_deg} & Dip difference between the two components, degrees \\
\texttt{arrangement} & end-to-end $\vert$ overlapping (overlap $>$ 2 km) $\vert$ oblique (strike or dip differs $>$ 30 deg) \\
\texttt{setting} & crustal $\vert$ interface $\vert$ intraslab (analysis/resources/tectonic\_setting.csv) \\
\texttt{calibration} & Eligible for the delay calibration: geometric, both points $\ge$ 5 km deep, distance $\le$ 15 km (all settings; the fit uses setting == crustal) \\
\texttt{within\_limit} & Inferred jump of this triggered component is $\le$ 15 km (jump-location comparison) \\
\bottomrule
\end{tabular}

%% file: analysis/out/supplement/table_darfield_edges.tex
\begin{tabular}{llrrc}
\hline
\textbf{Component} & \textbf{Component} & \textbf{$r$ (km)} & \textbf{$P$} & \textbf{Modal} \\
\hline
Greendale NW SE & Greendale central & 0.00 & 0.990 &  \\
Greendale NW SE & Charing Cross & 0.00 & 0.990 & $\bullet$ \\
Greendale central & Fault NE & 0.10 & 0.990 & $\bullet$ \\
Greendale East & Greendale central & 0.53 & 0.990 &  \\
Greendale East & Fault NE & 0.12 & 0.990 & $\bullet$ \\
Greendale central & Charing Cross & 0.00 & 0.990 & $\bullet$ \\
Charing Cross & Fault NNW & 0.00 & 0.990 & $\bullet$ \\
Greendale NW SE & Fault Near Hororata & 1.36 & 0.887 & $\bullet$ \\
Greendale central & Fault NNW & 4.75 & 0.287 &  \\
Greendale NW SE & Fault NNW & 4.84 & 0.278 &  \\
Fault NNW & Fault NE & 7.25 & 0.125 &  \\
Greendale East & Fault NNW & 7.52 & 0.114 &  \\
Charing Cross & Fault NE & 8.47 & 0.083 &  \\
Greendale East & Charing Cross & 8.62 & 0.079 &  \\
Greendale central & Fault Near Hororata & 10.17 & 0.047 &  \\
Fault Near Hororata & Charing Cross & 11.91 & 0.026 &  \\
Greendale NW SE & Fault NE & 12.02 & 0.025 &  \\
Greendale NW SE & Greendale East & 12.22 & 0.024 &  \\
\hline
\end{tabular}